\documentclass[a4paper,11pt]{article}

\usepackage{jheppub} 
                     
\usepackage{float}
\usepackage{ulem}
\usepackage{cancel}
\usepackage{enumitem}

\usepackage{color}
\usepackage{graphicx}

\usepackage{shuffle}

\usepackage{jheppub,psfrag,slashed,cancel,lscape,caption,array,graphicx,subcaption}
\usepackage[utf8]{inputenc}
\usepackage{amsmath}
\usepackage{bm}
\usepackage{mathtools}
\usepackage{mathrsfs}
\usepackage{multirow}
\usepackage{booktabs} 
\usepackage{caption}	
\usepackage{hyperref}

\usepackage{chngcntr}

\renewcommand{\thefootnote}{\fnsymbol{footnote}}
\renewcommand{\thanks}[1]{\footnote{#1}}

\newcommand{\bea}{\begin{eqnarray}}
\newcommand{\eea}{\end{eqnarray}}
\newcommand{\ee}{\end{equation}}
\newcommand{\be}{\begin{equation}}

\newcommand{\no}{\nonumber}

\def\eqn#1{eq.~(\ref{#1})}

\def\cA{{\cal A}}

\def\cD{{\cal D}}

\def\cF{{\cal F}}

\def\cL{{\cal L}}
\def\cM{{\cal M}}
\def\cN{{\cal N}}

\def\cP{{\cal P}}

\def\cR{{\cal R}}

\def\ep{\varepsilon}

\newcommand\nn{\nonumber}

\newcommand\spaq[1]{\langle #1\rangle}
\newcommand\spbq[1]{[#1]}

\def\spa#1.#2{\left\langle#1\,#2\right\rangle}
\def\spb#1.#2{\left[#1\,#2\right]}
\def\eqn#1{eq.~(\ref{#1})}

\def\be{\begin{equation}}
\def\ee{\end{equation}}
\def\bea{\begin{eqnarray}}
\def\eea{\end{eqnarray}}

\def\bep{\boldsymbol{\mathbf{\ep}}}
\def\ep{\varepsilon}
\def\te{\tilde \eta}

\def\Nima{\cite{Arkani-Hamed:2017jhn}}
\def\BernSpin{\cite{Bern:2020buy}}
\def\HSQED{\cite{Deser:2000dz}}
\def\HenrikAlex{\cite{Johansson:2019dnu}}
\def\HSlectures{\cite{Rahman:2013sta}}
\def\EMUnitarity{\cite{Ferrara:1992yc}}
\def\GravUnitarity{\cite{Cucchieri:1994tx}}
\def\Berends{\cite{Berends:1979rv}}

\def\SHFermion{\cite{Singh:1974rc}}

\def\angle#1{\langle #1 \rangle}
\def\Cspur{[ 4 | p_1 | 3 \rangle}
\def\Cres{[4 \bm1] \langle 3 \bm2 \rangle + [4 \bm2] \langle 3 \bm1 \rangle}
\def\Cresspurb{[\bm1 4] [\bm2 4] \langle \bm1 3 \rangle \langle \bm2 3 \rangle}
\def\feps{\mathfrak{F}}
\def\MSHquada{N_2}
\def\MSHquadb{N'_2}
\def\MSHquart{N_4}
\def\nA{Q}

\def\vmu{\vec{\mu}}
\def\vnu{\vec{\nu}}
\def\z{z}
\def\bH{{\overline H}}
\def\S{s}
\def\fermion{\lambda}
\def\AxxX{A_{0\oplus1/2}}
\def\MxxX{M_{0\oplus1/2}}
\def\AXXX{A_{1\oplus\,3/2}}
\def\w{w}
\def\Svec{S}
\def\Sop{\hat S}
\def\SOP{\mathbb{S}}
\def\xfactor{x}

\title{\boldmath
Compton Black-Hole Scattering for $\S\leq 5/2$}

\author[a]{Marco Chiodaroli,}
\author[a,b]{Henrik Johansson,}
\author[a]{and Paolo Pichini}

\affiliation[a]{Department of Physics and Astronomy, Uppsala University, \\ Box 516, 75120 Uppsala, Sweden}
\affiliation[b]{Nordita, Stockholm University and KTH Royal Institute of Technology, \\ Hannes Alfv\'{e}ns v\"{a}g 12, 10691 Stockholm, Sweden}

\emailAdd{marco.chiodaroli@physics.uu.se}
\emailAdd{henrik.johansson@physics.uu.se}
\emailAdd{paolo.pichini@physics.uu.se}

\abstract{Quantum scattering amplitudes for massive matter have received new attention in connection to classical calculations relevant to gravitational-wave physics. Amplitude methods and insights are now employed for precision computations of observables needed for describing the gravitational dynamics of bound massive objects such as black holes. An important direction is the inclusion of spin effects needed to accurately describe rotating (Kerr) black holes. Higher-spin amplitudes introduced by Arkani-Hamed, Huang and Huang at three points have by now a firm connection to the effective description of Kerr black-hole physics. The corresponding Compton higher-spin amplitudes remain however an elusive open problem. Here we draw from results of the higher-spin literature and show that physical insights can be used to uniquely fix the Compton amplitudes up to spin~$5/2$, by imposing a constraint on a three-point higher-spin current that is a necessary condition for the existence of an underlying unitary theory. We give the unique effective Lagrangians up to spin~$5/2$, and show that they reproduce the previously-known amplitudes.  For the multi-graviton amplitudes analogous to the Compton amplitude, no further corrections to our Lagrangians are expected, and hence such amplitudes are uniquely predicted.  
As an essential tool, we introduce a modified version of the massive spinor-helicity formalism which allows us to conveniently obtain higher-spin states, propagators and compact expressions for the amplitudes.  
}

\preprint{ UUITP-34/21 \\
\phantom{~} \hfill NORDITA 2021-013}

\begin{document}
\maketitle
\flushbottom

\renewcommand{\thefootnote}{\arabic{footnote}}

\section{Introduction}

More than a century after its inception, General Relativity continues to reap successes by accurately describing new physical phenomena, such as gravitational waves from inspiraling black holes~\cite{LIGOScientific:2016aoc} and neutron stars~\cite{LIGOScientific:2017vwq}. However, despite having received considerable attention, the classical scattering of massive spinning objects in General Relativity is still poorly understood. The classical scattering problem is closely related to bound-orbit dynamics, as both depend on the calculation of a two-body effective potential, which may include radiative effects. In the regime in which the compact objects are well separated, the effective potential is most naturally presented in a post-Minkowskian or post-Newtonian expansion.  
 
The seemingly-unrelated problem of quantum scattering amplitudes, accompanied by a classical limit procedure, has provided a versatile novel perspective to tackle these challenges
\cite{BjerrumBohr:2002kt,
Bjerrum-Bohr:2013bxa, 	
Neill:2013wsa, 
Akhoury:2013yua, 
Vaidya:2014kza,
Luna:2016due, 
Luna:2016idw, 
Luna:2017dtq,
Cachazo:2017jef, 
Damour:2017zjx,
Bjerrum-Bohr:2018xdl, 
Cheung:2018wkq, 
Bern:2019nnu, 
KoemansCollado:2019ggb, 
Arkani-Hamed:2019ymq,
Cristofoli:2019neg,
Bern:2019crd, 
Damgaard:2019lfh,
DiVecchia:2019kta, 
Damour:2019lcq,
Bjerrum-Bohr:2019kec,
Bern:2020gjj, 
Chung:2020rrz,
Parra-Martinez:2020dzs,
Aoude:2020mlg,
Bini:2020uiq, 
DiVecchia:2020ymx,
Damour:2020tta,
Bern:2020uwk,
Aoude:2020ygw, 
AccettulliHuber:2020dal, 
Bini:2020rzn, 
Bern:2021dqo,
Herrmann:2021lqe,
Kosmopoulos:2021zoq, 
Bjerrum-Bohr:2021vuf,
DiVecchia:2021bdo,
Herrmann:2021tct,
Cristofoli:2021vyo,
Bautista:2021wfy, 
Chen:2021huj}.\footnote{See also refs. \cite{Iwa,Oka,Donoghue:1993eb,Donoghue:1994dn} for earlier related work.}
Amplitude methods draw from ideas that originate from the  effective-field-theory (EFT) approach to quantum theories and are carried over to gravitational physics~\cite{Goldberger:2004jt, 
Goldberger:2007hy, 
Kol:2007bc,
Goldberger:2009qd,
Foffa:2013qca,
Porto:2016pyg,
Foffa:2016rgu,Levi:2018nxp}. Like in other analytical approaches, the goal is to describe the inspiral phase of a merging binary system, thus taking full advantage of the separation of scales inherent to the problem.  

Powerful techniques developed for amplitude computations over the past three decades are now being routinely employed for calculations in classical gravity, including Britto-Cachazo-Feng-Witten (BCFW) on-shell recursion~\cite{Britto:2005fq}, loop-level unitarity methods \cite{Bern:1994zx,Bern:1994cg,Bern:1995db,Bern:1997sc} and the double-copy construction~\cite{Kawai:1985xq,Bern:2008qj,Bern:2010ue} (see ref.~\cite{Bern:2019prr} for a comprehensive review). 
Most notably, scattering-amplitude methods have been instrumental in the calculation of the two-body effective potential up to the third and fourth post-Minkowskian 
orders $\mathcal{O}(G_N^3)$ and $\mathcal{O}(G_N^4)$ for non-spinning black-hole-like objects 
\cite{Bern:2019nnu,  
Bern:2019crd, 
Bern:2020gjj, 
Bern:2021dqo},  where $G_N$ is Newton's constant.\footnote{See also current state-of-the-art post-Newtonian calculations of the conservative effective potential up to fifth and sixth orders~\cite{Bini:2019nra,Blumlein:2020znm,Bini:2020wpo,Bini:2020nsb}. While the connection of new techniques to more traditional approaches deserves an in-depth discussion, this is beyond of the scope of this paper.
}
The two-body effective potential can in turn be used as analytical input for the calculation of gravitational-wave templates (see e.g. refs.~\cite{Antonelli:2019ytb,Buonanno:2014aza}). 
 
While many of the current applications of amplitude techniques concern scattering of objects without spin (e.g.~Schwarzschild black holes), an important research direction aims at quantifying the spin effects of rotating compact objects (e.g.~Kerr black holes).  
In the EFT framework, the study of spin effects was initiated in ref.
\cite{Porto:2005ac}
and further developed over the following years~\cite{Porto:2006bt,
Porto:2008tb,
Porto:2008jj,
Levi:2008nh, 
Porto:2010tr,
Porto:2010zg,
Levi:2010zu,
Levi:2011eq,
Porto:2012as,
Levi:2014gsa,
Levi:2014sba, 
Levi:2015msa, 
Levi:2015uxa, 
Levi:2015ixa, 
Levi:2016ofk,
Maia:2017yok, 
Maia:2017gxn,
Levi:2017oqx,
Levi:2017kzq,
Levi:2018nxp,
Levi:2019kgk,
Kalin:2019rwq,
Kalin:2019inp,
Levi:2020kvb,
Levi:2020uwu, 
Levi:2020lfn,
Liu:2021zxr,
Cho:2021mqw,
Jakobsen:2021lvp}.
More recently, quantum scattering amplitudes involving particles of arbitrary spin were re-analyzed by Arkani-Hamed, Huang and Huang using the massive spinor-helicity formalism~\cite{Arkani-Hamed:2017jhn}. Those results gave rise to an emergent research direction that seeks to establish a link between classical spin corrections to the two-body effective potential and quantum scattering amplitudes for spinning matter~\cite{Guevara:2017csg, 
Vines:2017hyw,
Vines:2018gqi,
Guevara:2018wpp, 
Chung:2018kqs, 
Guevara:2019fsj, 
Arkani-Hamed:2019ymq, 
Chung:2019duq,
Siemonsen:2019dsu, 
Guevara:2020xjx, 
Aoude:2020onz}
~(see also related work~\cite{Kosower:2018adc, 
Maybee:2019jus,Damgaard:2019lfh, Huang:2019cja, Chung:2019yfs, Haddad:2020tvs, Emond:2020lwi,Brandhuber:2021kpo}). 
In particular, certain three-point amplitudes found in ref.~\cite{Arkani-Hamed:2017jhn} were later shown to reproduce the classical scattering of spinning Kerr black holes via exponentiation of soft-graviton exchanges~\cite{Guevara:2018wpp}, as well as through other direct matchings~\cite{Chung:2018kqs,Arkani-Hamed:2019ymq, Aoude:2020onz}.

Quantum amplitudes for massive particles of spin~$\S$ can be used to access spin corrections of power $(\Svec^\mu)^{2\S}$ in the two-body effective potential, where $\Svec^\mu$ is the classical spin vector. Such calculations will in general need input from loop-level Feynman diagrams where the classical limit is taken such that the mass, momentum and spin of each scattered object become large macroscopic quantities. Alternatively, by using the factorization properties of loop diagrams and taking into account that  the large scale separation of the massive objects implies small graviton exchange momenta, it is possible to show that all the necessary information for computing the classical process is contained within tree-level scattering between a single black hole (counted as two states in an out-out formalism) and $n{-}2$ gravitons. 

The natural higher-spin three-point amplitudes were sewn together via BCFW on-shell recursion to produce simple candidates for the four-point Compton amplitudes in refs.~\cite{Arkani-Hamed:2017jhn,Johansson:2019dnu}. However, for spin~$\S\ge2$, spurious poles develop in the kinematic space of the Compton amplitudes, indicating that certain contact terms are not correctly included by the naive factorization.\footnote{BCFW recursion should, in principle, also include residues at infinity of the complex plane, which would remove the spurious poles. In general, these contributions cannot be a priori calculated.} Indeed, it is well-known that factorization properties alone do not completely fix higher-point tree amplitudes, which is especially true for generic spin.  In this case, the ambiguity in the Compton amplitude is linked to the proliferation of possible Lagrangian contact terms that respect all known symmetries of the problem.  Thus, it becomes important to find physical principles to constrain amplitudes that go beyond three-point factorization. 

A well-established approach for constraining amplitudes is to consider the high-energy limit, as done for the three-point amplitudes of ref.~\cite{Arkani-Hamed:2017jhn}, where they were deemed natural based on having a certain good high-energy behavior. A further interpretation that can be inferred from  ref.~\cite{Arkani-Hamed:2017jhn} is that these three-point amplitudes provide a ``minimal''  description of the higher-spin states,\footnote{In this paper, we use the term ``minimal coupling'' to describe interactions to matter that are induced by covariantizing the derivatives of the free-theory Lagrangian, and this should not  be confused with the similar term used to describe the amplitudes in ref.~\cite{Arkani-Hamed:2017jhn}, which require non-minimal Lagrangian terms for spin large enough.} and that the inclusion of other possible three-point contributions would correspond to having additional physical quantities (e.g.~Wilson coefficients) that further characterize the massive spinning states. While these are very reasonable properties that seem to resonate with what we know about black holes, it is important to establish the precise constraints that should be imposed on amplitudes at higher points. Furthermore, it would be desirable to have a covariant Lagrangian understanding of the properties of the amplitudes in ref. \cite{Arkani-Hamed:2017jhn} and, in general, determine what makes them special.

The higher-spin literature has long studied similar theories from a different angle 
\cite{Ferrara:1992yc,
Cucchieri:1994tx,
Klishevich:1997pd,
Giannakis:1998wi,
Buchbinder:2000ta,
Buchbinder:1999be,
Deser:2001dt,
Metsaev:2007rn,
Francia:2007ee,
Porrati:2008ha, 
Porrati:2008gv,
Francia:2008ac,
Sagnotti:2010at,
Hassan:2012wr,
Buchbinder:2012iz,
Cortese:2013lda,
Bernard:2015uic,
Rahman:2015pzl,
Fukuma:2016rru,
Bonifacio:2018aon,Afkhami-Jeddi:2018apj,Kaplan:2020ldi}
(for a review, see ref. \cite{Rahman:2013sta}). 
Theories with massive higher-spin fields and Minkowski background are understood as effective theories that exist below a certain energy scale. This occurs when the spin-$\S$ fields are coupled to gravity ($\S>2$), or to electromagnetism ($\S>1$), and thus the force carrier is not the field of highest spin. The presence of an intrinsic cutoff in these theories is linked to acausal behavior at high energy and, ultimately, leads to an obstruction to having a local fundamental theory with a finite number of elementary higher-spin fields. 
Cucchieri, Deser and Porrati showed in ref.~\cite{Cucchieri:1994tx} that a theory with elementary higher-spin fields of mass $m$ violates tree-level unitarity at energies well below the Planck scale. They attempted to restore tree-level unitarity by adding non-minimal higher-derivative terms $\sim \partial/m$ to the Lagrangian, and argued that a necessary condition for the theory to be well defined up to the Planck scale is that the longitudinal part of the higher-spin current $J^\mu$ vanishes in the $m\rightarrow 0$ limit,\footnote{This is equivalent to requiring higher-spin gauge invariance in the massless limit. For spin~$\S \geq 5/2$ this formula will be restricted to the traceless part, as we explain later.}
\begin{equation}
 \partial_\mu J^\mu = {\cal O} (m) \ .  \label{eqn1.1}
\end{equation}
Since the higher-spin current can be used to compute scattering amplitudes after dressing the free indices with on-shell polarizations, this condition has a direct link to the problem of finding natural and well-behaved Compton amplitudes. 

In this paper, we argue that the problem of fixing contact-term ambiguities of spinning black-hole Compton amplitudes is directly tied to established results in the higher-spin literature. We show that, given some standard physical assumptions on higher-derivative operators, the condition (\ref{eqn1.1}) uniquely fixes three-point and Compton amplitudes for massive spin~$\S \le 5/2$ fields in General Relativity and, analogously, for massive charged spin~$\S \le 3/2$ fields coupled to electromagnetism. The gravitational amplitudes that we find should be useful for computing $(\Svec^\mu)^{\le 5}$ corrections in the two-body effective potential for binary Kerr black holes. We also give the Lagrangians for the spin~$\S \le 5/2$ fields coupled to General Relativity, and argue that there exist no further terms that can be added to improve the tree-level unitarity properties (as addressed by \eqn{eqn1.1}). 

It should be emphasized that the kinematical regime in which classical black-hole calculations are carried out (mass and black-hole momenta large in Planck units) is quite different from the one originally envisioned in the higher-spin literature, which has masses well-below the Planck scale. This seems to require some mild reinterpretation of the precise connection between the tree-level unitarity argument and the (quantum) effective theory perspective.
In order to practically calculate observables relevant to classical black-hole physics, the Lagrangians given in this paper are ultimately meant to be used in a classical limit where the momentum transfer (carried by gravitons) is smaller than the inverse Schwarzschild radius, which is in turn is smaller than the inverse Compton length of the black hole.
This kinematical limit may be taken after obtaining the loop integrand of the amplitude under consideration, which can be done via Feynman diagrams or by unitarity-cut methods. In Section~\ref{Classical_limit_section}, the classical limit for tree-level amplitudes is briefly reviewed in our notation but, for a thorough exhibition of this procedure at the multi-loop level, we refer the reader to the unitarity-cut methods used in refs.~\cite{Vaidya:2014kza, Guevara:2017csg, Bern:2019crd,Bern:2019nnu,Bern:2020buy, Kosmopoulos:2021zoq,Herrmann:2021lqe,Herrmann:2021tct}.

In this paper, we make heavy use of a novel notation obtained as an extended version of the massive spinor-helicity formalism~\cite{Arkani-Hamed:2017jhn, Craig:2011ws}. In particular, the massive spinor-helicity formalism is simplified by dressing the little-group indices with auxiliary variables, which turn  expressions involving spinors or polarizations into polynomials, or in some situations, rational functions. Higher-spin states and amplitudes can be easily constructed in this notation, and through the completeness relation of such states one can  infer important details of the off-shell propagators for higher-spin fields. As a non-trivial application, we construct generating functions for all higher-spin projectors, as well as all covariant three-point amplitudes that match the known spinor-helicity expressions~\cite{Arkani-Hamed:2017jhn}.

The paper is structured as follows: in Section \ref{sec2}, we review our notation and introduce an extended version of the spinor-helicity formalism. In Section \ref{sec3}, we review the massive higher-spin three-point amplitudes in both spinor-helicity  and covariant notation, and show how to use the current constraint  (\ref{eqn1.1}) to build Ans\"{a}tze that uniquely reproduce these amplitudes up to spin~$5/2$. In Section \ref{SECT:msh}, we review the Compton amplitudes and display the new spurious-pole-free results. In Section \ref{sec5}, we consider Lagrangians that are compatible with the current constraint (\ref{eqn1.1}), and show that they are unique for the theories that we are considering. 

\section{Massive spinor-helicity for higher-spin states \label{sec2}}

In this section, we present an extension of the massive spinor-helicity formalism from ref.~\cite{Arkani-Hamed:2017jhn} (see also \cite{Craig:2011ws} for an early version) which is designed to conveniently describe higher-spin states and amplitudes. We recycle many of the conventions introduced in ref.~\cite{Arkani-Hamed:2017jhn}, such as using bold spinors for massive states with suppressed little-group indices. Hence, the  translation should be straightforward. 

\subsection{Massive spinor parametrization}

Consider a four-dimensional momentum $p^\mu$ that obeys the on-shell condition $p^2=m^2$. Let us decompose it in terms of the null vectors $k^\mu,q^\mu$, 
\be \label{massiveMom}
p^\mu = k^\mu+ \frac{m^2}{2 p \cdot q}q^\mu\,,
\ee
where we take $q^\mu$ to be an arbitrary reference vector, and $k^\mu$ is then defined by the decomposition. Note that the decomposition implies the identity $p \cdot q = k \cdot q$.

Using the fact that $k,q$ are null, we may now employ the massless spinor-helicity formalism. First, we rewrite~\eqn{massiveMom} into bi-spinors by contracting the momenta with $\sigma^\mu_{\alpha \dot \alpha}$ matrices,
\be
\sigma \cdot p = |k \rangle [k| +  \frac{m^2}{2 p \cdot q}  |q \rangle [q| \equiv | p^{a} \rangle  [ p_a |\,.
\ee
We intentionally suppress the $(\alpha, \dot \alpha)$ spinor indices of the Lorentz group $SL(2,\mathbb{C})\sim SO(1,3)$. We then recognize that the two terms can be reinterpreted as the contraction of two massive spinors that carry $a,b, \ldots$ indices of the little group $SU(2) \sim SO(3)$. 
The massive spinors can be identified as
\begin{align}\label{massiveWeyl}
|p^a\rangle=\left(\begin{matrix}
~ |q \rangle   \frac{m}{\spa{k}.{q}}  \\ |k\rangle~
\end{matrix}\right)\,, ~~~~~~~
|p^a]=\left(\begin{matrix}
  |k]~\\ ~ |q] \frac{m}{\spb{k}.{q}}
\end{matrix}\right)\,.
\end{align}
The mirrored spinors $\langle p^a|$ and $[ p^a|$ are obtained, as implied by the notation, by transposing the massless spinors: $ |k \rangle {\rightarrow} \langle k | $. Since the little group is $SU(2)$, we lower and raise those indices using the rules $|p_a\rangle=\epsilon_{ab}|p^b\rangle$ and  $|p^a\rangle=\epsilon^{ab}|p_b\rangle$; that is, we always multiply with the Levi-Civita symbols from the left. 
The antisymmetric Levi-Civita symbols are normalized as $\epsilon^{12}=\epsilon_{21}=1$. For real momentum $p$ with $E>0$, $m^2>0$, the angle and square spinors are complex conjugates of each other, up to a similarity transform. More specifically, $(|p^a])^* =\Omega | p_a \rangle$, $(|p_a\rangle)^* =\Omega^T | p^a ]$, where $\Omega$ is a 2-by-2 unitary matrix. Because the massive on-shell spinors are related by complex conjugation, one is justified to think of them as the two chiral components of a massive Majorana spinor. 

In general, we will consider amplitudes that depend on many momenta $p_i^\mu$, with particle labels $ i=1,\ldots,n$, which makes it convenient to simplify the notation by only indicating the particle label inside the spinor
\be
|i^a \rangle \equiv  |p_i^a \rangle\,,~~~  |i^a ] \equiv  |p_i^a]\,.
\ee
For each particle $i$ we have the associated reference vector $q_i^\mu$ and mass $m_i$. 

\subsection{Bookkeeping of little-group indices, polarizations and projectors}
It is convenient to define massive bosonic spinors that have no free little-group indices,
\be \label{weylcomp}
|\,{\bf i}\,\rangle   \equiv  |i^a \rangle \z_{i,a} \,, ~~~~~~ |\,{\bf i}\, ]   \equiv  |i^a ] \z_{i,a} 
\ee
where $\z_{i,a} =\epsilon_{ab} \,\z_i^b $ are complex Grassmann-even auxiliary variables that transform as spinors under the little group. Because we take the $\z$ variables to be complex, the two chiral spinors are no longer related by complex conjugation, and thus the spinors  (\ref{weylcomp}) can be interpreted as the chiral components of a massive Dirac spinor.

The spinor products are antisymmetric under label swaps,
\be
\spaq{{\bf1}{\bf2}} = -\spaq{{\bf2} {\bf1}}\,,~~~~\spbq{{\bf1} {\bf2}} = -\spbq{{\bf2} {\bf1}}\,,
\ee
implying that spinor products with repeated indices vanish, e.g. $\spaq{{\bf1}{\bf1}}=0$. Many other identities familiar from the massless spinor-helicity formalism still hold, such as Schouten and Fierz identities:
\be
\spaq{{\bf1}{\bf2}} \spaq{{\bf3}{\bf4}}+\spaq{{\bf2}{\bf3}} \spaq{{\bf1}{\bf4}}+\spaq{{\bf3}{\bf1}} \spaq{{\bf2}{\bf4}}=0\,,~~~~~  \langle {\bf 1}| \sigma^\mu | {\bf 2}]  \langle {\bf 3}| \sigma_\mu | {\bf 4}]= 2\spaq{{\bf1}{\bf3}} \spbq{{\bf4}{\bf2}}\,.
\ee
Because all indices are absent and the spinors are bosonic, we can now take arbitrary powers of the spinors, e.g.
\be
\spaq{{\bf1} {\bf2}}^{2\S} = \text{degree-}4\S\text{~polynomial in~} (\z_1^a, \z_2^a)\,,
\ee
which makes it possible to write down analytic functions with the spinors as arguments.  As a first example, consider the polarization vector for a massive vector boson, which we define as
\be
\bep_{i}^\mu = \frac{\langle {\bf i}| \sigma^\mu | {\bf i}]}{\sqrt{2}m_i} = \frac{[{\bf i}|  \bar \sigma^\mu | {\bf i} \rangle}{\sqrt{2}m_i}   = (\z_i^1)^2  \ep_{i,-}^\mu -\sqrt{2}\z_i^1\z_i^2 \ep_{i,L}^\mu - (\z_i^2)^2  \ep_{i,+}^\mu\,.
\ee
Here $\ep_{i,\pm}= \ep_{i,\pm}(k_i, q_i)$ are standard (massless) polarization vectors for the null momenta $k_i$, with $ q_i$ as the reference vector that appeared in~\eqn{massiveMom}, and $\ep_{i,L}$ is a longitudinal polarization. Explicit expressions can be given, e.g.
$\sqrt{2} \ep_{i,+}^\mu= \langle q_i|  \sigma^\mu |  k_i] / \langle q_i k_i \rangle$ and $\ep_{i,L}=k_i/m_i -m_i q_i/(2p_i \cdot q_i)$.  Note that the massive polarization $\bep_{i}^\mu$ is still a null vector, since $\bep_{i}^2 \propto \spa{\bf i}.{\bf i} \spb{\bf i}.{\bf i}$, and $\spa{\bf i}.{\bf i}= \spb{\bf i}.{\bf i}=0$.
Also, since the $\z$ variables are complex, the polarization $\bep_{i}^\mu$ naturally describes a complex massive vector boson.  In the massless limit, the longitudinal polarization will behave as $\ep_{L}^\mu \sim p^\mu/m$ and is thus singular, whereas the transverse polarizations $\ep_{\pm}$ are well defined. It is interesting to note that the need of an arbitrary reference vector $q$ to describe a massless polarization vector is easy to understand from the ambiguity of the parametrization~(\ref{massiveMom}) we used for the massive spinors.   

To check the completeness relation for the polarization vectors, we need to introduce polarizations that are complex conjugated,
\be
\bar \bep_{i}^\mu = \frac{\big([{\bf i}|  \sigma^\mu | {\bf i } \rangle\big)^*}{\sqrt{2}m_i}  \equiv - (\bar \z_i^1)^2  \ep_{i,-}^\mu +\sqrt{2}\bar \z_i^1 \bar \z_i^2 \ep_{i,L}^\mu + (\bar \z_i^2)^2  \ep_{i,+}^\mu\,,
\ee
where we have used $(\z_a)^* = \bar \z^a$, $(\z^a)^* = - \bar \z_a $ and $(\ep_{i,-}^\mu)^*=\ep_{i,+}^\mu$.
We then get the following non-zero Lorentz product
\be
\bep_i \cdot \bar \bep_{i} =-(\z_i^a \bar \z_{i,a})^2\,,
\ee
where the sign can be traced to the mostly-minus metric $\eta^{\mu\nu}={\rm diag}(1,-1,-1,-1)$. 

If we contract the little-group indices using a derivative operator, we get the completeness relation for the transverse part of the Lorentz group,
\be \label{spin1projector}
-\frac{1}{4}\left(\frac{\partial^2}{\partial \z_{i,a} \partial \bar \z_i^a}\right)^2 \bep_i^\mu  \bar \bep_{i}^\nu = \eta^{\mu \nu}- \frac{p^\mu_i p^\nu_i}{m^2_i}\,.
\ee 
The result should be familiar as the massive spin-$1$ projector, or as the tensor structure of the massive spin-$1$ propagator.  We will see that a direct generalization of the derivative operator introduced here will be convenient for computing state sums that contribute to amplitude factorization residues or loop-level unitarity cuts, for any spin states.  \\
 
\subsection{Bosonic higher-spin states}
Polarization tensors for bosonic spin-$\S$ fields are simply products of $\S$ polarization vectors
\be
\bep_{i}^{\mu_1\mu_2 \cdots \mu_\S}\equiv  \bep_{i}^{\mu_1} \bep_{i}^{\mu_2} \cdots \bep_{i}^{\mu_\S} = \text{degree-}2\S\text{~polynomial in~} \z_i^a\,.
\ee
Polarization tensors are automatically symmetric, traceless and transverse. Transversality $p_{i,\mu_1} \ep_{i}^{\mu_1\mu_2 \cdots \mu_\S}=0$ follows from the fact that $\langle i^a| p_i | i^b] \propto \epsilon^{ab}$, which vanishes after contracting with the symmetric object $\z_i^a \z_i^b$. 

Contracting two CPT-conjugate polarizations gives the little-group completeness relation
\be
\bep_{i}^{\mu_1\mu_2 \cdots \mu_\S} \bar \bep_{i,\mu_1\mu_2 \cdots \mu_\S} =(-1)^\S(\z_i^a \bar \z_{i,a})^{2\S}\,,
\ee
where again the sign is needed due to our mostly-minus signature.  For spin~$2$, we get the following completeness relation for the Lorentz structure:
\be \label{spin2projector}
\frac{1}{(4!)^2}\left(\frac{\partial^2}{\partial \z_{i,a} \partial \bar \z_i^a}\right)^4 \bep_i^{\mu\nu}  \bar \bep_{i}^{\rho\sigma}  = \frac{1}{2}\Big(\te^{\mu \rho}\te^{\nu \sigma}+ \te^{\mu \sigma}\te^{\nu \rho}-\frac{2}{3} \te^{\mu \nu}\te^{\sigma \rho}\Big) \,,
\ee 
where  $\te^{\mu \nu}\equiv  \eta^{\mu \nu}- \frac{p^\mu_i p^\nu_i}{m^2_i}$ is a shorthand notation for the spin-$1$ projector that appeared in \eqn{spin1projector}. The above \eqn{spin2projector} is the expected state projector for the five physical degrees of freedom of a massive spin-$2$ field (e.g.~massive graviton). 

For general bosonic spin~$\S$, we have the following state sum to evaluate:
\be
\frac{ (-1)^\S}{(2\S)!^{\,2}}\! \!\left(\frac{\partial^2}{\partial \z_{i,a} \partial \bar \z_i^a}\right)^{2\S} \!\!\!\!
\bep_{i}^{\mu_1\mu_2 \cdots \mu_\S} 
\bar \bep_{i}^{\nu_1\nu_2 \cdots \nu_\S} = \frac{1}{\S!}(\te^{\mu_1 \nu_1} \te^{\mu_2 \nu_2}\! {\ldots}  \te^{\mu_\S \nu_\S} + \text{perms}) +\ldots \equiv P_{(\S)}^{\vmu \vnu}\,.
\ee
Here $P_{(\S)}^{\vmu \vnu}$ is a compact notation for the state projector of an on-shell symmetric and traceless spin-$\S$ state. (See, e.g., ref.~\cite{Dobrev:1975ru} for early work on projectors.) Considered as a matrix, the projector should satisfy
\bea 
&&P_{(\S)}P_{(\S)} = P_{(\S)}\,, ~~ P_{(\S)}^T = P_{(\S)} \,,~~ {\rm tr} P_{(\S)} = 2\S{+}1\,,  ~~p_{\mu_i} P_{(\S)} =0 \,, \nn \\ \label{projector_properties}
&& P_{(\S)}\big|_{\mu_i \leftrightarrow \mu_j} = P_{(\S)} \,,~~~{\eta_{\mu_i \mu_{j}}}  P_{(\S)}=0\,,~~
{ \eta_{\mu_\S \nu_\S}}  P_{(\S)} = \frac{2\S{+}1}{2\S{-}1} P_{(\S{-}1)}\,,
\eea
where ${\rm tr} P_{(\S)} = 2\S{+}1$ is the number of degrees of freedom for a massive spin-$\S$ state.
From these properties, we can find a recursive formula that computes the projector:
\be
P_{(\S)}^{\vmu \vnu} = \frac{1}{(\S!)^2} \sum_{{\rm perm}\,\vmu} \sum_{{\rm perm}\, \vnu}\Big(\te^{\mu_\S \nu_\S} P_{(\S{-}1)}^{\vmu \vnu}- \frac{(\S-1)^2}{(2\S-1)(2\S-3)} \te^{\mu_\S \mu_{\S-1}}\te^{\nu_\S \nu_{\S-1}} P_{(\S{-}2)}^{\vmu \vnu} \Big)\,,
\ee
where $P_{(0)}=1$. It is convenient to simplify the notation by contracting the free Lorentz indices with auxiliary (symmetric) polarizations $\epsilon^{\vmu} \equiv \prod_i \epsilon^{\mu_i}$ and  $\bar \epsilon^{\vnu}  \equiv \prod_i \bar \epsilon^{\nu_i}$, where the complex vector $\epsilon^{\mu}$ is a dummy variable. We then get the simpler recursive formula for the scalar-valued projector $P_{(\S)}(\epsilon, \bar \epsilon) = \epsilon_{\vmu}  P_{(\S)}^{\vmu \vnu}  \bar \epsilon_{\vnu}\,$,
\be \label{recurs}
P_{(\S)}(\epsilon, \bar \epsilon) = \epsilon.{\bar \epsilon} P_{(\S{-}1)}(\epsilon, \bar \epsilon) - \frac{(\S-1)^2}{(2\S-1)(2\S-3)} \epsilon^2  {\bar \epsilon}^2 P_{(\S{-}2)}(\epsilon, \bar \epsilon)\,,
\ee
where all dot products are taken using the spin-$1$ projector $\te^{\mu\nu}$; that is, the ``lower dot product'' is defined as $\epsilon.{\bar \epsilon}\equiv\epsilon_\mu \te^{\mu\nu} {\bar \epsilon}_\nu$, with $\epsilon^2=\epsilon. \epsilon$ and  $\bar \epsilon^2=\bar \epsilon. \bar \epsilon$. This compact notation will only be used in the current section. The recursive formula (\ref{recurs}) can be solved in terms of the hypergeometric function ${}_2F_1$,
\bea 
P_{(\S)}(\epsilon, \bar \epsilon) &=&  \sum_{j=0}^\infty(\epsilon.{\bar \epsilon})^{\S - 2 j} \Big({-}\frac{\epsilon^2 {\bar \epsilon}^2}{2}\Big)^
  j \frac{\S! (2 \S - 2 j - 1)!!}{j! (\S - 2 j)!  (2 \S -1)!!} 
   \nn \\
&=&
(\epsilon.{\bar \epsilon})^\S {}_2F_1\Big(\frac{1 - \S}{2}, -\frac{\S}{2}, \frac{1}{2} - \S, \frac{\epsilon^2 {\bar \epsilon}^2}{(\epsilon.{\bar \epsilon})^2}\Big) \ .
\eea
Alternatively, the spin-$\S$ projector can be written directly in terms of Legendre polynomials $P_n(x)$,
\be
 P_{(\S)}(\epsilon, \bar \epsilon) = \frac{ \big(\epsilon^2 {\bar \epsilon}^2\big)^{\S/2}\S!}{(2\S-1)!!} P_\S(x)\,,
\ee
where $x^2 = \frac{(\epsilon.{\bar \epsilon})^2}{\epsilon^2 {\bar \epsilon}^2}$. Using Rodrigues' formula for Legendre polynomials, we can simplify the state projector so that it is manifestly local in the auxiliary vectors,
\be
 P_{(\S)}(\epsilon, \bar \epsilon) =
 \frac{\S!}{ (2 \S)!}  
 \Big(\frac{\partial}{\partial \epsilon.{\bar \epsilon}}\Big)^\S \big( (\epsilon.{\bar \epsilon})^2- \epsilon^2 {\bar \epsilon}^2 \big)^\S\,.
\ee
Since the spin-$\S$ projector is unique for on-shell states, we can use it to construct a propagator for a spin-$\S$ field, which takes the form
\be
\Delta_{(\S)}(\epsilon, \bar \epsilon) = i \frac{P_{(\S)}(\epsilon, \bar \epsilon) +\ldots}{p^2-m^2}\,,
\ee
where the ellipsis denote possible terms of ${\cal O}(p^2{-}m^2)$ that vanish on shell. Such off-shell terms depend on the precise details of the Lagrangian formulation used to describe the theory, and they can be altered by field redefinitions.  \\

\subsection{Fermionic higher-spin states}
Polarization spinors for fermionic spin-$(\S+\frac{1}{2})$ fields are built  analogously to the bosonic ones. From products of the massive chiral spinors and polarization vectors we get two chiral spin-$(\S+\frac{1}{2})$ polarizations,
\bea
|{\bf i}\rangle^{\mu_1\mu_2 \cdots \mu_\S} &\equiv&  |{\bf i} \rangle \bep_{i}^{\mu_1} \bep_{i}^{\mu_2} \cdots \bep_{i}^{\mu_\S}\,,  \nn  \\
|{\bf i}]^{\mu_1\mu_2 \cdots \mu_\S}& \equiv &  |{\bf i}] \bep_{i}^{\mu_1} \bep_{i}^{\mu_2} \cdots \bep_{i}^{\mu_\S}\,,
\eea
which are both degree $(2\S+1)$ polynomials in $\z_i^a$. The spinors are automatically symmetric and traceless in the Lorentz indices, but also gamma-traceless. They are gamma-traceless because $\sigma_{\mu_1}|{\bf i}\rangle^{\mu_1\mu_2 \cdots \mu_\S}= \spaq{{\bf i} {\bf i}} |{\bf i}]^{\mu_2 \cdots \mu_\S}$ and $\spaq{{\bf i} {\bf i}}=0$, with an analogous relation for the other spinor.

The chiral spinors can be assembled into a spin-$(\S+\frac{1}{2})$ Dirac tensor-spinor as follows,
\be
v^{\mu_1\mu_2 \cdots \mu_\S}(p_i) ~ \equiv  ~ v(p_i) \, \bep_{i}^{\mu_1} \bep_{i}^{\mu_2} \cdots \bep_{i}^{\mu_\S} ~ = ~\left(\begin{matrix}
  |{\bf i}\rangle^{\mu_1\mu_2 \cdots \mu_\S}   \\ - |{\bf i}]^{\mu_1\mu_2 \cdots \mu_\S} 
\end{matrix}\right) \,,
\ee
where $v$ is an on-shell Dirac spinor,
\be
v(p_i) = \left(\begin{matrix}
  |i^a\rangle   \\ - | i^a]
\end{matrix}\right) \z_{i,a}\,,
~~~~~~~ (\slash \!\!\! p_i +m_i ) v(p_i) =0\,,
\ee
where, as before, the auxiliary $\z_{i,a}$ soak up the $SU(2)$ little-group indices. The on-shell Dirac (tensor-)spinors are Grassmann even, and the fermionic statistical factors have to be added by hand in any calculation, which is the standard textbook treatment for fermionic polarizations.

Let us be precise and write out the Dirac conjugate in this notation,
\be
\bar v(p_i) = \bar \z_{i,a}  \left(\begin{matrix}
 \,  \langle i^a| \,,  & \, \, {}[i^a| \, \,
\end{matrix}\right)\,.
\ee
The inner product of two Dirac spinors of the same momentum is then
\be
\bar v(p_i)  v(p_i) = 2 m_i  \bar  \z^{a}_i \z_{i,a} \,.
\ee
The completeness relation is the familiar one,
\be
\frac{\partial^2}{\partial \z_{i,a} \partial \bar \z_i^a}  v(p_i)  \bar v(p_i) = 
\left(\begin{matrix} 
\, -m_i \, & \, \sigma \cdot p_i \, \\
 \, \bar \sigma \cdot p_i \,& \,- m_i \, 
\end{matrix}\right) = \slash \!\!\!p_i- m_i \,.
\ee
The corresponding $u(p_i)$ and $\bar u(p_i)$ Dirac spinors are defined in Appendix~\ref{appconv}. 

The inner product of two spin-$(\S+\frac{1}{2})$ tensor-spinors of the same momentum, but with conjugate auxiliary $\z$'s, is given by
\be
\bar v^{\mu_1\mu_2 \cdots \mu_\S}(p_i)   v_{\mu_1\mu_2 \cdots \mu_\S}(p_i)   =  2m_i\, (-1)^\S  (\bar \z_i^a \z_{i,a} )^{2\S+1}\,,
\ee
where $\bar v^{\mu_1\mu_2 \ldots \mu_\S}(p_i)  =     \bar \bep_{i}^{\mu_1}\bar  \bep_{i}^{\mu_2} \cdots \bar \bep_{i}^{\mu_\S} \bar v(p_i)$ is the standard Dirac conjugate. Similarly, the completeness relation for spin-$(\S+\frac{1}{2})$ is 
\be
\frac{ (-1)^\S}{(2\S+1)!^{\,2}}\left(\frac{\partial^2}{\partial \z_{i,a} \partial \bar \z_i^a}\right)^{2\S+1}  \!\! v^{\mu_1\mu_2 \cdots \mu_\S}(p_i) \bar v^{\nu_1\nu_2 \cdots \nu_\S}(p_i) \equiv  P_{(\S+1/2)}^{\vmu \vnu}\,,
\ee
where we have introduced the on-shell state projector. The state projector for fermionic higher-spin states can be worked out. It satisfies the same properties as the bosonic one (\ref{projector_properties}), but, in addition, it is gamma-traceless: $\gamma_{\mu_i} P_{(\S+1/2)}^{\vmu \vnu}=0$. 

By considering all the constraints on $P_{(\S+1/2)}^{\vmu \vnu}$ one can find a recursive formula that can be solved explicitly. The solution can be written down using the same compact notation as in the bosonic case, $P_{(\S+1/2)}(\epsilon, \bar \epsilon) = \epsilon_{\vmu}  P_{(\S+1/2)}^{\vmu \vnu}  \bar \epsilon_{\vnu}$,
\be \label{fermion-projector}
P_{(\S+1/2)}(\epsilon, \bar \epsilon) = (\slash \!\!\!p+m) \, A^{(\S)}  +   (\slash \!\!\!\epsilon + \frac{p\cdot\epsilon}{m} )(\slash \!\!\!p-m) (\slash \!\!\!\bar \epsilon + \frac{p\cdot\bar\epsilon}{m} ) \, B^{(\S)} \,,
\ee
where all objects that contain gamma matrices are explicitly displayed, and thus $A^{(\S)},B^{(\S)}$ are scalar-valued functions that depend on the dummy vector $\epsilon^\mu$ and momentum $p^\mu$. They satisfy the relations
\bea
A^{(\S)} &=& \epsilon. \bar \epsilon \, A^{(\S-1)} -\frac{\S+1}{ (2 \S-1) (2 \S+1)}  \epsilon^2  {\bar \epsilon}^2   A^{(\S-2)}\,, \nn \\
B^{(\S)}&=& \frac{A^{(\S-1)}}{2 \S+1}\,,
\eea
with $A^{(0)}=1$, $B^{(0)}=0$. The solution can again be given in terms of a hypergeometric function,
\bea
A^{(\S)}&=& \sum_{j=0}^\infty(\epsilon.{\bar \epsilon})^{\S - 2 j} 
(-\frac{\epsilon^2 {\bar \epsilon}^2}{2})^j \frac{\S! (2 \S - 2 j +1)!!}{j! ( \S-2 j)! (2 \S+1)!!} \nn \\
&=& (\epsilon.{\bar \epsilon})^\S {}_2F_1\Big(\frac{1 - \S}{2}, -\frac{\S}{2}, -\frac{1}{2} - \S, \frac{\epsilon^2 {\bar \epsilon}^2}{(\epsilon.{\bar \epsilon})^2}\Big)\,.
\eea
As before, the modified dot product in $\epsilon. \bar \epsilon, \epsilon^2, \bar \epsilon^2$ include the spin-$1$ state projector $\tilde\eta^{\mu \nu}$, whereas the dot products $p\cdot\epsilon$ and $p\cdot\bar\epsilon$ in \eqn{fermion-projector} are standard Lorentz products. 

The function $A^{(\S)}$ can be related to the associated Legendre functions $P_{\S+1}^1(x)$, and there is a Rodrigues' formula for it,
\be
A^{(\S)} =
 \frac{\S!}{ (2 \S+2)!}  
 \Big(\frac{\partial}{\partial \epsilon.{\bar \epsilon}}\Big)^{\S+2} \big( (\epsilon.{\bar \epsilon})^2- \epsilon^2 {\bar \epsilon}^2 \big)^{\S+1}\,,
\ee
which makes manifest that it is a polynomial.

Again, the fermionic spin-$(\S+1/2)$ projector is unique for on-shell states, and it can be used to construct a propagator
\be
\Delta_{(\S+1/2)}(\epsilon, \bar \epsilon) = i \frac{P_{(\S+1/2)}(\epsilon, \bar \epsilon) +\ldots}{p^2-m^2}\,,
\ee
where in general there are additional terms of ${\cal O}(p^2{-}m^2)$ in the numerator that contribute off shell. These terms depend on the details of the Lagrangian formulation of the theory. 

For the case of spin~$1/2$ and spin~$3/2$, additional terms are not expected, and the propagators are
\bea \label{Prop3/2}
\Delta_{(1/2)}(\epsilon, \bar \epsilon)& =& i\frac{\slash \!\!\!p+m}{p^2-m^2}\,,\nn \\
\Delta_{(3/2)}(\epsilon, \bar \epsilon) &=& i\frac{(\slash \!\!\!p+m) \, \epsilon. \bar \epsilon  +  \frac{1}{3} (\slash \!\!\!\epsilon + \frac{p\cdot\epsilon}{m} )(\slash \!\!\!p-m) (\slash \!\!\!\bar \epsilon + \frac{p\cdot\bar\epsilon}{m} )}{p^2-m^2} \,,
\eea
with $\epsilon. \bar \epsilon = \epsilon^\mu (\eta_{\mu \nu}-\frac{p_\mu p_\nu}{m^2}) \bar \epsilon^\nu$. The propagators with free Lorentz indices can be obtained by taking an appropriate number of derivatives $\frac{\partial}{\partial \epsilon^\mu}$ and $\frac{\partial}{\partial \bar \epsilon^\nu}$ that act on $\Delta_{(\S+1/2)}(\epsilon, \bar \epsilon)$. This will automatically symmetrize the Lorentz indices on each side of the propagator matrix.

\section{Higher-spin three-point amplitudes \label{sec3}}

We now consider amplitudes for a pair of spin-$\S$ particles using the massive spinor-helicity formalism. To avoid displaying unimportant overall normalization factors in the spinor-helicity formulae, we denote amplitudes with either straight or calligraphic symbols. The calligraphic ones, ${\cal A}(1,2,\ldots, n)$ for gauge theory and ${\cal M}(1,2,\ldots, n)$ for gravity, are more suitable for covariant formulae that use polarization vectors. The straight ones, $A(1,2,\ldots, n)$ and $M(1,2,\ldots, n)$, are more suitable for spinor-helicity expressions. Their relative normalizations are
\be
\begin{split}
{\cal A}(1,2,\ldots, n) &=  (-1)^{\lceil \S \rceil} \Big( \sqrt{2} e \Big)^{n-2} A(1,2,\ldots, n) \,, \\
{\cal M}(1,2,\ldots, n) &=  (-1)^{\lceil \S \rceil} \Big( \frac{\kappa}{2}\Big)^{n-2} M(1,2,\ldots, n) \, .
\end{split}
\ee
where $e$ is the gauge theory (electric) coupling,  $\kappa$ is the gravitational coupling, with $\kappa^2=32\pi G_N$. The ceiling function ${\lceil \S \rceil}$ takes into account phases that depend on the spin of the massive particle, which appear due to our mostly-minus metric signature choice.  Furthermore, sometimes it is convenient to set $e = 1$ or $\kappa = 1$, in which case the two normalizations simply differ by powers of $\sqrt{2}$ and signs.

\subsection{Spinor-helicity three-point amplitudes}

It was proposed by Arkani-Hamed, Huang and Huang~\cite{Arkani-Hamed:2017jhn} that the most natural three-point amplitudes between two massive higher-spin particles and a gauge boson should be the following maximally-chiral objects: 
\begin{equation} 
\label{AHH_amps}
A(1\phi^\S,2\bar \phi^\S, 3A^+) =  m \xfactor \frac{\spaq{{\bf 1} {\bf 2}}^{2\S}}{m^{2\S}}\,,~~~~~~
A(1\phi^\S,2\bar\phi^\S, 3A^-) =  \frac{m}{\xfactor} \frac{\spbq{{\bf 1} {\bf 2}}^{2\S}}{m^{2\S}}\,, 
\end{equation}
where $\phi^\S$ is a (complex) spin-$\S$ field (bosonic or fermionic) of mass $m$, and $A^\pm$ is a massless photon with helicity $\pm 1$. The $\xfactor$ notation was introduced~\cite{Arkani-Hamed:2017jhn} as a shorthand notation for the dimensionless scalar amplitude, and in our normalization convention it is given by
\be
 A(1\phi,2\bar \phi, 3A^\pm)= \frac{i}{\sqrt{2}}(p_1 -p_2) \cdot \ep_3^\pm \equiv m \xfactor^{\pm1}\,.
\ee
The amplitudes (\ref{AHH_amps}) were identified as being special due to their properties in the high-energy limit, and because they agree with well-known gauge-theory amplitudes for low spin~$\S=0,1/2,1$.

It is instructive to consider the high-energy limit in some detail. In this limit, the massive spinors degenerate into massless spinors, up to corrections ${\cal O}(m)$. We can choose the momenta such that $\spaq{{\bf 1} {\bf 2}} \sim p$, $\spbq{{\bf 1} {\bf 2}} \sim m$, and $\xfactor \sim m/p$, where $p\gg m$ is some scale that characterizes the momenta. (Note that three-point kinematics require complex momenta, thus the spinors are not related by complex conjugation). 
The amplitudes (\ref{AHH_amps}) behave in this limit as,
\be \label{AHH-scaling}
 m \xfactor \frac{\spaq{{\bf 1} {\bf 2}}^{2\S}}{m^{2\S}} \sim \frac{p^{2\S-1}}{m^{2\S-2}}\,,~~~~~~~~~~~~    \frac{m}{\xfactor} \frac{\spbq{{\bf 1} {\bf 2}}^{2\S}}{m^{2\S}} \ \sim p \,,
\ee
and we can compare this to generic amplitude terms
\be
A^{\rm generic}_{\S', \S''} (1\phi^\S,2\bar \phi^\S, 3A^\pm)  \sim  m \xfactor^{\pm 1} \frac{\spaq{{\bf 1} {\bf 2}}^{2\S'}}{m^{2\S'}} \frac{\spbq{{\bf 1} {\bf 2}}^{2\S''}}{m^{2\S''}} \ \sim \frac{p^{2\S'\mp1}}{m^{2\S'-1\mp1}} \,,
\ee
where $\S',\S''= 0,1/2,1, 3/2,\ldots$ are allowed powers and $\S=\S'+\S''$. We see that, in this high-energy limit, the most singular positive-helicity terms and least singular negative-helicity terms coincide with the amplitudes from ref. \cite{Arkani-Hamed:2017jhn}.\footnote{Alternatively, one can consider the opposite limit $\spaq{{\bf 1} {\bf 2}} \sim m$, $\spbq{{\bf 1} {\bf 2}} \sim p$ which is closer to the considerations of ref.~\cite{Arkani-Hamed:2017jhn}. However, the two limits are related by CPT conjugation.}  

It is interesting to note that some of the generic terms are both more singular and more soft than the ones from ref. \cite{Arkani-Hamed:2017jhn} (in a helicity averaged amplitude). For example, amplitude contributions  $\sim  \xfactor^{-1} \spaq{{\bf 1} {\bf 2}}^{2\S-1}\spbq{{\bf 1} {\bf 2}}$ are more singular than the terms $\sim  \xfactor \spaq{{\bf 1} {\bf 2}}^{2\S}$, and the terms $\sim  \xfactor \spbq{{\bf 1} {\bf 2}}^{2\S-1}\spaq{{\bf 1} {\bf 2}}$ are softer than $\sim  \xfactor^{-1} \spbq{{\bf 1} {\bf 2}}^{2\S}$ . 
 For $\S=1$, such amplitude contributions do appear in a spin-$1$ Lagrangian with non-minimal coupling $\Delta{\cal L} \sim  \alpha \overline W^\mu W^\nu F_{\mu \nu}  $, and they only cancel out for $\alpha=1$, which corresponds to a spontaneously-broken Yang-Mills theory, i.e. in a theory where the $W$-boson acquired its mass through the Higgs mechanism. See Section~\ref{Higgs} for further details on the spin-$1$ Lagrangian. 
 
 \subsection{Constraining covariant three-point amplitudes and currents}
 
In this paper, we argue that the three-point amplitudes (\ref{AHH_amps}) admit a covariant description in terms of off-shell higher-spin currents that appear to be associated to softly-broken higher-spin (gauge) symmetry. That is, we can find an off-shell current $J^{\mu}$, associated to a massive higher-spin field, that appropriately vanishes as $m \rightarrow 0$ when contracted with one power of its momentum, 
\be
\label{sect3pdotj}
p_{\mu} J^{\mu } \big{|}_{\rm traceless} = {\cal O}(m)\,.
\ee
We restrict the condition to hold for the traceless part (gamma-traceless part for fermions) of the contracted current. Since the current constraint holds off-shell, it is a powerful tool for constraining the higher-spin Lagrangian that gives the three-point amplitudes (\ref{AHH_amps}). See Section~\ref{sec5} for Lagrangian constructions based on this constraint, here we only focus on the current and amplitudes.  Naively, the property (\ref{sect3pdotj}) is somewhat surprising, since the higher-spin amplitudes (\ref{AHH_amps}) have arbitrarily high powers of the mass in the denominator, and the massless limits are expected to be singular rather than soft, at least beyond spin~$2$. 

We will not give the complete details of the off-shell currents here, since they are quite lengthy objects and contain ambiguities due to their off-shell nature. We will however show a few steps and then quote the final results for the three-point case. In Section~\ref{sec5}, more details are given using Lagrangians up to spin~$5/2$.

As a first step, we will find a covariant on-shell formula for the amplitudes (\ref{AHH_amps}). This formula will reveal the number of derivatives that are present in the three-point interactions, and also make manifest that all interactions are parity even. We will quote the formula now without further delay, and then prove it later. To give the covariant formula for all bosonic spin-$\S$ amplitudes, it is easiest to construct a generating series,
\be \label{covariant_AHH}
\sum_{\S=0}^{\infty}  A(1\phi^\S,2\bar \phi^\S, 3A)= A_{\phi\phi A} +\frac{A_{WWA} - (\bep_{1}\cdot \bep_{2})^2 \, A_{\phi \phi A} }{(1+\bep_{1}\cdot \bep_{2})^2 + \frac{2}{m^2}\bep_{1}\cdot p_{2}  \, \bep_{2}\cdot p_{1}}\,,
\ee
where the spin-$0$ and spin-$1$ amplitudes appear abbreviated on the right hand side, their explicit covariant forms are
\be
A_{\phi \phi A}\equiv i\sqrt{2}\, \ep_3 \cdot p_1\,, ~~~
A_{WWA}\equiv i\sqrt{2}\,( \bep_1 \cdot \bep_2 \, \ep_3 \cdot p_2+ \bep_2 \cdot   \ep_3\, \bep_1\cdot p_3+\ep_3 \cdot \bep_1 \, \bep_2\cdot p_1)\,.
\ee
Note that the massive polarization vectors are polynomials of $\z_i^a$ and hence the generating function is a rational function in these variables. Series expansion around $\z_i^a=0$ will return the amplitudes for different spins. 

To prove that \eqn{covariant_AHH} is the covariant formula for the resummed amplitudes~(\ref{AHH_amps}), we simply evaluate the polarization vectors in terms of the spinor-helicity variables. The denominator evaluates to
\be \label{denominator_ev}
(1+\bep_{1}\cdot \bep_{2})^2 + \frac{2}{m^2}\bep_{1}\cdot p_{2}  \, \bep_{2}\cdot p_{1} = \Big(1-\frac{\spaq{{\bf 1} {\bf 2}}^2}{m^2}\Big) \Big(1-\frac{\spbq{{\bf 1} {\bf 2}}^2}{m^2}\Big)\,,
\ee
and the numerator becomes $ A_{WWA} - (\bep_{1}\cdot \bep_{2})^2  A_{\phi\phi A} = \xfactor \spaq{{\bf 1} {\bf 2}}^2 (m^2-\spbq{{\bf 1} {\bf 2}}^2)/m^{3}$ for positive helicity $\ep_3^+$ (the negative helicity case is obtained by swapping square and angle spinors, and letting $\xfactor\rightarrow 1/\xfactor$). The factor with square brackets cancels out between the denominator and numerator, and the generating function simplifies to
\be
\sum_{\S=0}^{\infty}  A(1\phi^\S,2\bar \phi^\S, 3A^+)= \frac{m\xfactor}{1-\frac{\spaq{{\bf 1} {\bf 2}}^2}{m^2}}\,,
\ee
which is indeed the geometric series that describes $A(1\phi^\S,2\bar \phi^\S, 3A^+)$ in \eqn{AHH_amps}. The negative helicity amplitudes $A(1\phi^\S,2\bar \phi^\S, 3A^-)$ are obtained analogously. This proves the assertion that \eqn{covariant_AHH} gives the covariant bosonic amplitudes. 

Let us address the question of uniqueness of \eqn{covariant_AHH}. It is straightforward to check at low spin that there exist no other covariant and parity-even formulae with lower number of momentum powers (i.e. derivatives in the Lagrangian). Also, one can confirm that it is not possible to write down Gram-determinant expressions for the five independent vectors $\{\bep_1, \bep_2, \ep_3, p_1, p_2\}$, since such terms would necessarily be quadratic in the massless polarization vector $\varepsilon_3$ and hence describe interactions of a graviton rather than a gauge boson. Finally, we note that, given that we assume parity-even and gauge-invariant interactions, there are only four  independent dimensionless variables that can be used to construct the amplitude: two of them involve $\ep_3$, and can be chosen to be the amplitudes $A_{\phi\phi A}/m$ and $A_{WWA}/m$, and the other two can be chosen as $\bep_1 \cdot \bep_2$ and $ \bep_{1}\cdot p_{2}  \, \bep_{2}\cdot p_{1}/m^2$. The spinor-helicity expressions appear to only involve three dimensionless variables: $\spaq{{\bf 1} {\bf 2}}/m$, $\spbq{{\bf 1} {\bf 2}}/m$, and $\xfactor$. However, because the amplitudes in \eqn{AHH_amps} give two equations in these variables, there is a unique covariantization of the spinor-helicity amplitudes. 

It turns out that the generalization to fermionic gauge-theory amplitudes is almost identical to the bosonic case. We only need to identify two low-spin amplitudes, say, spin~$1/2$ and spin~$3/2$, and then the other fermionic amplitudes are linear combinations of these. In fact, we observe that it is exactly the same linear combinations that appeared in the bosonic generating function  (\ref{covariant_AHH}). Hence, the fermionic amplitudes can be formally resummed as
\be \label{covariant_AHH_fermion}
\sum_{\S=0}^{\infty}  A(1\phi^{\S+1/2},2\bar \phi^{\S+1/2}, 3A)= A_{\lambda \lambda A} +\frac{A_{\psi \psi A}  - (\bep_{1}\cdot \bep_{2})^2 \, A_{\lambda \lambda A} }{(1+\bep_{1}\cdot \bep_{2})^2 + \frac{2}{m^2}\bep_{1}\cdot p_{2}  \, \bep_{2}\cdot p_{1}}\,,
\ee
where $A_{\lambda \lambda A} = \frac{i}{\sqrt{2}}\bar u_2 \slashed{\ep}_3  v_1$ is the standard electron-photon three-point amplitude in quantum electrodynamics, and $A_{\psi \psi A}$ is a well-known gravitino-photon amplitude that is unique in supergravity with spontaneously-broken supersymmetry. The details of this spin-$3/2$ amplitude are given in Section~\ref{gravitino_section}; for the current purpose we only need to know that $A_{\psi \psi A}$ comes from parity-even interaction terms that are linear in the momenta. That the covariant formula \eqn{covariant_AHH_fermion} is correct follows from analogous arguments to the ones used below \eqn{denominator_ev}. The uniqueness is less clear since Gram-determinant expressions can be constructed using the vectors and higher-rank forms that come from the fermion bilinears $\bar u_2 \gamma^{\mu_1 \ldots \mu_n}  v_1$. Nevertheless, at spin~$1/2$ and spin~$3/2$, the covariant amplitudes are unique as will be discussed in Section~\ref{gravitino_section}.

Let us now switch the discussion to gravitational higher-spin amplitudes. In ref.~\cite{Arkani-Hamed:2017jhn} Arkani-Hamed, Huang and Huang gave the following three-point amplitudes:
\begin{align} \label{Gnima3pt}
&M(1\phi^\S,2\bar\phi^\S,3h^+) =  i   m^2 \xfactor^{2} \frac{ \angle{\bm1 \bm2}^{2\S}}{m^{2\S}}\, \nn ,\\
&M(1\phi^\S,2\bar\phi^\S,3h^-) =  i    \frac{m^2}{\xfactor^2}\frac{ [\bm1 \bm2]^{2\S}}{m^{2\S}}\, .
\end{align}
It is clear from the spinor-helicity expressions that they are related to the gauge-theory amplitudes by the double copy~\cite{Kawai:1985xq},
\be
M(1\phi^\S,2\bar\phi^\S,3h^\pm) =  i   A(1\phi^{\S_{\rm L}},2\bar\phi^{\S_{\rm L}},3A^\pm)  A(1\phi^{\S_{\rm R}},2\bar\phi^{\S_{\rm R}},3A^\pm)\,,
\ee
where $\S=\S_{\rm L}+\S_{\rm R}$, and equivalent formulae are obtained for any $\S_{\rm L},\S_{\rm R}\ge 0$.  Since this relation is insensitive to the helicity of the massless state, it follows that the covariant gravitational amplitudes are also given by a double copy of the covariant gauge-theory amplitudes. 

Using the double copy, it is a small step to show that a generating function for the covariant gravitational amplitudes can be constructed out of the gauge-theory generating functions. However, because the spin~$\S$ can be decomposed into $\S_{\rm L}+\S_{\rm R}$ in multiple different ways, there is some ambiguity on how to write it. We find that it is convenient to write a combined generating function for bosonic and fermionic covariant gravitational amplitudes, which has the form
\be  \label{covariant_AHH_gravity}
\sum_{2\S=0}^{\infty}  M(1\phi^{\S},2\bar \phi^{\S}, 3h)= \MxxX+ i A_{WW A} \Big(\AxxX +\frac{\AXXX  - (\bep_{1}\cdot \bep_{2})^2 \, \AxxX }{(1+\bep_{1}\cdot \bep_{2})^2 + \frac{2}{m^2}\bep_{1}\cdot p_{2}  \, \bep_{2}\cdot p_{1}}\Big)\,,
\ee
where the sum runs over both integer and half-integer spin~$\S$. Here we use the shorthands $\AxxX= A_{\phi \phi A}+A_{\lambda \lambda A}$, $\AXXX= A_{WW A}+A_{\psi\psi A}$ and $\MxxX= i A_{\phi \phi A} \AxxX $,  which combine the independent low-spin amplitudes that appeared in the bosonic  (\ref{covariant_AHH}) and fermionic (\ref{covariant_AHH_fermion}) generating functions. The reason for multiplying the gauge-theory generating function by $A_{WW A}$, rather than by $A_{\phi \phi A}$, is that this way of writing it exposes the correct maximal momentum power counting of the gravity amplitudes. 

As an aside, it is interesting to note that we have already identified the gauge-theory amplitudes $A_{WW A}$ and $A_{\psi\psi A}$, as well-known three-point amplitudes in theories with spontaneously broken gauge symmetry ($\S=1$) and supersymmetry $(\S=3/2)$, respectively. From this we should expect that some of the low-spin gravitational amplitudes also have an interesting physical interpretation. Indeed, it is known from previous work on the double copy~\cite{Chiodaroli:2017ehv,Chiodaroli:2018dbu} that these massive spin-$1$ and spin-$3/2$ gravitational amplitudes, $A_{\phi \phi A} \times A_{WW A}$ and $A_{\lambda \lambda A} \times A_{WW A}$, are precisely those that appear in R-symmetry gauged supergravity, and the massive spin-$2$ amplitude $(A_{WW A})^2$ is precisely the one that appears in Kaluza-Klein gravity~\cite{Chiodaroli:2015rdg, Momeni:2020vvr, Johnson:2020pny, Momeni:2020hmc, Hang:2021fmp, Chi:2021mio}. Indeed, in Section \ref{KK_section}, we obtain a spin-$2$ Lagrangian from Kaluza-Klein compactification, and show that it is compatible with the expected  spin-$2$ current and Compton amplitude. 

 \subsection{Off-shell current construction}

Inspecting the covariant formulae (\ref{covariant_AHH}),  (\ref{covariant_AHH_fermion}) and (\ref{covariant_AHH_gravity}), we learn that the gauge theory higher-spin three-point interactions are parity invariant, and they have an upper bound for the number of derivatives. For integer $\S\ge 1$, the bosonic and fermionic covariant gauge-theory amplitudes have a maximal power of momentum (derivatives) that scale as
\be
A(1\phi^{\S},2\bar \phi^{\S}, 3A) \sim p^{2\S-1}\,,~~~~ A(1\phi^{\S+1/2},2\bar \phi^{\S+1/2}, 3A) \sim p^{2\S-1}\,.
\ee
For integer $\S\ge 2$, the bosonic and fermionic covariant gravity amplitudes have a maximal power of momentum (derivatives) that scale as
\be
M(1\phi^{\S},2\bar \phi^{\S}, 3h) \sim p^{2\S-2}\,,~~~~ M(1\phi^{\S+1/2},2\bar \phi^{\S+1/2}, 3h) \sim p^{2\S-2}\,.
\ee
This matches the scaling that we found in \eqn{AHH-scaling} for gauge-theory bosons. For fermions, an extra power of momentum can be attributed to the wave functions, but it is not counted here since we now focus on the momentum which is explicit in the covariant formulae, as this will correspond to the derivatives in a putative Lagrangian. The gravity momentum scaling is also compatible with \eqn{AHH-scaling} after accounting for the extra factor of $\xfactor\sim m/p $ in the gravitational amplitudes~(\ref{Gnima3pt}). Note that the covariant amplitudes have a slightly different scaling in the gauge-theory $\S<1$ and gravity $\S<2$ cases. In these cases, the massive particle is not the highest spin state, and thus the gauge and gravity theories are expected to be completely standard, that is, with minimally-coupled matter. 

Let us now discuss off-shell versions of the covariant amplitudes. Consider constructing an Ansatz for the off-shell correlation function between two bosonic higher-spin fields $\phi^{\S},\bar\phi^{\S}$ and a gauge boson $A^\w$ of spin~$\w$, that is we combine the discussion of gauge theory ($\w=1$) and gravity ($\w=2$), so that $A^\w=(A^\mu, h^{\mu \nu})$. The discussion simplifies if we saturate all the free Lorentz indices with dummy vectors $\epsilon^\mu_i$, in analogy with the projector discussion in Section~\ref{sec2}.  The off-shell correlation function for bosonic higher-spin fields then takes the form
\be \label{poly}
\langle \phi^{\S}(p_1) \bar \phi^{\S}(p_2) A^\w(p_3) \rangle = m^\w \, {\rm Poly} (\epsilon_i \cdot \epsilon_j, \, \epsilon_i \cdot p_j /m, \, p_i \cdot p_j/m^2)\,,
\ee
where ${\rm Poly}(\ldots)$ is a polynomial Ansatz of the Lorentz invariants that appear in the arguments. The range of leg labels is $i,j=1,2,3$; we assume momentum conservation, $p_1+p_2+p_3=0$, but no other relations exist among the arguments. All the arguments are dimensionless, and we have factored out $\w$ powers of the mass, so that the correlator is of the correct dimension for a gauge/gravity theory. For fermionic higher-spin fields, the correlation function is almost the same, except that the polynomial Ansatz (\ref{poly}) also contains contractions between the dimensionless vectors $p_i^\mu/m, \epsilon_i^\mu$ and the dimensionless fermion bilinears $\bar u_2 \gamma^{\mu_1 \ldots \mu_n}  v_1 /m$, where $\bar u_2,   v_1$ are temporarily treated as unconstrained off-shell spinors. Note that every term in the polynomial Ansatz must contain exactly one fermion bilinear, and since the polarizations and momenta constitute five independent vectors, the tensor rank of the fermion bilinear is at most five.   

We will now give a list of constraints on the off-shell correlator that is compatible with the covariant amplitudes (\ref{covariant_AHH}), and hence compatible with the  Arkani-Hamed, Huang and Huang amplitudes~\cite{Arkani-Hamed:2017jhn}. In fact, the constraints will uniquely pin down these amplitudes up to spin-$3/2$ gauge theory, and spin-$5/2$ gravity.  
We constrain the polynomial  (\ref{poly}) such that: 
\begin{itemize}
\item Each term contains $\S$ powers of the higher-spin polarizations, $(\epsilon_{1})^\S$, $(\epsilon_{2})^\S$, and $\w$ powers of the gauge boson $(\epsilon_3)^\w$;
\item Each term have at most $p^{2\S-\w}$ powers of momenta;
\item Gauge invariance is imposed for the gauge current $p_{3} \cdot J_{\rm gauge}=0$;
\item The current constraint is imposed on the higher-spin currents; for leg 1 this means $p_{1} \cdot J\big|_{\rm traceless} = {\cal O}(m)$,  and similarly for leg 2;
\item Standard kinetic-term behavior is imposed on interactions.
\end{itemize}
Let us clarify what this list of constraints means. Firstly, the gauge current and higher-spin current are defined as
\bea
J_{\rm gauge}^{\mu} &=& \frac{\partial}{\partial \epsilon_3^\mu} \langle \phi^{\S}(p_1) \bar \phi^{\S}(p_2) A^\w(p_3) \rangle \Big|_{1\& 2~\text{on shell}}\,, \nn \\
J^{\mu} &=& \frac{\partial}{\partial \epsilon_1^\mu} \langle \phi^{\S}(p_1) \bar \phi^{\S}(p_2) A^\w(p_3) \rangle \Big|_{2\&3~\text{on shell}} \,,
\eea
and the higher-spin current for leg 2 is obtained by swapping $1\leftrightarrow 2$. The currents are simply defined as the correlator where a Lorentz index is exposed corresponding to an off-shell leg, and then the remaining two legs are subject to on-shell conditions. The on-shell conditions for the higher-spin legs $ i=1,2$ are $\{p_i^2=m^2, \epsilon_i\cdot p_i =0, \epsilon_i^2=0 \}$ and for the gauge-field leg 3 they are $\{p_3^2=0, \epsilon_3\cdot p_3 =0, \epsilon_3^2=0 \}$. For fermions, the on-shell conditions are supplemented by $\{(\slashed p_1+m) v_1 =  \bar u_2  (\slashed p_2-m) =  \slashed \epsilon_1 v_1 = \bar u_2 \slashed \epsilon_2 =0  \}$.  Note that tracelessness, $\epsilon_i^2=0$, and gamma-tracelessness, $\slashed \epsilon_1 v_1 = \bar u_2 \slashed \epsilon_2 =0 $, are not imposed for the off-shell legs in the current. However, after contracting the higher-spin current into its momentum, we remove the trace components from the current constraint: $p_{1} \cdot J\big|_{\rm traceless} = {\cal O}(m)$. Without removing the trace components, the current constraint is too restrictive.  

Finally, we need to impose the existence of a standard kinetic term. This may sound strange since the correlator Ansatz (\ref{poly}) only contains the three-point  interactions. It is clear that some of the terms in the three-point correlator will originate from the non-linear terms of the kinetic term, whereas other terms originate from higher-derivative gauge-invariant cubic operators (involving the gauge-boson field strength). To have a standard kinetic term, we need to forbid interaction terms that originate from possible higher-derivative kinetic terms.  This can be done by demanding that all higher-derivative terms obey gauge invariance for the full off-shell correlator (hence they originate from the field strength, and not from the covariant derivatives of the kinetic term). The standard kinetic term behavior is imposed on the off-shell correlator by the constraint
\be
p_3^\mu \frac{\partial}{\partial \epsilon_3^\mu} \langle \phi^{\S}(p_1) \bar \phi^{\S}(p_2) A^\w(p_3)  \rangle = {\cal O}(m^0)\,.
\ee
The right hand side of this equation involves no inverse powers of the mass, since such terms come from higher-derivative operators, which should be fully gauge invariant off-shell. A standard kinetic term would have no poles in $m$, and the corresponding interaction terms need not be gauge invariant for an off-shell correlator.  

After having applied the above constraints on the correlator Ansatz (\ref{poly}), all legs can be placed on-shell to obtain a covariant amplitude. Without any input from Lagrangians or spinor-helicity formulae, this gives unique tree-point amplitudes for spin-$1$ and spin-$3/2$ gauge theory, as well as spin-$2$ and spin-$5/2$ gravity, precisely agreeing with the covariant amplitudes (\ref{covariant_AHH_fermion}) and (\ref{covariant_AHH_gravity}).\footnote{The Ansatz procedure also leads to unique amplitudes for the low spin cases, $s<1$ gauge theory and $s<2$ gravity, but the derivative counting needs to be adjusted upwards to be compatible with the kinetic term; hence these are all minimal-coupling theories.} For higher spin, beyond spin-$3/2$ gauge theory and beyond spin-$5/2$ gravity, the procedure does not result in unique amplitudes, but within the parameter space of the allowed solution one can always find the amplitudes (\ref{covariant_AHH_fermion}) and (\ref{covariant_AHH_gravity}). We explicitly constructed the correlator and currents up to spin~5 for both gravity and gauge theory, and found them to be compatible with the covariant amplitudes. Thus there is strong evidence that the spinor-helicity amplitudes of ref.~\cite{Arkani-Hamed:2017jhn} originate from higher-spin theories that obey the current constraint.

Thus far, we have ignored the subtleties that come from four-dimensional kinematics, so let us comment on this. For spin-$5/2$ gravity, one can find a unique solution to the current constraint, but this requires that we use the following Gram-determinant identity:
\be
\label{Gram5o2}
G = \epsilon_{1\mu} f_{3 \nu \rho}  f_{2  \sigma \kappa} \bar u_2 \gamma^{[\mu \nu} v_1  f^{\rho \sigma}_3 \epsilon_2^{\kappa]}  ~ \stackrel{D=4}{=} ~ 0\,,
\ee
where the five upper Lorentz indices are antisymmetrized, and  $f_i^{ \mu \nu}=p_i^\mu\epsilon_{i}^\nu-p_i^\nu\epsilon_{i}^\mu$ is a field strength of particle $i$ (the corresponding linearized ``Riemann tensor'' is the square $f_i^{ \mu \nu} f_i^{ \rho \sigma}$).
The current constraint then holds modulo this four-dimensional identity
\be
p_1 \cdot J_{5/2}\Big|_{\rm traceless} ~\sim~ \frac{G}{m} +  {\cal O}(m)\,.
\ee
Note that $G$ is one of the first Gram determinants that can appear for low-spin correlators. In fact, for three-point correlators in a bosonic gauge theory one cannot write down any Gram determinant compatible with our Ansatz. Similarly, in a bosonic gravity theory one can write down only one, $G(\epsilon_1,\epsilon_2,\epsilon_3, p_1,p_2)$, which is quadratic in all three polarizations. This object can first show up in a current constraint for spin-$3$ particles; however, we find that the current constraint can be solved without imposing the vanishing of the Gram determinant. For fermionic gauge theory correlators, one could in principle write down $G' =f_{3 \rho \sigma}  f_{2 \kappa \lambda} \epsilon_{1\nu}\bar u_1 \gamma^{[\nu \rho \sigma \kappa} v_2  \epsilon_2^{\lambda]}$, which also corresponds to spin~$5/2$. However, surprisingly, we find that we do not need to make use of such identities. It is curious that it is only for the fermionic gravity correlators that it seems necessary to use the Gram determinants to satisfy the current constraint.

\subsection{Classical limit}
\label{Classical_limit_section}

The gravitational amplitudes discussed in Section~\ref{sec3} have been shown to match the classical energy-momentum tensor of a Kerr black hole~\cite{Guevara:2018wpp,Arkani-Hamed:2019ymq}. Here we briefly review that argument using our notation, and we will also make an interesting comparison to the higher-spin classical amplitudes recently computed in ref.~\BernSpin.

The linearized energy-momentum tensor for a Kerr black hole can be given in momentum space as \cite{Vines:2017hyw, Guevara:2018wpp}
\be \label{KerrT}
T^{\mu \nu} (-k) = 2 \pi \, \delta (p \cdot k) \, p^{(\mu} {\exp (m^{-1} \Svec * i k)^{\nu)}}_\rho \, p^\rho\,,
\ee
where $p^\mu$ and $\Svec^\mu$ are the black hole momentum and classical spin vector (Pauli-Lubanski vector), respectively, and ${(\Svec * k)^\mu}_{\nu} = {\epsilon^\mu}_{\nu \rho \sigma} \Svec^\rho k^\sigma$. 

We can contract the tensor with an on-shell graviton polarization to obtain what one may call a classical amplitude~\cite{Guevara:2018wpp},
\be
\label{Kerr3pt}
\ep_{\mu \nu} (k)  T^{\mu \nu} (-k) = (2 \pi)^2 \, \delta (p \cdot k) \, \delta (k^2) (\varepsilon \cdot p)^2 \exp \! \Big(\!- i \frac{k_{\mu} \varepsilon_{\nu} \Svec^{\mu \nu}}{\varepsilon \cdot p} \Big) \,,
\ee
where $\ep_{\mu \nu} (k) = \varepsilon_{\mu}(k) \varepsilon_{\nu}(k)$ and $\Svec^{\mu \nu} = m^{-1} \epsilon^{\mu \nu \rho \sigma} p_\rho \Svec_{\sigma}$.

One can compare the above classical amplitude  to the quantum three-point amplitude obtained by Arkani-Hamed, Huang and Huang. Let us focus on a positive-helicity graviton with momentum $k$ and polarization $\ep_{\mu \nu} (k) $ that scatters against two bosonic spin-$\S$ particles with momenta $p_1$ and $p_2$. We express a $\langle \bm1 \bm2 \rangle^2$ factor in terms of massive polarizations to get an expression reminiscent of an exponential, 
\be
\label{nima3ptcov}
  (-1)^\S i \frac{m^2 \xfactor^2}{2} \frac{\langle \bm1 \bm2 \rangle^{2\S}}{m^{2\S}} =  -i(\varepsilon \cdot p_1)^2 \left( \bep_2 \cdot \Big(1 + \frac{1}{m} k \cdot \Sop + \frac{1}{m^2} (k \cdot \Sop)^2 \Big) \cdot \bep_1 \right)^\S \,,
\ee
where $\Sop^\mu = (2m)^{-1} \epsilon^{\mu \nu \rho \sigma} p_{1 \nu} M_{\rho \sigma}$ is a spin operator, with $(M^{\mu \nu})_{\alpha \beta} = 2 i \delta^{[\mu}_\alpha \delta^{\nu]}_\beta$ being the spin-$1$ Lorentz generator.
The positive-helicity amplitude is similar, and it can be obtained by the replacement $\Sop \rightarrow - \Sop$ in the above expression.

We will now take the classical limit, which we can do in two steps. First take the momentum of the graviton $k$ to be small compared to momenta of the higher-spin particles. Using momentum conservation and imposing that polarizations are null and transverse, one can relate $\varepsilon_2$ and $\bar \varepsilon_1$ through a finite power series in $k$,\footnote{We thank Lucile Cangemi for highlighting the importance of this relation.}
\be
\bep_2^\mu = - \bar{\bep}_1^\mu +p_1^\mu \,\frac{ \bar\bep_1 \cdot k}{m^2} +k^\mu \, \frac{ \bar\bep_1 \cdot k}{2m^2} \,.
\ee
Next, we identify the classical spin vector with the spin-operator expectation value as $-\bar \bep_1  \cdot  \Sop^\mu  \cdot \bep_1  \rightarrow  \Svec^\mu/\S$. We also assume that the polarizations are normalized such that $\bar \bep_1 \cdot \bep_1=-1$ (which imposes $(\bar \z^a\z_a)^2=1$). The second step is to take the large-spin limit. Combining both steps, we get the result
\be
~~ \stackrel{k \ll p} {\longrightarrow}~~  (\varepsilon \cdot p)^2  \Big(1 + \frac{1}{m \S} k \cdot \Svec + {\cal O}(S^2/s^2) \Big)^\S ~~ \stackrel{\S \rightarrow \infty} {\longrightarrow}~~ (\varepsilon \cdot p)^2  \exp\Big(\frac{1}{m} k \cdot \Svec  \Big)   \,.
\ee
The suppressed quadratic term corresponds to a quantum contribution and vanishes in the large-$\S$ limit.  Finally we have to make use of the on-shell relation $ i k_{\mu} \varepsilon_{\nu}^\pm \Svec^{\mu \nu} = \mp k \cdot \Svec \, \varepsilon^\pm {\cdot\,} p_1/m$ and then the match to \eqn{Kerr3pt} works up to the overall normalization. 

It is instructive to compare to the results from ref.~\BernSpin, where a higher-spin Lagrangian formalism was used to obtain amplitudes that reproduces the linearized energy-momentum tensor~(\ref{KerrT}) in the classical limit.  The three-point amplitude, with the same momentum and polarization labeling as above, can be written as \BernSpin\ 
\begin{equation}
\label{bern3pt}
\cM(1\phi^\S,2\bar\phi^\S,3h) =  - i (\varepsilon \cdot p_1)^2 \bep_2^\S \cdot \left( \exp \Big(\frac{1}{m} k  \cdot \SOP \Big) - \frac{i}{m^2} p_1 \cdot \mathbb{M} \cdot k \right) \cdot \bep_1^\S \, ,
\end{equation}
where $\mathbb{M}$ and $\SOP$ are the spin-$\S$ representations of the operators that we called $M$ and $\Sop$ above. 
Even before taking the classical limit, we can note that the exponential term can be expanded out as
\bea
\bep_2^\S \cdot  \exp \left(\frac{1}{m} k \cdot \SOP \right)  \cdot \bep_1^\S & =& \left(  \bep_2 \cdot \exp \Big(\frac{1}{m} k \cdot \Sop \Big) \cdot \bep_1 \right)^\S \\
&=& \left(  \bep_2 \cdot \Big(1 + \frac{1}{m} k \cdot \Sop + \frac{1}{2 m^2} (k \cdot \Sop)^2 \Big) \cdot \bep_1 \right)^\S\,,\nn 
\eea
where we used that the exponent is a nilpotent matrix $(k \cdot \Sop)^3 = 0$. This expression looks very similar to the quantum amplitude (\ref{nima3ptcov}), except that the term quadratic in $\Sop$ has a numerical coefficient smaller by a factor of $1/2$. Since we already mentioned that this term vanishes in the classical limit, it can be thought of as a quantum contribution. The same is true of the second term in \eqn{bern3pt} that contains  $p_1 \cdot \mathbb{M}  \cdot k$.  

Let us check if the two contributions responsible for the quantum mismatch between \eqn{nima3ptcov} and \eqn{bern3pt} are perhaps related. We find the relation
\be
\label{bern3ptextra}
\bep_2^\S \cdot \Big( \frac{i}{m^2} p_1 \cdot \mathbb{M} \cdot k \Big) \cdot \bep_1^\S 
=  \S (\bep_1 \cdot \bep_2)^{\S-1} \bep_2 \cdot \Big( \frac{1}{m^2} (k \cdot \Sop)^2 \Big) \cdot \bep_1 \,,
\ee
and for $\S=1$ the two expressions indeed conspire in \eqn{bern3pt} with numerical coefficients $1/2-1$. However, this still does not add up to the unit coefficient of this term in \eqn{nima3ptcov}, which through $\S\le 5/2$ should give the unique theories that satisfy tree-level unitarity. That said, the terms proportional to $p_1 \cdot \mathbb{M}  \cdot k$ or to $\bep_2 \cdot (k \cdot \Sop)^2 \cdot \bep_1$ are  subleading in the classical limit and thus the quantum difference is irrelevant for the purpose of describing astrophysical black holes. In conclusion, this analysis confirms that \eqn{bern3pt} and \eqn{nima3ptcov} are classically equivalent and match the Kerr black-hole dynamics.

\section{Spinor-helicity Compton amplitudes for $\S\le5/2$ }
\label{SECT:msh}

In ref.~\Nima, three-point higher-spin amplitudes, which we discussed in Section~\ref{sec3}, were used together with BCFW recursion~\cite{Britto:2004ap,Britto:2005fq} to construct candidates for the Compton amplitudes with opposite-helicity photons/gravitons. In a later reference the equal-helicity Compton amplitudes were obtained in the same way~\HenrikAlex. Let us start by inspecting the photon amplitudes
\begin{subequations}
\label{comptonnima}
\begin{align}
A(1\phi^\S,2\bar\phi^\S,3A^+,4A^+) &= i \frac{\angle{\bm1 \bm2}^{2\S} [3 4]^2}{m^{2\S-2} t_{13} t_{14}} \,, \label{equalHelPhoton}\\
A(1\phi^\S,2\bar\phi^\S,3A^-,4A^+) &=- i \frac{\Cspur^{2-2\S} (\Cres)^{2\S}}{t_{13} t_{14}} \,,
\end{align}
\end{subequations}
 where $s_{1 2}= (p_1 + p_2)^2$ and $t_{i j} = (p_i + p_j)^2 - m^2$. 
As was discussed in ref.~\Nima, the opposite-helicity amplitude is well behaved for $\S\le1$, and starting at $\S=3/2$ it develops a spurious pole corresponding to the factor $\Cspur^{2-2\S}$. This pole is unphysical, and must be canceled by adding a contact interaction to the Compton amplitude, such that it has a compensating spurious pole. Exactly how to do this in a unique way has not yet been firmly established. In contrast, we see that the equal-helicity Compton amplitude does not have a spurious pole for any spin. And this suggests that it should not be corrected by contact terms, although a priori it cannot be ruled out that it receives corrections that are manifestly free of momentum poles. 

Next, let us quote the corresponding candidate Compton amplitudes for gravity, which can be obtained via BCFW recursion in the same way~\cite{Arkani-Hamed:2017jhn,Johansson:2019dnu}, 
\begin{subequations}
\label{Gcomptonnima}
\begin{align}
M(1\phi^\S,2\bar\phi^\S,3h^+,4h^+) &= i \frac{\angle{\bm1 \bm2}^{2\S} [3 4]^4}{m^{2\S-4} s_{12} t_{13} t_{14}} \,, \label{equalHelGrav} \\
M(1\phi^\S,2\bar\phi^\S,3h^-,4h^+) &= i  \frac{\Cspur^{4-2\S} (\Cres)^{2\S}}{s_{12} t_{13} t_{14}} \,.
\end{align}
\end{subequations}
Again, the opposite-helicity amplitudes develop a spurious pole $\Cspur^{4-2\S}$ for spin~$\S > 2$. The naive application of BCFW recursion does not yield a physical answer, and an unknown contact term needs to be added to cancel the spurious pole. The results are perhaps not surprising, since naive application of BCFW recursion assumes a good high-energy behavior with an absent residue at infinity.  However, higher-spin theories are effective theories, and one may expect that their high-energy behavior requires a careful analysis. A possible choice for the residue at infinity was discussed in ref.~\Nima, but it is not unique and thus further physical criteria are needed for pinning down these contributions, and by extension, finding the Compton amplitudes that describe a Kerr black hole.

In Section~\ref{sec5}, we will show the explicit details of using the $P \cdot J = {\cal O}(m)$ condition to uniquely obtain the three-point and Compton amplitudes up to spin-$3/2$ gauge theory and spin-$5/2$ gravity. 
These exactly agree with the spurious-pole-free amplitudes of refs.~\cite{Arkani-Hamed:2017jhn,Johansson:2019dnu}. In addition, we obtain unique contact term corrections to the spin-$3/2$ gauge theory and spin-$5/2$ gravity Compton amplitudes. As a spoiler of that discussion, here we will briefly quote the new spurious-pole-free amplitudes, and in Section \ref{secRevisit} we will revisit the details of the correction terms.

First, let us pick a shorthand notation for some recurring spinor combinations,
\begin{align} \label{Ndefs}
&\MSHquada = \Cres \, , \nn \\
&\MSHquadb = [4 \bm1] \langle 3 \bm2 \rangle - [4 \bm2] \langle 3 \bm1 \rangle \, , \\
&\MSHquart = \Cresspurb \, , \nn
\end{align}
where the subscripts are indicating the dimension counting. In the opposite-helicity case, we obtain the following Compton amplitude for spin-$3/2$ matter coupled to photons:
\begin{equation}
\label{finalcompton3/2}
A(1\phi^{3/2}\!,2\bar\phi^{3/2}\!,3A^-\!,4A^+) = 
\frac{i \MSHquada}{m^4 t_{13} t_{14}} ( m^3 \MSHquada (\angle{\bm2 \bm1} + [\bm2 \bm1]) + m^2 \Cspur \angle{\bm2 \bm1} [\bm2 \bm1] + [ 3 | p_1 | 4 \rangle \MSHquart ) \, .
\end{equation}
The corresponding equal-helicity Compton amplitude for spin-$3/2$ matter receives no corrections, and hence \eqn{equalHelPhoton} remains valid.

In the case of spin-$5/2$ matter coupled to gravitons, we obtain a corrected opposite-helicity Compton amplitude given by
\begin{equation}
\label{Gfinalcompton5/2}
M(1\phi^{5/2},2\bar\phi^{5/2},3h^-,4h^+) = i \MSHquada \frac{m^4 \nA_4 + m^3 \nA_3 + m^2 \nA_2 + \nA_0}{m^6 s_{12} t_{13} t_{14}}\,,
\end{equation}
where we have decomposed the contributions according to powers of the mass; these are
\begin{align}
&\nA_0 = - s_{12} \MSHquart^2 [ 3 | p_1 | 4 \rangle \,, \nn \\
&\nA_3 = 2 \MSHquada  (\angle{\bm1 \bm2} + [\bm1 \bm2])  \big(\Cspur^2 (\angle{\bm1 \bm2} - [\bm1 \bm2])^2+ s_{12}  \MSHquart  \big)\, , \\
&\nA_4 = - 2 \MSHquada^2 \Cspur (\angle{\bm1 \bm2} - [\bm1 \bm2])^2  \nn \,, 
\end{align}
and 
\begin{multline}
\nA_2 = 
\frac{1}{2} \Cspur \left\{
 2 \Cspur^2 \angle{\bm1 \bm2}[\bm2 \bm1] \left( \angle{\bm1 \bm2}^2 +  [\bm1 \bm2]^2 + 3 \angle{\bm1 \bm2}[\bm2 \bm1] \right) \right. \\
\left. \null + \Cspur (\angle{\bm1 \bm2}^2 + [\bm1 \bm2]^2)  \big( [\bm2 | p_{3, 4} | \bm1\rangle (\MSHquadb + \MSHquada) + [\bm1 | p_{3, 4} | \bm2\rangle (\MSHquadb - \MSHquada) \big) \right. \\
\left.  \null + 2 \MSHquart s_{12} \left( 3 \angle{\bm1 \bm2}^2 + 3 [\bm1 \bm2]^2 + 4 \angle{\bm1 \bm2}[\bm2 \bm1] \right)
\right\}\,,
\end{multline}
where $p_{3, 4} \equiv p_3 - p_4$.  Again, the equal-helicity Compton amplitude for spin-$5/2$ matter receives no corrections, and hence \eqn{equalHelGrav} remains valid. Note that, although the opposite-helicity Compton amplitude is much more complicated than the equal-helicity one, they are both given by the same covariant expression obtained via the Lagrangian that we give in Section \ref{section:Lagrangian5o2}.

Before moving on to derive these expressions, let us comment about the possibility of including additional contact terms. We first note that the gauge theory amplitude in \eqn{finalcompton3/2} is dimensionless, and thus any additional terms must be dimensionless and free of any kinematic poles. This means that all the  spinors need to appear in the numerator; for example, the schematic minimal expression $|{\bf 1}\rangle^3 |{\bf 2}\rangle^3 |3\rangle^2 |4]^2/m^5$ has the right little-group weights and dimension to be a contact term corresponding to an operator quadratic in field strength $\sim (F_{\mu\nu})^2$. However, it clearly has a different $1/m$ behavior than the terms already present in \eqn{finalcompton3/2}, hence it seems implausible that such contact terms would conspire to further improve the high-energy behavior or the tree-level unitarity properties of the Compton amplitude. 

The same analysis can be done for the gravitational amplitude (\ref{Gfinalcompton5/2}), which is of dimension two.  A schematic minimal expression that has the correct little-group weights and dimension is $|{\bf 1}\rangle^5 |{\bf 2}\rangle^5 |3\rangle^4 |4]^4/m^7$, which corresponds to an operator quadratic in the Riemann tensor $\sim (R_{\mu\nu\rho \sigma})^2$. However, again it has a higher power of $1/m$ than the terms already present in \eqn{Gfinalcompton5/2} and thus there is no reason to think that the addition of such terms can improve the high-energy behavior or the tree-level unitarity properties of the Compton amplitude. For the theories we consider in the next section, we will see that operators that are quadratic in the field strength or Riemann tensor do not appear in  Lagrangians that have a minimal number of derivatives and satisfy the current constraint.


\section{Lagrangians and Compton amplitudes from $P \cdot J = {\cal O}(m)$ \label{sec5}}

It is well known from the higher-spin literature that the minimal-coupling prescription, i.e. only replacing ordinary derivatives with covariant ones, is not the most natural choice for higher-spin interactions in gauge theory and gravity. For instance, ref.~\EMUnitarity\ shows the necessity of non-minimal terms to avoid tree-level unitarity violations, and how this imposes the familiar value of $g = 2$ for the gyromagnetic ratio.

In this section we will discuss the natural Lagrangians for spinning massive matter, and how they are uniquely fixed by the current constraint $P \cdot J = {\cal O}(m)$ and counting of derivatives. We start with matter with low spin and work our way up in spin until it exceeds the spin of the gauge bosons (photon/graviton). While these Lagrangians are far from new, the purpose of this exercise is to connect the Lagrangians to the three-point and four-point amplitudes in Sections \ref{sec3} and \ref{SECT:msh}, which have been argued to describe the dynamics of Kerr black holes, as well as ``root-Kerr'' gauge theory solutions~\cite{Monteiro:2014cda,Arkani-Hamed:2019ymq}.

 Effective Lagrangians appropriate for describing black holes and other massive objects will in general contain additional higher-derivative terms that behave as $(\partial/m)^w$ for some positive power $w$. The scattering processes that the effective Lagrangians are geared towards describing include the Compton amplitude, but also any $n$-point process involving two massive higher-spin states and $n{-}2$ gravitons (or gluons). This implies that we are interested in higher-derivative operators that involve two higher-spin fields and $n{-}2$ factors of field strengths $R_{\mu \nu \rho \sigma}$ (or $F_{\mu \nu}$) for the massless gauge bosons. Thus, there is a minimum number of derivatives that an $n$-point amplitude must contain in order for such an operator to be present. In gravity it is $(\partial/m)^{2n-4}$ and in gauge theory it is $(\partial/m)^{n-2}$ derivatives. Whether or not one should add such $n$-point operators to the Lagrangian depends on the derivative counting that is already present in the amplitude from previously-added operators, as well as whether there are problems with the minimal-coupling Lagrangian that the new operators will cure. In the cases that we consider, the tree-level unitarity violation will be ameliorated by the addition of three-point operators, which then dictate the derivative counting such that it forbids possible $n{>}3$ higher-derivative operators. The details will be shown below.
Of course, one could still consider including such operators, but they are not expected to help with the tree-level unitarity considerations, and instead they could potentially re-introduce the tree-level unitarity problem. Thus, in this section we will make the minimalistic choice to not consider operators that are beyond linear in the field strength, up to spin-$5/2$ gravity and spin-$3/2$ gauge theory.

\subsection{Gauge-theory Lagrangians}

As a warm-up, we here consider massive spinning fields in a gauge theory, and we use the current constraint to find the allowed theories with the lowest number of derivatives. In particular, in Section~\ref{sec3}, we showed that the current constraint can be solved at cubic order with at most $\partial^{2\S-1}$ derivatives for a spin-$\S$ boson. Assuming that additional quartic interactions introduce no worse momentum scaling in the four-point amplitude, this implies an expectation of  $\partial^{2\S-1}\frac{1}{\Box} \partial^{2\S-1}=\partial^{4\S-4}$ derivatives for the quartic terms. Here we assumed that the propagator schematically behave as $\frac{1}{\Box}$ (in the high-energy limit), which can only be true if the current constraint is obeyed, as we will elaborate on shortly. Similarly, for a spin-$\S$ fermion, the current constraint can be solved at cubic order with at most $\partial^{2\S-2}$ derivatives, which implies the expectation of $\partial^{2\S-2}\frac{1}{\partial}\partial^{2\S-2}=\partial^{4\S-5}$ derivatives at quartic order.

More generally, we can do the same derivative counting in terms of schematic new $n$-point operators, under the assumption that their existence is tied to the non-minimal interactions at three points. For multiplicity $n\ge2$ and bosonic spin~$\S\ge 1$, the derivative counting of Section~\ref{sec3} permits operators
\be \label{non-min_L}
\Delta {\cal L}_n^{(\S)} ~ \sim ~ D^{2(n-2) (\S - 2) +2} \, \bar  \phi^\S \Big(\frac{F_{\mu\nu}}{m^{2 \S - 2}}\Big)^{n - 2} \phi^\S \,,
\ee
or with fewer derivatives. Here $D$ is the covariant derivative, $\phi^\S$ is the higher-spin field, and $F_{\mu\nu}$ is the photon field strength. For fermions, the same schematic formula is obtained except that the operators have exactly one derivative less. We see that the number of covariant derivatives becomes negative for $\S\le 3/2$ and $n\ge 4$, and thus such operators are not considered. 

\subsubsection{Spin~$1$ -- the $W$ boson}
\label{Higgs}

Let us begin from the simplest case where a current for the matter field can be first defined, that of a massive charged spin-$1$ field $W^\mu \equiv \phi^1$, coupled to the photon $A^\mu$. From the general considerations in Section~\ref{sec3}, we expect that the current constraint can be satisfied by cubic interactions with at most one derivative, and quartic interactions with zero derivatives. The Lagrangian Ansatz is then
\be
\label{Seq1QED}
\cL = 2 D_{[\mu} \overline{W}_{\nu]} D^{[\mu} W^{\nu]} - m^2 \overline{W}_{\mu} W^{\mu} + i e \alpha   \overline{W}^{\mu} W^{\nu} F_{\mu \nu} \,,
\ee
with covariant derivative $D_\mu = \partial_\mu - i e A_\mu$ and field strength $F_{\mu \nu} = 2 \partial_{[\mu} A_{\nu]}$. The free parameter $\alpha$ multiplies the non-minimal interaction, and the choice $\alpha = 0$ gives the minimally-coupled theory. 
Since quartic operators must come with derivatives in the form of $F^2$ terms, such operators are not allowed by our derivative counting. For convenience, we work with $e = 1$ in the rest of this section.

Let us compute the Compton amplitude, as well as the three-point current. To do so, we will need the massive spin-$1$ propagator,
\be
\Delta^{\rho \sigma}_{(1)}(p) =  \frac{i}{p^2 - m^2} \Big( \eta^{\rho \sigma} - \frac{p^\rho p^\sigma}{m^2} \Big).
\ee
Following ref.~\EMUnitarity, one notes that this propagator may introduce a mass singularity in the Compton amplitude. This can be phrased as tree-level unitarity problem: for momentum $p \gg m$, the amplitude may grow larger than unity. The problem can be avoided by constraining the off-shell current that is contracted into the propagator.

Let $J^{\mu}(-P,p_2,p_3)$ be the three-point current of the massive spin-$1$ field, $\mu$ its free Lorentz index and $P = p_2 + p_3$ its off-shell momentum. The momenta $p_2$, $p_3$ correspond to an on-shell spin-$1$ boson and an on-shell photon, respectively. Suppose that the current satisfies
\be
\label{PdotJ}
P_\mu J^{\mu} (-P,p_2,p_3) = m\, X(-P,p_2,p_3)\,,
\ee
with $X(-P,p_2,p_3)$ bounded as $m \to 0$. With these conditions, the potentially dangerous mass singularity in the propagator factor $P^\mu / m$ is neutralized when contracted into the current as one builds up the Feynman diagrams. Hence, the current constraint is a necessary condition for avoiding the tree-level unitarity violation.

Explicit computation using the Lagrangian (\ref{Seq1QED}) yields
\be
P_\mu J^{\mu} (-P,p_2,p_3) = i m^2 \bep_2 \cdot \varepsilon_3 + i (1 - \alpha) (\varepsilon_3 \cdot P \, \bep_2 \cdot p_3 - \bep_2 \cdot \varepsilon_3 \, p_3 \cdot P)\,,
\ee
which confirms that tree-level unitarity requires $\alpha = 1$. Not surprisingly, this is precisely the value that is required for $W$-bosons in a spontaneously-broken Yang-Mills theory, such as the Standard Model. 

We may confirm that $\alpha = 1$ reproduces the expected amplitudes. The three-point amplitude, with $P \to -p_1$ on shell, is given by
\be
\cA(1W,2\overline W,3A) =  - 2 i \bep_1 \cdot \bep_2 \, \varepsilon_3 \cdot p_1 - i (1 + \alpha) \bep_{2 \mu} f_3^{\mu \nu} \bep_{1 \nu} \,,
\ee
where $f_3^{\mu \nu} = p_3^\mu \varepsilon_3^\nu - p_3^\nu \varepsilon_3^\mu$. With photon helicity $\varepsilon_3 = \varepsilon_3^-$, we have
\be
A(1W,2\overline W,3A^-) =  -\frac{(1+\alpha) \xfactor^{-1}}{2 m} [\bm1 \bm2]^2 - \frac{(1-\alpha) \xfactor^{-1}}{2 m} [\bm1 \bm2] \angle{\bm1 \bm2} \,,
\ee
which reduces to \eqn{AHH_amps} for $\alpha = 1$. Furthermore, at four points, one finds that $\alpha = 1$ reproduces exactly the spin-$1$ Compton amplitude in~\eqn{comptonnima}. Of course, we could have directly matched the amplitudes to \eqn{AHH_amps} or \eqn{comptonnima}, which would have given the same $\alpha = 1$ solution. However, as we will demonstrate with more examples, the current constraint  pins down both the Lagrangians and the amplitudes with very little input.

Before moving on to higher spin, let us discuss some salient points. In the $m \to 0$ limit, the current constraint  (\ref{PdotJ}) becomes the Ward identity corresponding to gauge invariance of a massless spin-$1$ particle. It is not surprising that the resulting Lagrangian (\ref{Seq1QED}) is built out of the standard interaction terms for $W$-bosons, which in the massless limit reduces to non-abelian Yang-Mills theory. Here we are not interested in the $W^4$ contact terms and the Higgs field that are also present in such a theory because, with an eye towards the application of classical scattering with very massive compact objects (black holes in gravity), such terms would correspond to quantum corrections that are irrelevant. Thus, we are only interested in Lagrangians that are quadratic in the massive spinning field.

\subsubsection{Spin~$3/2$ -- charged gravitino} 
\label{gravitino_section}

We now discuss the Lagrangian for photons coupled to a massive charged spin-$3/2$ field $\psi_\mu$, which for convenience we often call ``gravitino''. We follow the approach of ref.~\HSQED\ and consider the most general Lagrangian linear in the field strength, which precisely matches our expectation on the maximum number of allowed derivatives, as discussed in the beginning of this section. Although the considerations in Section~\ref{sec3} ruled out parity-odd terms, we include them  here for illustrative purposes. The Lagrangian Ansatz consists of the Rarita-Schwinger Lagrangian~\cite{Rarita:1941mf} plus non-minimal terms,
\begin{multline}
\label{gravitinolagrangian}
\cL = \bar{\psi}^\mu \gamma_{\mu \nu \rho} \Big( i D^\nu - \frac{1}{2} m \gamma^\nu \Big) \psi^\rho 
- \frac{i e}{m} \Big(
l_1 \bar{\psi}_\mu F^{\mu \nu} \psi_\nu 
+ l_2 \bar{\psi}_\mu F_{\rho \sigma} \gamma^\rho \gamma^\sigma \psi^\mu + l_3 F^{\mu \nu} ( \bar{\psi}_\mu \gamma_\nu \gamma \cdot \psi  \\
+ \bar{\psi} \cdot \gamma \gamma_\mu \psi_\nu )
+ l_4 \bar{\psi} \cdot \gamma F_{\rho \sigma} \gamma^\rho \gamma^\sigma \gamma \cdot \psi 
+ i l_5 F^{\mu \nu} ( \bar{\psi}_\mu \gamma_\nu \gamma \cdot \psi - \bar{\psi} \cdot \gamma \gamma_\mu \psi_\nu ) 
\Big) \,,
\end{multline}
with the same covariant derivative and field strength as defined in the previous subsection. The gravitino is described by a vector-spinor field, with the free equation of motion $\gamma \cdot \psi = (i \slashed\partial - m) \psi_\mu = 0$.  

It is known that the parameter values $l_1 = -2, l_2 = 1/2, l_3 = 1,l_4 = -1/2, l_5=0$ uniquely match a truncation of gauged $\cN = 2$ supergravity~\cite{Freedman:1976aw}. As a spoiler, this choice precisely gives the three-point amplitude (\ref{AHH_amps}), as well as the Compton amplitudes (\ref{comptonnima}) and (\ref{finalcompton3/2}). Furthermore, these values are uniquely fixed  by the property $P \cdot J = {\cal O}(m)$. With this said, let us clean up the Lagrangian corresponding to this parameter choice,
\be
\label{gravitinosugralagrangian}
\cL = \bar{\psi}^\mu \gamma_{\mu \nu \rho} \Big( i D^\nu - \frac{1}{2} m \gamma^\nu \Big) \psi^\rho + \frac{i e}{m} \bar\psi_{\mu} \cF^{\mu \nu} \psi_\nu\,,
\ee
where $\cF^{\mu \nu} \equiv F^{\mu \nu} - i/2 \,\gamma^5 \epsilon^{\mu \nu \rho \sigma} F_{\rho \sigma}$ is a gamma-matrix-augmented field strength, which is parity even.

Let us confirm the result. Consider the three-point amplitude in the Lagrangian Ansatz, and note that $l_3$, $l_4$ and $l_5$ do not contribute due to the gamma-tracelessness of on-shell gravitinos. Assuming $\varepsilon_3 = \varepsilon_3^-$, we have
\begin{equation}
\cA(1\phi^{3/2},2\bar\phi^{3/2},3A^-) = i \bar{u}_{2 \rho} \slashed\varepsilon_3 v_1^\rho + i \frac{l_1}{m} \bar{u}_{2 \mu} f_3^{\mu \nu} v_{1 \nu} + i \frac{l_2}{m} \bar{u}_{2 \mu} f_{3 \rho \sigma} \gamma^\rho \gamma^\sigma v_1^\mu\,,
\end{equation}
which leads to the spinor-helicity expression 
\be
A(1\phi^{3/2} ,2\bar\phi^{3/2} ,3A^-) = - \frac{\xfactor^{-1}}{2 m^2} [\bm1 \bm2]^3 \left( \! -2 \frac{\angle{\bm1 \bm2}}{[\bm1 \bm2]} + l_1 \left(1 \! - \! \frac{\angle{\bm1 \bm2}^2}{[\bm1 \bm2]^2}\right) + 4 l_2 \left(\frac{\angle{\bm1 \bm2}}{[\bm1 \bm2]} \! - \! \frac{\angle{\bm1 \bm2}^2}{[\bm1 \bm2]^2}\right) \! \right)\! .
\ee 
One sees that  $l_1 = -2$ and $l_2 = 1/2$ reproduce \eqn{AHH_amps}. 

Next, consider the calculation of the Compton amplitude. We need the massive gravitino propagator that we derived in \eqn{Prop3/2}; here it is in full detail:
\be
\label{Prop3/2b}
\Delta^{\rho \sigma}_{(3/2)}(p) = 
\frac{i}{p^2 - m^2 } \left(    \Big( \eta^{\rho \sigma} - \frac{p^\rho p^\sigma}{m^2} \Big) \left( \slashed{p} + m \right) + \frac{1}{3} \Big( \frac{p^\rho}{m} + \gamma^\rho \Big) \left( \slashed{p} - m \right) \Big( \frac{p^\sigma}{m} + \gamma^\sigma \Big)  \right) .
\ee
Similar to the spin-$1$ case, the above propagator introduces mass divergences in the Compton amplitude. Note that, with the field content considered, we will not be able to find a special theory that has a finite massless limit. If this were possible, it would result in a massless spin-$3/2$ field coupled to the photon, and this is forbidden by various no-go results~\cite{Johnson:1960vt, Velo:1969bt}, including the Weinberg-Witten theorem~\cite{Weinberg:1980kq}. The only known completion of a massless spin-$3/2$ theory is within the context of supergravity, but that would require that we add a graviton as well as other fields to the Lagrangian~\cite{Das:1976ct, Deser:1977uq,Freedman:1976aw}. For a review on these topics, see ref.~\HSlectures.

Nonetheless, there is a notion of unitarity violation in the Lagrangian~(\ref{gravitinolagrangian}) that we can cure. A detailed argument can be found in ref.~\GravUnitarity, here we give a brief summary. Imposing the constraint $P \cdot J = {\cal O}(m)$ ensures that the mass divergences due to the longitudinal part of the current drop out. However, the addition of non-minimal terms in the Lagrangian will in general introduce  explicit poles $1/m$, as can be seen in \eqn{gravitinolagrangian}, and hence give new mass divergences in the transverse part of the current $J_{\bot}$, which automatically obeys $P \cdot J_{\bot} = 0$. One can argue that this is a separate problem that needs to be solved via other methods, such as introducing new fields, including possibly an infinite tower of higher-spin fields~\GravUnitarity. Such an approach, however, cannot solve the unitarity violation in the longitudinal current, and hence the current constraint is yet again to be interpreted as a \textit{necessary} condition for tree-level unitarity. 

Now we follow the same procedure as in the spin-$1$ case. Let us define the three-point current $J^{\mu}(-P,p_2,p_3)$, where $P$ is off shell, and $p_2$ and $p_3$ are on-shell. From \eqn{gravitinolagrangian} we compute the current
\begin{multline}
\label{3o2curr}
J^{\mu} (-P,p_2,p_3) v_{P \mu} = \\
i \bar{u}_{2 \mu} \gamma^{\mu \nu \rho} \varepsilon_{3 \nu} v_{P \rho}
+ \frac{i l_1}{m} f_3^{\mu \nu} \bar{u}_{2 \mu} v_{P \nu}
+ \frac{2 i l_2}{m} \bar{u}_{2 \mu} \slashed{p}_3 \slashed{\varepsilon}_3 v_P^\mu
+ \frac{i l_3 - l_5}{m}  f_3^{\mu \nu} \bar{u}_{2 \mu} \gamma_\nu \gamma \cdot v_P\,,
\end{multline}
where $\bar{u}_{2 \mu}$ is an on-shell vector-spinor, and $v_{P \mu}$ is an auxiliary vector-spinor for the off-shell leg. To check the current constraint, we replace $v_{P \mu}\rightarrow P_\mu v_P  $,  where $v_P$ is an off-shell spinor, giving
\begin{multline}
 J^{\mu } (-P,p_2,p_3) P_\mu v_P  = 
i m \varepsilon_3 \cdot \bar{u}_2 v_P
- i ( 1 - l_3 - i l_5 ) f_3^{\mu \nu} \bar{u}_{2 \mu} \gamma_\nu v_P \\
+ \frac{i}{m} ( l_1 + 2 l_3 + 2 i l_5 ) f_3^{\mu \nu} p_{2 \mu} \bar{u}_{2 \nu} v_P
+ \frac{i}{m} ( 2 l_2 - l_3 - i l_5 ) k \cdot \bar{u}_2 \slashed{\varepsilon_3} \slashed{p}_3 v_P \,. 
\end{multline}
To have $P \cdot J = {\cal O}(m)$, we need all terms except the first one to vanish. Since the $l_i$ are real parameters, this implies $l_1 = -2, l_2 = 1/2, l_3 = 1,l_4 = -1/2, l_5 = 0$, as promised.

We proceed to compute the Compton amplitude for the truncated supergravity theory. The ingredients required are the three-point vertex (\ref{3o2curr}) and the propagator (\ref{Prop3/2}). The final result is best presented in terms of spinor-helicity notation, and it exactly gives the two formulae already quoted in eqs.~(\ref{equalHelPhoton}) and (\ref{finalcompton3/2}) for the equal- and opposite-helicity cases, respectively.  This confirms that the equal-helicity amplitude is not modified by the contact term that cures the spurious-pole problem of the opposite-helicity Compton amplitude. Also, by the analysis of Lagrangian properties (current constraint and derivative counting)  we can rule out the presence of $F^2$ operators in a well-behaved and hairless spin-$3/2$ gauge theory.  By ``hairless" we mean that there are no physical properties (e.g.~Wilson coefficients) that characterize the spin-$3/2$ particle other than its mass, spin and charge.

\subsection{Gravity Lagrangians}

We now proceed to analyze gravitational Lagrangians and amplitudes for spinning matter. We find that imposing $P \cdot J = {\cal O}(m)$ on the currents uniquely fixes the Lagrangians up to spin~5/2, which in turn gives the tree-point amplitudes in \eqn{Gnima3pt}, as well as the unique Compton amplitudes that we presented in Section \ref{SECT:msh}.

In analogy with the gauge-theory discussion around \eqn{non-min_L}, we will work with gravitational theories that can satisfy the current constraint with the fewest number of derivatives in the non-minimal interactions. Given that we already know the three-point derivative counting from the discussion in Section~\ref{sec3}, we can infer the multiplicity-$n$ counting by assuming that they give rise to no worse behavior.  For spin-$\S$ bosons we get that the derivative counting is compatible with $n$-point operators of the schematic form
\be
\Delta {\cal L}_n^{(\S)} ~ \sim ~ \nabla^{2(n-2) (\S - 3) +2} \, \bar  \phi^\S \Big(\frac{R_{\mu\nu\rho\sigma}}{m^{2 \S - 4}}\Big)^{n - 2} \phi^\S \,,
\ee
where $\nabla$ is the covariant derivative, $\phi^\S$ is the higher-spin field, and $R_{\mu\nu\rho\sigma}$ is the Riemann tensor. For fermions, the same schematic formula is obtained except that the operators have exactly one derivative less. The number of derivatives becomes negative for $\S\le5/2$ and $n\ge4$, thus they are not considered. For further details on the non-minimal terms, see ref.~\cite{Chung:2019duq} where an on-shell action was given for arbitrary spin up to $n=3$, which matches with the three-point amplitudes of ref.~\Nima.

\subsubsection{Spin~$1/2$ -- standard fermion} 
We start by considering the simplest example of a spin-$1/2$ massive field $\fermion\equiv  \phi^{\S=1/2} $ coupled to gravity. While this case is somewhat trivial, it gives a warm-up that highlights how fermions couple to gravity. Since the spin-$1/2$ current will not have any free Lorentz indices, the current constraint does not apply. However, it then automatically follows that the minimally-coupled theory is free of tree-level unitarity problems. For convenience, we work with $\kappa = 1$ in this section.

The minimally-coupled Lagrangian is given by the fermion kinetic and mass terms, $\cL =  \sqrt{- g} \, \bar\fermion( i \slashed{\nabla} - m) \fermion$, which we expand out for convenience,
\be
\label{spin1o2lagrangian}
  \sqrt{- g}\, \bar\fermion( i \slashed{\nabla} - m) \fermion = \bar\fermion( i \slashed{\partial} - m) \fermion - \frac{i}{2}\bar\fermion \gamma^a h_a^\mu \partial_\mu  \fermion +  \frac{i}{8} \bar\fermion \gamma^\mu \omega^{a b}_\mu [\gamma_a,\gamma_b] \fermion + {\cal O}(h^2, \eta^{\mu \nu} h_{\mu \nu}) .
\ee
 We here distinguish between Greek indices $\{\mu,\nu,\dots\}$ used for curved coordinate indices, and Latin indices $\{a,b,\dots\}$ used for flat indices, such as in the tetrad $e_a^\mu$. However, when there is no potential for confusion, such as in amplitude and current expressions, we resort to using Greek indices for flat space contractions.  

The covariant derivative is written explicitly as
\begin{equation}
 \nabla_\mu \fermion = \partial_\mu \fermion + {1 \over 8} \omega^{a b}_\mu [\gamma_a,\gamma_b] \fermion \,,
\end{equation}
in terms of the spin connection $({\omega^{a b}})_\mu \equiv e^a_\nu \nabla_\mu e^{b \nu} = (1/2) (\eta^{b c} \Gamma^a_{c \mu} - \eta^{a c} \Gamma^b_{c \mu}) + {\cal O}(h^2)$. 
Gamma matrices with curved indices are defined in terms of the flat-space gamma matrices and the tetrad as  $\gamma^\mu = \gamma^a e_a^\mu$. 
In turn, the tetrads are expanded in terms of the fluctuations of the metric $h_{\mu\nu}$ as $e_a^\mu = \delta_a^\mu - (1/2) h_a^\mu + {\cal O}(h^2)$. 
Hence, there will be additional contributions to the vertices coming from the expansion of the tetrad fields which dress the flat-space gamma matrices. 
This corresponds to the second term in \eqn{spin1o2lagrangian}.

Now we wish to compute the three-point amplitude. The connection one-form term does not contribute if the graviton is on shell. 
This is because the symmetry of the Christoffel symbol and the Clifford algebra relations can be used to rewrite it in terms of $\eta^{\mu \nu} \Gamma^\lambda_{\mu \nu}$ and $\Gamma^\mu_{\mu \nu}$, both vanishing on-shell. The second term instead gives
\begin{equation}
\label{spin1o23pt}
\cM(1\phi^{1/2},2\bar\phi^{1/2},3h) = \frac{i}{2} \varepsilon_3 \cdot p_1 \, \bar{u}_2 \slashed{\varepsilon}_3 v_1 \, .
\end{equation}
Assuming the helicity choice $\varepsilon_3 = \varepsilon_3^-$, we have
\begin{equation}
\cM(1\phi^{1/2},2\bar\phi^{1/2},3h^-)
=  - i \frac{m \xfactor^{-2}}{2} [\bm1 \bm2] \,,
\end{equation}
in agreement with \eqn{Gnima3pt}. 

The Compton amplitude can be computed by expanding the kinetic term to second order in $h^{\mu\nu}$, and also considering the cubic graviton self-interactions coming from the Einstein-Hilbert action. We will not give the details of the calculation here, as it has little bearing on the higher-spin current argument. We can instead rely on BCFW recursion or double copy to argue that the Compton amplitude must agree with \eqn{Gcomptonnima}. For minimally coupled matter of spin~$s\le 2$, BCFW recursion is expected to give correct Compton amplitudes with no contributions from infinity of the complex plane (see e.g.~refs.~\cite{Arkani-Hamed:2008bsc,Cohen:2010mi,Arkani-Hamed:2017jhn,Johansson:2019dnu}). The double copy is also expected to hold for Compton amplitudes of minimally coupled massive matter with spin~$s\le 2$ (see e.g.~refs.~\cite{Chiodaroli:2015rdg,Chiodaroli:2017ehv,Chiodaroli:2018dbu,Ochirov:2018uyq,Johansson:2019dnu,Bautista:2019evw,Edison:2020ehu}). 

\subsubsection{Spin~$1$ -- Proca boson }

Like the previous subsection, the spin-$1$ case is somewhat trivial, but for the sake of completeness we discuss it briefly. Consider a massive spin-$1$ field $W_\mu  \equiv \Phi^{\S=1} $ coupled to gravity. The Lagrangian is obtained by covariantizing the Proca Lagrangian, which is essentially the quadratic piece of the Lagrangian (\ref{Seq1QED}),
\be \label{Lag1}
\cL = \sqrt{-g}\,\big(2 \nabla_{[\mu} \overline{W}_{\nu]} \nabla^{[\mu} W^{\nu]} - m^2 \overline{W}_{\mu} W^{\mu}\big) \ .
\ee
Here the covariant derivative is given by
\begin{equation}
\nabla_\mu W_\rho = \partial_\mu W_\rho  - \Gamma^\lambda_{\rho \mu} W_\lambda \ .
\end{equation}
The current constraint is automatically satisfied in this case. This is not surprising given that the Proca field can be thought of as a $W$ boson in a spontaneously-broken U(1) gauge theory in the St{\"u}ckelberg formulation.

Explicit computation using the Lagrangian  (\ref{Lag1}) yields the amplitude
\be
\cM(1W,2\overline{W},3h) =  i \bep_1 \cdot \bep_2 \ p_1 \cdot  \varepsilon_{3}\, \varepsilon_{3} \cdot   p_2 -i p_1 \cdot \bep_2 \  \bep_1 \cdot  \varepsilon_{3} \,\varepsilon_{3} \cdot   p_2 - i p_2 \cdot \bep_1 \ p_1  \cdot  \varepsilon_{3}\, \varepsilon_{3} \cdot  \bep_2\,.
\ee
If we take $\varepsilon_3 = \varepsilon_3^-$, we obtain the amplitude
\be
\cM(1W,2\overline{W},3h^-) = - {i \over 2} \xfactor^{-2} [\bm1 \bm2]^2 \ ,
\ee
in agreement with \eqn{Gnima3pt}. We will not work out the Compton amplitude, since it was already checked in ref.~\cite{Johansson:2019dnu} using Feynman rules that it agrees with the amplitudes~(\ref{Gcomptonnima}). That reference also showed that the double copy provides a streamlined method for computing multi-graviton scattering with massive spin-1 vectors, by means of computing the much simpler quark-gluon diagrams of QCD~\cite{Johansson:2015oia,Johansson:2019dnu}.

\subsubsection{Spin~$3/2$ -- massive gravitino}

We consider the minimally-coupled theory of a massive spin-$3/2$ field $\psi_\mu \equiv \phi^{\S=3/2}$ coupled to gravity. The Lagrangian can be read off from the minimal part of \eqn{gravitinolagrangian}, and this is simply the covariantization of the Rarita-Schwinger Lagrangian,
\begin{multline}
\label{spin3o2lagrangian}
\cL = \sqrt{- g} \bar\psi_\mu \gamma^{\mu \rho \sigma} \Big( i \nabla_\rho - \frac{1}{2} m \gamma_\rho \Big) \psi_\sigma
\equiv \cL_{free} + \cL_e + \cL_\omega + {\cal O}(h^2,\eta^{\mu \nu} h_{\mu \nu})\,,
\end{multline}
where
\begin{align}
\cL_e &= - \frac{1}{2} \bar\psi_\mu ( {h^\mu}_a \gamma^{a \rho \sigma} +  {h^\rho}_a \gamma^{\mu a \sigma} +  {h^\sigma}_a \gamma^{\mu \rho a} ) i \partial_\rho \psi_\sigma
- \frac{m}{2} \bar\psi_\mu ( {h^\mu}_a \gamma^{a \rho} + {h^\rho}_a \gamma^{\mu a} ) \psi_\rho \, , \nn \\
\cL_\omega &= \frac{i}{8} \bar\psi_\mu \gamma^{\mu \rho \sigma} \omega^{a b}_\rho [\gamma_a,\gamma_b]  \psi_\sigma
=  \frac{i}{8} \bar\psi_\mu \gamma^{\mu \rho \sigma} \Gamma^a_{b \rho} [\gamma_a,\gamma^b]  \psi_\sigma  \,.
\end{align}
Note that, at ${\cal O}(h)$, there is no difference between the Latin and Greek indices in $\cL_e$ and $\cL_\omega$, since ${e^\mu}_a = {\delta^\mu}_a$ at zeroth order. Also note that the above Lagrangian, in the massless limit, becomes the gravitino Lagrangian in supergravity~\cite{Freedman:1976xh, Deser:1976eh}. The covariant derivative here is given by 
\begin{equation}
\nabla_\mu \psi_\rho = \partial_\mu \psi_\rho + {1  \over 8} \omega^{a b}_\mu [\gamma_a,\gamma_b] \psi_\rho - \Gamma^\lambda_{\rho \mu} \psi_\lambda \ ,
\end{equation}
but the last term drops out due to the antisymmetry of $\gamma^{\mu \rho \sigma}$. 

We wish to compute $\cM(1\phi^{3/2},2\bar\phi^{3/2},3h)$, but, having the Compton amplitude in mind, we can start by keeping the gravitinos off-shell and derive the three-point Feynman rules in momentum space. 
The three-point current contribution from $\cL_e$ is given by
\begin{multline}
\label{3ptoffshell1}
-\frac{i}{2} \{  \varepsilon_3 \cdot \bar{u}_2 \gamma (\varepsilon_3 , P , v_P) +  \varepsilon_3 \cdot P \gamma (\bar{u}_2 , \varepsilon_3 , v_P) + \gamma (\bar{u}_2 , P , \varepsilon_3) \varepsilon_3 \cdot v_P  \\
 + m \varepsilon_3 \cdot \bar{u}_2 \gamma(\varepsilon_3 , v_P) + m \gamma(\bar{u}_2 , \varepsilon_3) \varepsilon_3 \cdot v_P \} ,
\end{multline}
which for on-shell gravitinos reduces to
\begin{equation}
\label{3ptonshell1}
\frac{i}{2} \varepsilon_3 \cdot p_1 \bar{u}_{2 \rho} \slashed{\varepsilon}_3 v_1^\rho + \frac{i}{2} p_3 \cdot \bar{u}_2 \slashed{\varepsilon}_3 \varepsilon_3 \cdot v_1\, .
\end{equation}
Similarly we can find the contribution from $\cL_\omega$,
\begin{equation}
\label{3ptoffshell2}
\frac{i}{8} \bar{u}_{2 \mu} ( \gamma^\mu \varepsilon_3^\nu - \gamma^\nu \varepsilon_3^\mu ) f_{3 a b} \gamma^a \gamma^b v_{P \nu} \,,
\end{equation}
which on-shell reduces to
\begin{equation}
\label{3ptonshell2}
- \frac{i}{2} \varepsilon_3 \cdot \bar{u}_2 \slashed{\varepsilon}_3 p_3 \cdot v_1 \, .
\end{equation}

The graviton is on-shell in all the above expressions. Summing the contributions in \eqn{3ptonshell1} and \eqn{3ptonshell2} yields
\begin{equation}
\label{spin3o23pt}
\cM(1\phi^{3/2},2\bar\phi^{3/2},3h) = \frac{i}{2} \varepsilon_3 \cdot p_1 \bar{u}_{2 \rho} \slashed{\varepsilon_3} v_1^\rho + \frac{i}{2} f_{3 \rho \sigma} \bar{u}_2^\rho \slashed{\varepsilon}_3 v_1^\sigma \, .
\end{equation}
Assuming $\varepsilon_3 = \varepsilon_3^-$, we have
\begin{equation}
\label{}
\frac{i}{2} \varepsilon_3 \cdot p_1 \bar{u}_{2 \rho} \slashed{\varepsilon_3} v_1^\rho + \frac{i}{2} f_{3 \rho \sigma} \bar{u}_2^\rho \slashed{\varepsilon}_3 v_1^\sigma
=  i\frac{\xfactor^{-2}}{2 m} [\bm1 \bm2]^3\,,
\end{equation}
matching \eqn{Gnima3pt}. Now we wish to check the property $P \cdot J = {\cal O}(m)$. We consider the current $J_\mu (-P,p_2,p_3)$ with two massive gravitinos of momenta $-P$ and $p_2$ and one graviton of momentum $p_3$. From \eqn{3ptoffshell1} and \eqn{3ptoffshell2} we have:
\begin{multline}
\label{}
J^\mu (-P,p_2,p_3) v_{P \mu} = 
\frac{i}{2} \Big(
f_{3 \mu \nu} \bar{u}_2^\mu \gamma^\nu \varepsilon_3 \cdot v_P 
- \varepsilon_3 \cdot p_2 \bar{u}_{2 \rho} \slashed{\varepsilon}_3 v_P^\rho
+ \varepsilon_3 \cdot \bar{u}_2 \slashed{\varepsilon}_3 P \cdot v_P \\
- \varepsilon_3 \cdot \bar{u}_2 \slashed{\varepsilon}_3 \slashed{p_3} \slashed{v}_P
- \frac{1}{2} \varepsilon_3 \cdot \bar{u}_2 \gamma^\nu \slashed{p}_3 \slashed{\varepsilon}_3 v_{P \nu}
\Big) \,,
\end{multline}
where $v_P^\nu$ is a reference vector-spinor, and hence:
\begin{equation}
\label{Gspin3o2PdotJ}
J^\mu (-P,p_2,p_3) P_\mu v_{P} = \frac{i m}{4} \varepsilon_3 \cdot \bar{u}_2 ( \slashed{P} + m ) \slashed{\varepsilon}_3 v_P \,,
\end{equation}
where $v_P$ is a reference spinor. 

The remaining task is to compute the Compton amplitude, and show that it equals \eqn{Gcomptonnima}. However, this is a standard text-book calculation, equivalent to that of considering a massive gravitino amplitude in a spontaneously broken supergravity, so we will not do it explicitly. We use an indirect argument instead. Due to the property (\ref{Gspin3o2PdotJ}), the gravitino propagator is not responsible for any $1/m$ poles in the amplitude. Any such poles must come from the longitudinal part of the polarization vectors, at most one $1/m$ factor for each external gravitino. Thus the Compton amplitude obtained from the Lagrangian~(\ref{spin3o2lagrangian}) at worst scale as $1/m^2$. Any local ambiguity of the Compton amplitude must correspond to a gauge-invariant operator, which on shell is a polynomial in the spinors. For example, consider a schematic local expression $|{\bf 1}\rangle^3 |{\bf 2}\rangle^3 |3\rangle^4 |4]^4/m^5$, which has the correct dimension at the expense of too many powers of $1/m$. Thus~\eqn{Gcomptonnima} is the only possible answer compatible with the Lagrangian~(\ref{spin3o2lagrangian}). As can be seen, the Compton amplitudes (\ref{Gcomptonnima}) have in fact no $1/m$ poles, because in the $m\rightarrow 0$ limit there is a hidden symmetry that protects it, namely, supersymmetry.

\subsubsection{Spin~$2$ -- Kaluza-Klein graviton}
\label{KK_section}
In this subsection, we discuss a convenient way to obtain the Lagrangian for a massive spin-2 field $H_{\mu\nu}\equiv \phi^{\S=2} $. The starting point is the Einstein-Hilbert action in one dimension higher,
\begin{equation}
S_{EH} =  -2  \!  \int \! d^5x\, \sqrt{-g}\, R \, .   
\end{equation}
The action is expanded in terms of the graviton field as $g_{\hat \mu \hat \nu} = \eta_{\hat \mu \hat \nu} + h_{\hat \mu \hat \nu}$. Upon dimensional reduction to four dimensions, the five-dimensional graviton yields a rank-two symmetric tensor, a vector  and a scalar,
\begin{equation}
h_{\hat \mu \hat \nu} = \big\{ h_{ \mu  \nu} \ , \ a_\mu \equiv h_{ \mu 5} \ , \ \phi \equiv h_{55} \big\}  \ .
\end{equation}
Masses are associated to the five-dimensional components of momenta. Massive and massless modes can be separated as
\begin{equation}
h_{\hat \mu \hat \nu} = h^0_{ \mu  \nu} + \sum_{m} H^{(m)}_{\mu \nu}  \, .
\end{equation}
The summation runs over the whole Kaluza-Klein tower of states and involves an infinite number of massive modes. Having Compton amplitudes in mind, we only keep terms in the Lagrangian that have two massive fields with equal mass (up to a sign).\footnote{Note that this is not a consistent truncation in the conventional sense as general amplitudes from the truncated Lagrangian cannot be obtained from the full Kaluza-Klein theory by simply inserting projectors on the external states. However, for the subsector that contains the Compton amplitude, as well as the multi-graviton generalizations of the Compton amplitude, the truncation can be done consistently.} We now indicate the massive spin-$2$ fields as $\bH_{\mu \nu}$,$H_{\mu \nu}$ and massive vectors and scalars as $ A_{\mu }, \bar A_{\mu }$ and $\Phi, \overline \Phi$.

Using the equivalent of the $R_\xi$ unitarity gauge, we introduce a gauge-fixing term of the form
\be
{\cal L}_{\text{gf}} = -{2 \over \xi} \Big| \partial_{\nu} H^{ \nu}_{\mu} - \partial_\mu H^{\nu}_\nu -  i m \xi A_\mu - {1 \over 2} \partial_\mu \Phi \Big|^2 - {2 \over \xi} \Big| \partial_\mu A^{\mu} + {3 \over 2}i  m H^{ \mu}_\mu - {i \over 2} m \xi \Phi \Big|^2 \,.
\ee
Additionally, the spin-2 field is redefined as 
\begin{equation}
H_{\mu \nu} \rightarrow H_{\mu \nu} + {1 \over 2} \eta_{\mu \nu} \Phi  \,.
\end{equation}
With these, the kinetic term becomes diagonal in the different fields. In the $\xi \rightarrow \infty$ limit, vector and scalar become infinitely massive and can be integrated out (in fact, these fields contain no physical states, they are pure gauge). By carrying out an explicit calculation,
we find that the Compton amplitude from Kaluza-Klein gravity is reproduced by the simple Lagrangian
\begin{eqnarray}{\cal L} &=& \sqrt{-g}\, \big(\nabla_\mu \bH_{\nu \rho} \nabla^\mu H^{\nu \rho}   - 2 \nabla_\nu \bH_{\mu}^{\nu} \nabla^\rho H^{\mu}_{ \rho}   -   \bH^\rho_{ \rho} \nabla_\mu \nabla_\nu H^{\mu \nu}  - H^\rho_{ \rho}  \nabla_\mu \nabla_\nu \bH^{\mu \nu}  \no \\
&& ~~~~~~~~ - \nabla_\mu \bH^{ \nu}_\nu \nabla^\mu  H_{\rho}^{\rho} - m^2  \bH_{\mu \nu} H^{\mu \nu}   +  m^2  \bH^{ \mu}_{\mu } H^{ \nu}_\nu   -
2 R^{\mu \nu \rho \sigma} \bH_{\mu \rho} H_{\nu \sigma} \big) \, . \label{FierzPauli+R}
\end{eqnarray} 
This is a complex version of the Fierz-Pauli Lagrangian \cite{Fierz:1939ix} with the addition of a non-minimal coupling proportional to the Riemann tensor. It is instructive to consider the case in which the non-minimal term has a free coefficient,
\begin{equation}
 \cL_\text{non-min} = -2\alpha   \sqrt{-g}\, R^{\mu \nu \rho \sigma} \bH_{\mu \rho} H_{\nu \sigma} \, .
\end{equation}
A direct calculation of the amplitude at three points leads to
\begin{equation}
\cM(1 H, 2 \bH, 3 h^-) =  i  {\xfactor^{-2} \over 2 m^2} [{\bf 12}]^4 \Bigg\{ 1 + (\alpha-1) \Big( 1 - {\langle {\bf 12} \rangle \over [{\bf 12}]} \Big)^2 \Bigg\} \, .
\end{equation}
This reproduces the three-point amplitude from Arkani-Hamed, Huang and Huang provided that $\alpha=1$ \cite{Arkani-Hamed:2017jhn}.

The same Lagrangian, up to total-derivative terms, can be obtained by calculating the longitudinal part of the off-shell current 
\begin{equation}
P^\nu J_{\mu\nu} ( -P, p_2, p_3) =i  {( \alpha-1  )} \big( P^2 \, {\bep}_2 \cdot \varepsilon_3 + 2 P \cdot  {\bep}_2 \, P \cdot  \varepsilon_3 \big)
\big(  p_3 \cdot {\bep}_2 \ \varepsilon_{3\mu} -p_{3\mu} \, {\bep}_2 \cdot \varepsilon_3 \big)+ {\cal O} (m^2)  .
\end{equation}
Imposing that the current vanishes as $m^2$ in the $m \rightarrow 0$ limit fixes again $\alpha=1$. 
Starting with a more general Ansatz in which all terms in the Lagrangian (\ref{FierzPauli+R}) have free coefficients,
 this condition fixes completely the theory up to terms proportional to the mass. 

The Compton amplitude can be calculated using the Feynman rules coming from the Lagrangian~(\ref{FierzPauli+R}). But it is easier to use the Kaluza-Klein trick directly on a massless graviton amplitude. Starting from a covariant four-graviton amplitude, two of the external momenta are chosen as $p^{\hat \mu}_1= (p^\mu_1,m)$, $p^{\hat \mu}_2= (p^\mu_2,-m)$, and the remaining momenta $p_3, p_4$, as well as all the polarizations, are chosen to only have four-dimensional components. After plugging in these momenta and polarizations into the amplitude, and converting to spinor-helicity notation, the answer agrees with~\eqn{Gcomptonnima}.

\subsubsection{Spin~5/2}
\label{section:Lagrangian5o2}

The Lagrangian for a spin-$5/2$ field minimally coupled to gravity can be written in terms of a symmetric tensor-spinor $\psi_{\mu\nu} \equiv \phi^{\S=5/2} $ and an auxiliary spinor $\lambda$,
\begin{multline}
\label{spin5o2minlagrangian}
e^{-1}\cL_{\rm min} = \bar\psi_{\mu \nu}( i \slashed{\nabla} - m) \psi^{\mu \nu}
+ 2 \bar\psi_{\mu \nu} \gamma^\nu ( i \slashed{\nabla} + m) \gamma^\rho \psi_\rho^\mu
- \frac{1}{2} \bar\psi_\mu^\mu ( i \slashed{\nabla} - m) \psi_\rho^\rho \\
~~~~~~  - (  2  \bar\psi^{\rho \mu} i \nabla_\rho \gamma^\nu \psi_{\mu \nu} + 2 \bar\psi_{\mu \nu} \gamma^\nu i \nabla_\rho \psi^{\rho \mu})
+ (  \bar\psi_\mu^\mu i \nabla_\rho \gamma_\sigma \psi^{\rho \sigma} +  \bar\psi^{\rho \sigma} \gamma_\sigma i \nabla_\rho \psi_\mu^\mu ) \\
+ m (\bar\psi_\mu^\mu \lambda + \bar\lambda \psi_\mu^\mu)
- \frac{12}{5} \bar\lambda (i \slashed{\nabla} + 3m) \lambda \, ,
\end{multline}
where $e= \sqrt{-g}$ and the covariant derivatives are
\begin{eqnarray}
\nabla_\mu \psi_{\rho \sigma} &=& \partial_\mu \psi_{\rho \sigma} + {1 \over 8} \omega^{a b}_\mu [\gamma_a,\gamma_b] \psi_{\rho \sigma} - \Gamma^\lambda_{\rho \mu} \psi_{\lambda \sigma} - \Gamma^\lambda_{\sigma \mu} \psi_{\rho \lambda} \ , \\
\nabla_\mu \lambda &=& \partial_\mu \lambda + {1 \over 8} \omega^{a b}_\mu [\gamma_a,\gamma_b] \lambda \, .
\end{eqnarray} 
This Lagrangian can be derived from the one by Singh and Hagen \SHFermion\ via a field redefinition.
The free equations of motion are given by $( i \slashed{\partial} - m ) \psi_{\mu \nu} = \gamma^\mu \psi_{\mu \nu} = \partial^\mu \psi_{\mu \nu} = \lambda = 0$. In addition, by taking $m \to 0$ in the free Lagrangian one recovers a gauge symmetry $\delta \psi_{\mu \nu} = \partial_{(\mu} \xi_{\nu)}$. Note that this is a \textit{restricted} gauge symmetry, since the spinor-vector gauge variable $\xi_{\mu}$ must satisfy the conditions $\gamma \cdot \xi = 0$. This implies a modification of the $P \cdot J = {\cal O}(m)$ condition in the gravitational current, as we will see shortly.

The propagator can be written as
\be
\label{prop5o2}
\epsilon_\mu \epsilon_\nu \Delta_{(5/2)}^{\mu \nu, \rho \sigma} \bar\epsilon_\rho \bar\epsilon_\sigma =
 \frac{i}{p^2 - m^2} \left( P_{(5/2)}+ (p^2-m^2) \Delta^{(2)} \right)\,,
\ee
where for convenience we have contracted the Lorentz indices with auxiliary polarizations, following the discussion in Section~\ref{sec2}. The first term of the propagator corresponds to the on-shell spin-$5/2$ projector that we computed in that section; we spell it out here in full detail
\def\seps{\slash \!\!\!\bar \epsilon}
\be
P_{(5/2)}\! =\!
(\slashed{p} + m) \Big(\! P_{(1)}(\epsilon,\!\bar\epsilon)^2 
- \frac{1}{5} P_{(1)}(\epsilon,\!\epsilon)P_{(1)}(\bar\epsilon,\!\bar\epsilon)\! \Big)  \!
+ \frac{2}{5} P_{(1)}(\epsilon,\!\bar\epsilon) \Big(\!\slashed{\epsilon} \!+\! \frac{\epsilon \! \cdot \! p}{m}\Big) \!  (\slashed{p} - m)  \!  \Big(\!\seps\!+\! \frac{\bar\epsilon \! \cdot \! p}{m}\Big),
\ee
where $P_{(1)}(\epsilon,\epsilon') \equiv \epsilon \cdot \epsilon' - (\epsilon \cdot p) (\epsilon' \cdot p)/m^{2} $ is the spin-$1$ projector. As expected, the propagator becomes proportional to the spin-$5/2$ projector for $p^2 = m^2$. The second term depends on the details of the Lagrangian (\ref{spin5o2minlagrangian}), and in our case it is given by
\bea
\Delta^{(2)}& =& 
\frac{\epsilon^2  \bar\epsilon^2 (\slashed{p} + 13m)}{1280 m^2}
+ \frac{\epsilon \cdot p \, \bar\epsilon \cdot p\, \slashed{\epsilon} (\slashed{p} + 5m) \seps}{80 m^4}
- \frac{\epsilon \cdot p \, \bar\epsilon^2 {\slashed\epsilon} (\slashed{p} + 9m)}{160 m^3}
+ \frac{(\bar\epsilon \cdot p)^2 \epsilon^2}{40 m^3} \nn \\
&& \null - \frac{(\bar\epsilon \cdot p)^2 \, \epsilon \cdot p {\slashed\epsilon}}{10 m^4} 
+ \big\{  \bar\epsilon \leftrightarrow \epsilon, \gamma^{\mu_1} {\cdots} \gamma^{\mu_n} \leftrightarrow \gamma^{\mu_n} {\cdots} \gamma^{\mu_1}  \big\} \, .
\eea
The $\Delta^{(2)}$ term has important implications on the current constraint: contrary to all the previous cases, not all mass divergences in the propagator come with a factor of $p^\mu$ contracting into the vertices. Hence, we might expect a modification of the $P \cdot J = {\cal O}(m)$ condition. The correct prescription turns out to be $P_\mu J^{\mu \nu} v_{P \nu} = {\cal O}(m)$, where $v_{P \nu}$ is a reference vector-spinor satisfying $\gamma \cdot v_P = 0$, recovering the Ward identity for \textit{restricted} gauge invariance in the massless limit \GravUnitarity.

Therefore, we wish to add a non-minimal term to the Lagrangian (\ref{spin5o2minlagrangian}), such that we get $P_\mu J^{\mu \nu} v_{P \nu} = {\cal O}(m)$ with $\gamma \cdot v_P = 0$. As we will see, this requires \GravUnitarity
\be \label{L_non_min_spin5o2}
\cL_\text{non-min} =  - \frac{1}{m} \sqrt{-g}\, \bar\psi_{\mu \rho} \cR^{\mu \nu \rho \sigma} \psi_{\nu \sigma}\,, 
\ee
where $\cR^{\mu \nu \rho \sigma} = R^{\mu \nu \rho \sigma} - (i/2) \gamma^5 \epsilon^{\rho \sigma \alpha \beta} {R^{\mu \nu}}_{\alpha \beta}$ is a gamma-matrix-augmented Riemann tensor, and ${R^\mu}_{\nu \rho \sigma}$ is the standard Riemann tensor. The full Lagrangian is thus given by $\cL = \cL_\text{min} + \cL_\text{non-min}$. Interestingly, the non-minimal term is analogous to the one that appeared in gauge theory, \eqn{gravitinosugralagrangian}, which is consistent with the observed double-copy structure of the three-point amplitudes.

To check that the above is the theory we want, we start by computing the three-point amplitude. Note that only the first term in \eqn{spin5o2minlagrangian} contributes, as well as the non-minimal term, since the others are zero on-shell. The result is
\be
\label{spin5o23ptA}
\cM(1\phi^{5/2},2\bar\phi^{5/2},3h) = \frac{i}{2} \varepsilon_3 \cdot p_1 \, \bar{u}_{2 \mu \kappa} \slashed{\varepsilon}_3 v_1^{\mu \kappa} + i f_3^{\kappa \lambda} \bar{u}_{2 \mu \kappa} \slashed{\varepsilon}_3 {v_1^\mu}_{\lambda} - \frac{i}{2 m} f_3^{\kappa \lambda} \bar{u}_{2 \mu \kappa} \feps_3^{ \mu \nu} v_{1 \nu \lambda}\,, 
\ee
where $\feps^{\mu \nu}_j \equiv f^{\mu \nu}_j - (i/2) \gamma^5 \epsilon^{\mu \nu \rho \sigma} f_{j \rho \sigma}$. 

Using the on-shell identity $\bar{u}_{2 \mu} \feps_3^{ \mu \nu} v_{1 \nu} \, \bep_1 {\cdot} \bep_2 \, \varepsilon_3 {\cdot} p_1 {=} - m \, \bar{u}_{2 \rho} \slashed{\varepsilon}_3 v_1^\rho \, \bep_{2 \mu} f_3^{\mu \nu} \bep_{1 \nu}$, this can be written as
\bea
\cM(1\phi^{5/2},2\bar\phi^{5/2},3h) &=&
\frac{i}{2} ( \bar{u}_{2 \rho} \slashed{\varepsilon}_3 v_1^\rho - \frac{1}{m} \bar{u}_{2 \mu} \feps_3^{ \mu \nu} v_{1 \nu} ) ( \bep_1 \cdot \bep_2 \, \varepsilon_3 \cdot p_1 + \bep_{2 \mu} f_3^{\mu \nu} \bep_{1 \nu} ) \nn \\
&=&  -{i \over 2} A(1\phi^{3/2},2\bar\phi^{3/2},3A) A(1\phi^{1},2\bar\phi^{1},3A) \,.
\eea
The right-hand side is a product of the amplitudes of \eqn{AHH_amps} for spins 1 and $3/2$, hence by double-copy the $\cM(1\phi^{5/2},2\bar\phi^{5/2},3h)$ amplitude will match \eqn{Gnima3pt}.

As we did in the spin-2 case, we can also consider a one-parameter family of Lagrangians where the non-minimal term is given by
\be
\cL_\text{non-min} =  - \frac{\alpha}{m} \sqrt{-g}\, \bar\psi_{\mu \rho} \cR^{\mu \nu \rho \sigma} \psi_{\nu \sigma}\, .
\ee
In this case, the three-point amplitude is 
\begin{equation}
\cM(1\phi^{5/2},2\bar\phi^{5/2},3h) = - {i \over 2} {\xfactor^{-2} \over m^3} [{\bf 12}]^5 \left\{ 1 + (\alpha-1) \Big( 1 - {\langle {\bf 12} \rangle \over [{\bf 12 }]}  \Big)^2 \right\} \, .
\end{equation} 
 Then we proceed to checking the $P_\mu J^{\mu \nu} v_{P \nu} = {\cal O}(m)$ property. The current is easiest written with indices contracted with an off-shell spinor-tensor $v_{P \mu \nu}$,
\begin{multline}
\label{fullcurr5o2}
J^{\mu \nu} (-P,p_2,p_3) v_{P \mu \nu} = 
- \frac{i}{2} \varepsilon_3 \cdot p_2 \, \bar{u}_2^{\mu \nu} \slashed{\varepsilon}_3 v_{P \mu \nu}
- \frac{i}{2} p_{3\mu} \varepsilon_{3 \nu} \bar{u}_2^{\mu \nu} \slashed{\varepsilon}_3 v_{P \rho}^\rho
+ i \varepsilon_3 \cdot \bar{u}_2^{\mu} \slashed{\varepsilon}_3 P \cdot v_{P \mu} \\
\hskip1.9cm + 2 i \varepsilon_3 \cdot p_2 \, \varepsilon_3 \cdot \bar{u}_2^{\mu} \gamma \cdot v_{P \mu}
- i \varepsilon_{3 \mu} \varepsilon_{3 \nu} \bar{u}_2^{\mu \nu} \gamma_\sigma v_P^{\rho \sigma} p_{2 \rho}
- i \varepsilon_3 \cdot \bar{u}_2^{\mu} \slashed{p}_3 \varepsilon_3 \cdot v_{P \mu} \\
\hskip1.1cm  + \frac{i}{2} \varepsilon_{3 \mu} \varepsilon_{3 \nu} \bar{u}_2^{\mu \nu} \slashed{p}_3 v_{P \rho}^\rho 
- i m \varepsilon_3 \cdot \bar{u}_2^{\mu} \slashed{\varepsilon}_3 \gamma \cdot v_{P \mu}
- i \varepsilon_3 \cdot \bar{u}_2^{\mu} \slashed{\varepsilon}_3 \slashed{P} \gamma \cdot v_{P \mu} \\
\hskip2.6cm- \frac{i}{2} \varepsilon_3 \cdot \bar{u}_2^{\mu} \slashed{p}_3 \slashed{\varepsilon}_3 \gamma \cdot v_{P \mu}
 + i f_{3}^{\rho \sigma} \bar{u}_{2 \mu \rho} \slashed{\varepsilon}_3 v_{P \sigma}^\mu 
 - \frac{i}{2 m} f_3^{\mu \nu} \bar{u}_{2 \mu \kappa} \feps_3^{ \kappa \lambda} v_{P \nu \lambda} \,,
\end{multline}
with $P=p_2+p_3$.
We should now check the behavior of the current contracted into its momentum, $P_\mu J^{\mu \nu} v_{P \nu}$.  Let us quote an identity that we need to use,
\be
\frac{1}{m} f_3^{ \kappa \lambda} P_\lambda \bar{u}_{2 \mu \kappa} \feps_3^{\mu \nu} = f_3^{\kappa \lambda} f_3^{\mu \nu} \bar{u}_{2 \mu \kappa} \gamma_\lambda \ ,
\ee
which is valid for on-shell $p_2$ and $p_3$. 
This identity follows from the Gram-determinant expression given in \eqn{Gram5o2}, and hence it is an identity specific to four dimensions. We get that the contracted current is
\begin{multline}
\label{}
J^{\mu \nu} (-P,p_2,p_3) P_\mu v_{P \nu} = 
- \frac{i m}{2} \varepsilon_{3 \mu} \varepsilon_{3 \nu} \bar{u}_2^{\mu \nu} p_2 \cdot v_P
- \frac{i m}{4} \varepsilon_3 \cdot \bar{u}_2^{\mu} \slashed{\varepsilon}_3 (\slashed{p}_3{-}\, 2m) v_{P \mu}\,,
\end{multline}
provided that the residual off-shell vector-spinor is gamma-tracless, $\gamma \cdot v_P = 0$. Therefore the theory satisfies the required current constraint.

\begin{figure}[t]
	\begin{center}
		\includegraphics[width=10cm]{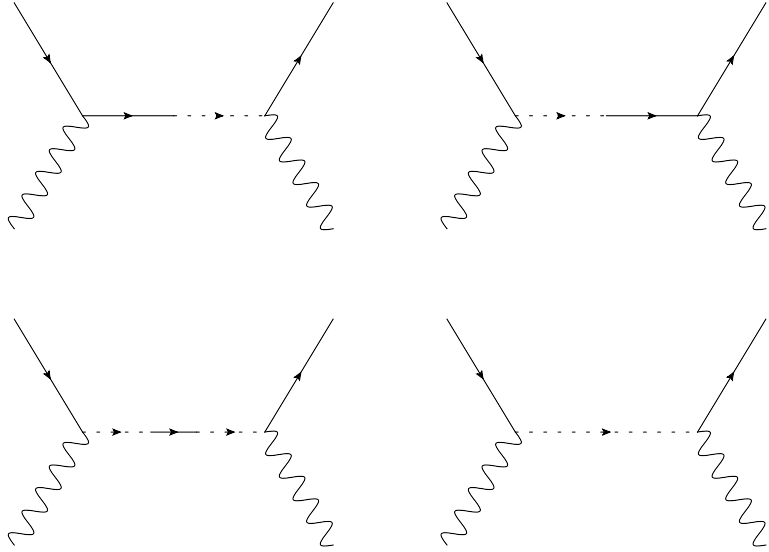
		}
	\end{center}
	\caption{\label{fig:diagram2345}Massive exchange diagrams that involve non-diagonal propagators. The solid line is the physical spin-$5/2$ field and the dotted line the auxiliary spin-$1/2$  field.}
\end{figure}

All that is left to do is to compute the Compton amplitude. We start from the massive channel diagrams. The first one is the exchange of the physical spin-$5/2$ field shown in Figure~\ref{fig:diagram1}, and it can be computed via the Feynman rules in \eqn{prop5o2} and \eqn{fullcurr5o2}. There are also diagrams involving the auxiliary field $\lambda$, shown in Figure~\ref{fig:diagram2345}. Here we need the auxiliary field propagator $\Delta_{\rm aux}$, the three-point current $J_{\rm aux}$ between the fields $\{h, \psi, \lambda\}$, and the two-point vertex $V^{(2)}_{\rm aux}$ between $\psi$ and $\lambda$.\footnote{The auxiliary-field propagator is simply the inverse of the term quadratic in $\lambda$ in the Lagrangian. This is different from \eqn{prop5o2}, which is obtained by resumming all auxiliary-field insertions.} These are given below:
\vspace{2mm}
\begin{figure}[t]
	\begin{center}
		\includegraphics[width=4.2cm]{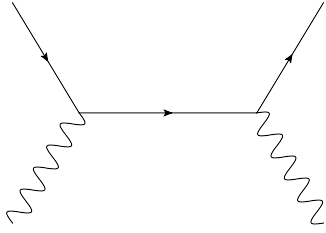}
	\end{center}
	\caption{\label{fig:diagram1} Feynman diagram with an exchange of the massive spin-$5/2$ field.}
\end{figure}

\begin{equation}
\begin{split}
\label{propaux5o2}
&\Delta_{\rm aux}(p) = - \frac{5 i}{12} \frac{\slashed{p} - 3m}{p^2 - 9m^2}\, ,\\
&J_{\rm aux}(-P,p_2,p_3) = - i m \varepsilon_{3 \mu} \varepsilon_{3 \nu} \bar{u}_2^{\mu \nu} v^{(\lambda)}_P\, ,\\
&V^{(2)}_{\rm aux}(p) = i m (\bar{u}_p)_\rho^\rho v^{(\lambda)}_p\, .
\end{split}
\end{equation}
Note that the propagator $\Delta_{\rm aux}$ introduces spurious poles; however, these cancel out in the sum of all diagrams. 

Next, we compute the graviton-exchange diagram. This is simpler, since the $\psi$ field is on-shell and the $\psi$ propagator does not appear. Aside from the usual graviton Feynman rules, we need the three-point current $J_{\rm grav}$ between the fields $\{\bar\psi, \psi, h\}$, the latter being off-shell with momentum $q^\mu = -p_1^\mu-p_2^\mu$,
\be
J_{\rm grav}(p_1,p_2,q) = 
- \frac{i}{4} \varepsilon_q \cdot (p_2 - p_1) \, \bar{u}_{2 \mu \nu} \slashed{\varepsilon}_q v_1^{\mu \nu} 
+ i f_q^{\mu \nu} \bar{u}_{2 \mu \rho} \slashed{\varepsilon}_q v_{1 \nu}^\rho
- \frac{i}{2 m} f_q^{\mu \nu} \bar{u}_{2 \mu \kappa} \feps_q^{ \kappa \lambda} v_{1 \nu \lambda} \,.
\ee

Finally, we need to include the four-point contact diagram, which for our purpose is easy to do by requiring that we add local terms which restore gauge invariance of the full amplitude. This is possible to do without mixing in new operators into the contact terms, because in the current Lagrangian there is only one explicit power of $1/m$ in the vertices, whereas the simplest independent contact operator would be of the schematic form $\bar\psi R^2 \psi/m^{3}$. 
Putting everything together, we obtain the amplitude in \eqn{Gfinalcompton5/2} for opposite-helicity gravitons, and \eqn{Gcomptonnima} in the same-helicity case. {Note that \eqn{Gfinalcompton5/2} is ${\cal O}(m^{-6})$ in the $m \to 0$ limit, whereas $ \bar\psi R^2 \psi/m^{3}$ is ${\cal O}(m^{-7})$, hence, as before, the theory with no $R^2$ contributions has the smallest mass singularity}.

\vspace{2mm}

The final result is best presented in terms of spinor-helicity notation, and it exactly gives the two formulae already quoted in eqs.~(\ref{equalHelGrav}) and (\ref{Gfinalcompton5/2}) for the equal- and opposite-helicity cases, respectively.  This confirms that the equal-helicity amplitude is not modified by the contact term that cures the spurious-pole problem of the opposite-helicity Compton amplitude. Also, by the combined analysis of Lagrangian properties (current constraint and derivative counting) and the spinor-helicity amplitude properties (little-group weights and $1/m$ scaling), we can rule our the presence of $R^2$ operators in a well-behaved and hairless spin-$5/2$ gauge theory. By ``hairless" here we mean that there are no physical properties (e.g. Wilson coefficients) that characterize the spin-$5/2$ particle other than its mass, spin and (zero) charge. We conclude that this Lagrangian is well suited for describing a Kerr black hole of spin~$\S=5/2$.

\subsection{Compton amplitude revisited}
\label{secRevisit}

Having completed the analysis for Lagrangians and currents, we revisit the construction of the opposite-helicity Compton amplitudes (\ref{finalcompton3/2}) and (\ref{Gfinalcompton5/2}). Based on their specific forms one can infer that there is an alternative route for finding these.  Let us start with spin-$3/2$ gauge theory. As discussed, the candidate expression (\ref{comptonnima}) has a simple pole that is unphysical, and one can try to proceed naively by subtracting its residue from the original expression.  Of course, whether or not this works, depends on the details of how we choose to write the residue. 

Omitting some overall factors, and considering general spin, we may write the relevant pieces of the residue as
\begin{equation}
\label{spurres}
\lim_{\Cspur \to 0} \frac{\MSHquada^{2\S}}{t_{13} t_{14}} = 
\begin{cases} 
      m^{-2-2\S} s_{12}^{\S-1} \MSHquart^\S  & \S~\text{integer}\, , \\
      m^{-2-2\S} s_{12}^{\S-3/2} m \MSHquada \MSHquart^{\S-1/2} & \S~\text{half-integer}\,,
\end{cases}
\end{equation}
where the $N_i$ variables are defined in \eqn{Ndefs}.
Since the spin-3/2 amplitude only has a simple pole, we can remove it by a single subtraction, and get the Compton amplitude
\begin{equation}
\label{fixedcompton3/2}
A(1\phi^{3/2},2\bar\phi^{3/2},3A^-,4A^+) =- i \frac{\MSHquada}{\Cspur} \left( \frac{\MSHquada^{2}}{t_{13} t_{14}} - \frac{\MSHquart}{m^4} \right) .
\end{equation}
This expression is now spurious-pole-free (although not manifestly), and by construction it matches \eqn{AHH_amps} on the physical factorization channels, and it has at most a mass singularity $1/m^4$. Earlier, we have argued that \eqn{finalcompton3/2} is the unique amplitude with these properties, and indeed a simple check confirms that the two expressions are equivalent.

We can repeat the same procedure in the case of spin-$5/2$ gravity, since it also has a simple pole that is unphysical. Recycling \eqn{spurres} for gravity, we construct a new Compton amplitude that is spurious-pole-free,
\begin{equation}
\label{Gfixedcompton5/2}
M(1\phi^{5/2},2\bar\phi^{5/2},3h^-,4h^+) = i \frac{\MSHquada}{\Cspur} \left( \frac{\MSHquada^{4}}{s_{12} t_{13} t_{14}} - \frac{\MSHquart^2}{m^6} \right) .
\end{equation}
By construction this expression matches \eqn{Gnima3pt} on the factorization channels, and it has a mass singularity of at most $1/m^6$. These properties are also true for the amplitude (\ref{Gfinalcompton5/2}), and this uniquely identifies the two amplitudes. Indeed, a simple check confirms that the two expressions are equivalent.

While the amplitudes (\ref{fixedcompton3/2}) and (\ref{Gfixedcompton5/2}) are not related by the usual four-point KLT formula~\cite{Kawai:1985xq}, is interesting to note that there is a clearly visible echo of the double copy in this way of writing the amplitudes. In fact, before subtracting out the $N_4/m^4$ and $N_4^2/m^6$ terms, the two amplitudes are exactly related by the four-point  KLT formula. However, the subtraction terms spoils this relation (see ref.~\HenrikAlex~for a detailed discussion on this issue). Since the subtraction terms are also related by a squaring procedure, it may suggest the existence of a modified double-copy formula that perhaps can be generalized to other higher-spin Compton amplitudes. 

The expressions obtained by a direct subtraction of the simple spurious pole are very compact, not much more complicated than the BCFW-constructed original ones~\Nima. It suggest that direct subtraction may be a useful approach; and indeed, it has been previously used in ref.~\cite{Chung:2018kqs} for general spin, and in ref.~\cite{Falkowski:2020aso} for spin-5/2 gravity. We find agreement with the spin-5/2 Compton amplitude in ref.~\cite{Falkowski:2020aso}, but not with the corresponding Compton amplitudes of ref.~\cite{Chung:2018kqs}.  We believe that the current condition $P \cdot J = {\cal O}(m)$ is a powerful approach with the potential to overcome ambiguities related to subtraction schemes. Without using the three-point amplitudes (\ref{AHH_amps}) and (\ref{Gnima3pt}) as input, it provided an independent derivation of those results up to spin-3/2 gauge theory and spin-5/2 gravity.  Also the new insights presented in this paper establish a link to known theories in the higher-spin literature. Since we have the Lagrangian descriptions, it is easy to directly compute the covariant amplitudes and show that they match the equal-helicity cases \eqn{comptonnima} and \eqn{Gcomptonnima}, as well as the opposite-helicity amplitudes given by the above subtractions.

\section{Conclusion}

In this work, we used spinor-helicity methods as well as Lagrangian approaches to analyze amplitudes and currents of massive spin-$\S$ particles in gauge theory and gravity. Physically-motivated conditions on the higher-spin currents uniquely fix the three-point and Compton amplitudes up to spin~$3/2$ in gauge theory and up to spin~$5/2$ in gravity. The three-point amplitudes agree with those of  Arkani-Hamed, Huang and Huang~\Nima, and the Compton amplitudes we obtain extend their results by eliminating certain spurious poles. By our combined on-shell and off-shell approach, we learn that the natural higher-spin amplitudes from ref.~\Nima, which are believed to describe Kerr black holes~\cite{Guevara:2018wpp,Chung:2018kqs,Arkani-Hamed:2019ymq, Aoude:2020onz}, are constrained to be compatible with tree-level unitarity arguments.

Tree-level unitarity considerations are important in the higher-spin literature~\cite{Ferrara:1992yc,Porrati:1993in}, where Lagrangians are constrained by demanding that off-shell currents obey
\begin{equation}
\partial_\mu J^\mu \Big|_{\rm traceless} = {\cal O} (m) \, .  
\end{equation}
We used this condition together with standard physical assumptions on higher-derivative operators to show that Lagrangians are uniquely determined up to spin~$3/2$ in the gauge theory and spin~$5/2$ in gravity. Crucially, the Lagrangians for the higher-spin fields are assumed to contain the lowest number of derivatives for which there exists a solution to the current constraint. This implies that there is little room for new higher-derivative operators to contribute at higher multiplicity, since the derivative counting prevents them from mixing with the terms already included. 

The restriction on the number of derivatives rules out, for example, interaction terms proportional to $F^2$ in spin~$\S \le 3/2$ gauge theory, as well as $R^2$ terms in spin~$\S\le 5/2$ gravity. Such interaction terms would correspond to independent contact terms at four points, and, if included in the Lagrangians, they would likely come with their own independent Wilson coefficient. In the situation in which we aim at describing objects whose only invariant attributes are mass, spin and charge, the expectation is that such Wilson coefficients are not needed. For example, recently it has been shown that Kerr black holes are best described by a vanishing Love number~\cite{Poisson:2020mdi,Chia:2020yla,LeTiec:2020bos,Goldberger:2020fot}, which controls the tidal deformability via $R^2$ interactions. While the details may be different, this result is in line  with our general conclusions. 

We note that the Compton amplitudes and Lagrangians up to spin~$2$ in gravity and spin~$1$ in gauge theory correspond to familiar theories. They can all be obtained from the corresponding minimally-coupled massless five-dimensional theories, by compactifying to four dimensions and keeping the lowest massive Kaluza-Klein mode as well as the graviton or the photon (in the gravity or gauge-theory case, respectively). We illustrated this in detail for the case of massive spin~$2$, which gives a Kaluza-Klein graviton described by a non-minimally coupled Lagrangian. The truncation is straightforward for tree-level amplitudes with two massive and $n{-}2$ massless states. As a further practical simplification, massive spin-$2$ and spin-$3/2$ gravity theories can be obtained as double copies of spin-$1$ gauge theory (spontaneously-broken Yang-Mills theory) with itself~\cite{Chiodaroli:2015rdg} or with massive spin-$1/2$ gauge theory~\cite{Chiodaroli:2017ehv,Chiodaroli:2018dbu}, respectively. The same holds for lower spin, massive spin-$1$ gravity can, for example, be obtained via the square of QCD~\cite{Johansson:2015oia}. See also refs.~\cite{Johansson:2019dnu,Bautista:2019evw}, where the double-copy connection to low-spin Kerr black holes was first discussed in some detail. Altogether, this suggests an efficient route for obtaining massive spin~$\S\le2$ tree amplitudes with any number of gravitons: Recycle the corresponding massless spin~$\S\le1$ gauge theory amplitudes and use the double copy together with compact momentum for the mass parameter; see e.g. refs.~\cite{Edison:2020ehu,Bjerrum-Bohr:2020syg} for such calculations.

\begin{table}
\begin{center}
\begin{align}
\begin{array}{|>{\centering $}p{3cm}<{$}|>{\centering $}p{2.2cm}<{$}|>{\centering $}p{2.2cm}<{$}|>{\centering $}p{2.2cm}<{$}|>{\centering\arraybackslash $}p{2cm}<{$}|}
\hline
\text{theory} & \text{coupling}& \text{Compton}  & \text{Lagrangian}  & P \cdot J   \\ \hline \hline
\multirow{3}{*}{spin-3/2 gauge} &  \text{min} &m^{-4} &1 &  1  \\ \cline{2-2}  \cline{3-3} \cline{4-4} \cline{5-5} 
&\text{generic}& m^{-6} & m^{-1} & 1 \\  \cline{2-2}  \cline{3-3} \cline{4-4} \cline{5-5} 
&\text{special}& m^{-4} & m^{-1} & m  \\ \hline \hline
\multirow{3}{*}{spin-5/2 gravity} &  \text{min} &m^{-6} &1 &  1  \\ \cline{2-2}  \cline{3-3} \cline{4-4} \cline{5-5} 
&\text{generic}& m^{-8} & m^{-1} & 1 \\  \cline{2-2}  \cline{3-3} \cline{4-4} \cline{5-5} 
&\text{special}& m^{-6} & m^{-1} & m  \\ \hline
\end{array}\nn
\end{align}
\caption{\label{masstable} Details of the $m \to 0$ scaling for the parallel cases of spin-$3/2$ gauge theory and spin-$5/2$ gravity. Minimal coupling is compared with generic or special non-minimal couplings linear in the field strength. The special theories satisfy the current constraint, with non-minimal couplings~(\ref{gravitinosugralagrangian}) and (\ref{L_non_min_spin5o2}).
Note that the longitudinal components of the polarizations dominate the behavior of the Compton amplitudes.}
\end{center}
\end{table}

Let us further elaborate on the validity of our four-point results for $\S= 5/2$ gravity and $\S=3/2$ gauge theory. As mentioned, the current-constraint method permits us to resolve the contact-term ambiguity~\Nima~in the opposite-helicity Compton amplitude, giving a unique result up to spin~$5/2$. Furthermore, our covariant amplitude exactly reproduces the spurious-pole-free equal-helicity Compton amplitude that was obtained through BCFW on-shell recursion in ref. \HenrikAlex, and thus this is a strong check that our approach is consistent. As a further supporting argument, we note that the same corrected Compton amplitudes have recently appeared in the appendix of ref.~\cite{Falkowski:2020aso}, where an approach was used that relied on improving the amplitude behavior in the $m \rightarrow 0$ limit. While this approach can naively be suspected to be related to the current constraint method, there is a clear distinction if we look at the general behavior of different deformed Lagrangian theories. Table~\ref{masstable} summarizes the interesting $m \rightarrow 0$ behavior of some of quantities that we have considered.  It is interesting to note that the minimally-coupled theories and the special theories (i.e.~the theories studied in this work) have equally good behavior in the $m \rightarrow 0$ limit for the Compton amplitude, whereas a generic deformation linear in the field strength has worse behavior. At the Lagrangian level, the special theories look worse than the minimally-coupled theories, but the special  theories shine when it comes to the current constraint, which is necessary for tree-level unitarity.

Beyond $\S>5/2$ gravity (and $\S>3/2$ gauge theory), our method constrains the tree-point amplitudes such that they are compatible with the ones in ref.~\Nima; however, they are no longer unique. One can, of course, impose by hand agreement at three points, but the non-uniqueness hints that the problem of fixing contact-term ambiguities for Compton amplitudes may become harder starting with spin-$3$ gravity (and spin-$2$ gauge theory). We also observe that for these spins the straightforward derivative counting, implied by the counting for three-point amplitudes, leads to a proliferation of new higher-point higher-derivative operators, assuming that no further considerations forbid them.

In this paper, we have developed an extended version of the massive spinor-helicity formalism~\Nima,  optimized for complicated calculations involving higher-spin asymptotic states. We use auxiliary parameters to soak up the free little-group indices of the higher-spin states, which implies that any such state is an ordinary product of elementary spin-$1/2$ and spin-$1$ polarizations. The on-shell conditions (symmetric, traceless, transverse) are automatic and the amplitudes become polynomials that are easy to manipulate. With this notation, we can write down general analytic functions of the spinors and polarizations, as we demonstrate by writing down two all-spin generating functions for the covariant three-point amplitudes.

The massive spinor-helicity formalism~\Nima~has  provided both valuable insights and practical means for new calculations, which bode well for the future. The treatment of spin effects is instrumental for understanding gravitational radiation from astrophysical black holes, and applying amplitude methods to calculations is a powerful complement to more traditional approaches~\cite{Porto:2005ac,Porto:2006bt,
	Porto:2008tb,
	Porto:2008jj,
	Levi:2008nh, 
	Porto:2010tr,
	Porto:2010zg,
	Levi:2010zu,
	Levi:2011eq,
	Porto:2012as,
	Levi:2014sba, 
	Levi:2015msa, 
	Levi:2015uxa, 
	Levi:2015ixa, 
	Levi:2016ofk, 
	Maia:2017yok,
	Maia:2017gxn,
	Levi:2017oqx,
	Levi:2017kzq,
	Levi:2018nxp,
	Levi:2019kgk,
	Kalin:2019rwq,
	Kalin:2019inp,
	Levi:2020kvb,
	Levi:2020uwu, 
	Levi:2020lfn,
	Liu:2021zxr,
	Cho:2021mqw}.
	A natural next step is to use the amplitude (\ref{Gfinalcompton5/2}) for computing observables for binary black-hole scattering (or bound systems). These could in turn be compared to the EFT results in the literature that include spin effects, or to standard General Relativity considerations where the Kerr metric is systematically perturbed using the  framework of the Teukolsky equation. Alternatively, closer to the amplitudes approach, it would be interesting to compare calculations using the Lagrangians given here to those obtained through the general-spin EFT of ref. \BernSpin, which has a more straightforward classical limit. 
	
A promising direction for follow-up work is to systematically extend our work beyond spin~$5/2$.   The long-term goal of this line of research is to find a Lagrangian description for any spin~$s$, which can produce the correct tree amplitudes for a Kerr black hole scattering against $n{-}2$ gravitons. The first attempt in this direction used BCFW recursion to compute the spin-$s$ Compton amplitudes, which gives spurious poles for higher-spin particles~\Nima. The elimination of such spurious poles  introduces in turn a proliferation of contact-term ambiguities~\cite{Chung:2018kqs}, many of which may contribute to the classical limit.  
The methods presented here provide a new strategy for constraining such ambiguities.

The rationale behind the $P \cdot J = {\cal O}(m)$ constraint comes from an EFT picture where the interactions of the massive higher-spin fields are demanded to be consistent up to an improved cutoff. In this regard, it was originally introduced as a necessary condition for restoring tree-level unitarity~\cite{Ferrara:1992yc,Porrati:1993in} in the minimally-coupled theories. The constraint can be also thought of as the requirement that the higher-spin gauge symmetry of the free theories is restored in the $m \rightarrow 0$ limit, even if this limit is not expected to exist for finite-spectrum theories. It is unclear why these considerations are relevant to classical black-hole scattering. Indeed, the appropriate kinematical limit that one considers for classical black-hole  EFT calculations is very different from the one probing the high-energy properties of a subatomic theory. A better understanding of this issue may uncover a deeper connection between black holes and high-energy properties of higher-spin theories.

Since the current condition $P \cdot J = {\cal O}(m)$ is expected to be valid in any theory that respects tree-level unitarity, and certainly valid in a consistent theory of quantum gravity, it would be useful to compare to known consistent theories that have massive higher-spin excitations. An obvious candidate is string theory, which contains a multitude of different massive states of the same spin. Initial checks show that three-point amplitudes in string theory differ from \eqn{Gnima3pt} for higher spins, and thus further work is needed to study how string theory can be of assistance.  Further work is ongoing, and we hope that additional physical requirements, to be imposed on amplitudes, currents and Lagrangians, will point us in a direction that ultimately fully constrains the classical Kerr black hole scattering process.

\vspace{0.5mm}

\noindent
\textbf{Note:} during the completion of this work, an updated version of ref.~\cite{Falkowski:2020aso} appeared that contains an equivalent spin-$5/2$ Compton amplitude to the one obtained here.

\section*{Acknowledgments}

We would like to thank Paolo di Vecchia, Lucile Cangemi, Carlo Heissenberg, Yu-tin Huang, Mich\`{e}le Levi, Alexander Ochirov, Radu Roiban, Oliver Schlotterer, Justin Vines for enlightening discussions related to this work.~We also thank Gregor K\"{a}lin, Gustav Mogull and Bram Verbeek for collaboration on related topics. 
This research is supported in part by the Knut and Alice Wallenberg Foundation under grants KAW 2018.0116 ({\it From Scattering Amplitudes to Gravitational Waves}) and KAW 2018.0162, the Swedish Research Council under grant 621-2014-5722, and the Ragnar S\"{o}derberg Foundation (Swedish Foundations' Starting Grant). The work of MC is also supported by the Swedish Research Council under grant 2019-05283. Computational resources (project SNIC 2019/3-645) were provided by the Swedish National Infrastructure for Computing (SNIC) at UPPMAX, partially funded by the Swedish Research Council through grant no. 2018-05973.

\clearpage

\appendix

\section{More on spinor-helicity, conventions and identities \label{appconv}}

We work in mostly-minus signature $\eta^{\mu\nu}={\rm diag}(1,-1,-1,-1)$. Massless Weyl spinors satisfy the identities
\begin{equation}
\begin{split}
& | k ] \langle k | = k \cdot  \bar \sigma \,, \\
& | k \rangle [ k | = k \cdot \sigma \,,  \\
&k \cdot \sigma | k ] =  k \cdot \bar{\sigma} | k \rangle = 0  \,,  \\
&k^\mu = \frac{1}{2}\langle k| \sigma^\mu |k] \,,
\end{split}
\end{equation}
where $k^\mu$ is a massless momentum. 
Similarly, for massive momentum $p^2=m^2$, massive Weyl spinors satisfy the identities
\begin{equation}
\begin{split}
&p \cdot \sigma | p^a ] =  m  | p^a \rangle \,,  \\
&p \cdot \bar{\sigma} | p^a \rangle = m | p^a ] \,,  \\
& | p^a \rangle [ p_a | =  \epsilon_{a b} |  p^a \rangle [ p^b | = p \cdot \sigma\, , \\
& | p_a ]\langle p^a | = \epsilon_{b a} |  p^a ]\langle p^b | = p \cdot \bar \sigma \,,  \\
&| p_a \rangle_{\alpha} \langle p^a |^\beta =  m \delta^\beta_\alpha \,,  \\
&| p^a ]^{\dot\alpha}[p_a |_{\dot\beta} = m \delta^{\dot\alpha}_{\dot\beta}\, ,  \\
&[p^a p^b]  =- \langle p^a p^b \rangle= m \epsilon^{ab}  \,,  \\
&p^\mu = \frac{1}{2}\langle p^a| \sigma^\mu |p_a] \,,
\end{split}
\end{equation}
where $a, b=1,2$ are the $SU(2)$ little-group indices, and $\epsilon^{a b}$ is the antisymmetric Levi-Civita tensor with normalization $\epsilon^{1 2}=\epsilon_{2 1} = 1$. The indices $\alpha, \beta=1,2$,  $\dot \alpha, \dot \beta=1,2$ are left and right SU(2) indices of the Lorentz group.

Dirac spinors for particles and antiparticles can be written as
\begin{equation}
\begin{split}
u^a(p) &= \left(\begin{matrix}
|p^a\rangle   \\  | p^a]
\end{matrix}\right) \,, ~~~~
\bar u^a(p) =  \left(\begin{matrix}
 \,  \langle p^a| \,,  & \, \, -[p^a| \, \,
\end{matrix}\right)\,,
\\
v^a(p) &= \left(\begin{matrix}
  \phantom{-} |p^a\rangle   \\  -  | p^a]
\end{matrix}\right) \,, ~~~~
\bar v^a(p) =  \left(\begin{matrix}
 \,  \langle p^a| \,,  & \, \, [p^a| \, \,
\end{matrix}\right)\,.
\end{split}
\end{equation}
The Dirac spinors are related by $v^a(p) = u^a(-p)$ and $\bar v^a(p) = \bar u^a(-p)$, given that we use the standard convention for extracting an overall momentum sign out of the spinors
\be
 |{-}p^a\rangle =  |p^a\rangle \,,~~~~~~   |{-}p^a] =  - |p^a]\,.
\ee
This rule guarantees that the spinor-helicity variables behave well under global transformations (P,T) of the Lorentz group. The completeness relation for the $u^a$ and $v^a$ spinors are
\be
u_a(p) \bar u^a(p) = \slash \!\!\! p + m\,,~~~~~~ v^a(p) \bar v_a(p) = \slash \!\!\! p - m \,.
\ee 
The boldface notation for massive spinors mean that we have soaked up the little-group indices using auxiliary variables $\z_a$,
\begin{equation}
| \boldsymbol{p} ] =  | p^a ]\z_a \, ,~~~~~~| \boldsymbol{p} \rangle =| p^a \rangle \z_a \, .
\end{equation}
It allows us to define massive $(p^2=m^2)$ and massless $(k^2 = 0)$ polarization vectors as follows:
\begin{equation}
\begin{split}
\bep (p) & = \sqrt{2}\frac{| \bm{p}  \rangle [\bm{p} |}{m} \,, ~~~~~~~~~  \bep^\mu (p) =  \frac{\langle \bm{p}| \sigma^\mu | \bm{p}]}{\sqrt{2}m}\,,\\
\varepsilon^+ (k,q) &= \sqrt{2} \frac{| q  \rangle [k |}{\spa{q}.{k}} \,, ~~~~~~~~   \varepsilon_+^\mu (k,q) = \frac{\langle q|  \sigma^\mu |  k]}{\sqrt{2}\spa{q}.{k}} \,,\\
\varepsilon^- (k,q) & = \sqrt{2} \frac{| k \rangle [q |}{\spb{k}.{q}}\,, ~~~~~~~~   \varepsilon_-^\mu (k,q) = \frac{\langle k|  \sigma^\mu |  q]}{\sqrt{2}\spb{k}.{q}} \,,
\end{split}
\end{equation}
where $q$ is an arbitrary reference null vector. The polarizations are null vectors $\bep^\mu \bep_\mu = 0$, $ \varepsilon_\pm^\mu  \varepsilon^\pm_\mu=0$, and the normalization is $ \varepsilon_+^\mu  \varepsilon^-_\mu=-1$.

The Dirac spinors in boldface notation are 
\begin{equation}
\begin{split}
u(p) &= \left(\begin{matrix}
|\overline{\bm{p}}\rangle   \\   | \overline{\bm{p}}]
\end{matrix}\right) \,, ~~~~
\bar u(p) =  \left(\begin{matrix}
 \,  \langle \bm{p} | \,,  & \, \, - [ \bm{p} | \, \,
\end{matrix}\right)\,,
\\
v(p) &= \left(\begin{matrix}
  | \bm{p} \rangle   \\ -  | \bm{p}]
\end{matrix}\right) \,, ~~~~
\bar v(p) =  \left(\begin{matrix}
 \,  \langle \overline{\bm{p}}| \,,  & \, \, [\overline{\bm{p}}| \, \,
\end{matrix}\right)\,,
\end{split}
\end{equation}
where the bared Weyl spinors are defined as $| \overline{\bm{p}} \rangle = | p^a \rangle \bar \z_a$, $| \overline{\bm{p}}] = | p^a ]\bar \z_a$. For real momentum, with $E>0$ and $m^2>0$, they are related to the unbarred spinors as $| \overline{\bm{p}}\rangle=\big([ \bm{p}|\big)^\dagger $ and $ | \overline{\bm{p}}]= -\big(\langle  \bm{p}| \big)^\dagger$. The auxiliary variables satisfy  $(\z_a)^* = \bar \z^a$, $(\z^a)^* = - \bar \z_a $ under complex conjugation.
Together this implies that the Dirac conjugates are related in the standard way, $\bar u =u^\dagger\gamma^0$ and $\bar v =v^\dagger\gamma^0$.  

Note that the $\z$ variables simply describe outgoing particles, and $\bar \z$ variables describe incoming particles. If amplitudes are computed in the out-out formalism, they will only be functions of $\z$'s. Conversely, if the in-in formalism is used, the amplitudes will be functions of only $\bar \z$'s. We use the former, hence only $\z$-dependent spinor-helicity variables appear in the amplitudes of this paper. Note that this interpretation explains why all inner products between identical massive states are zero, e.g. $\spa{\bm{p}}.{\bm{p}}=0$, $\spb{\bm{p}}.{\bm{p}}=0$ and $\bep^\mu \bep_\mu = 0$. Since the massive states considered in this paper are complex (charged) they can only have a non-zero overlap with the CPT conjugate states, hence it follows that $\spa{\bm{p}}.{\overline{\bm{p}}} \neq 0$, $\spb{\bm{p}}.{\overline{\bm{p}}} \neq 0$ and $\bep^\mu \bar \bep_\mu  \neq 0$. 

We define the \textit{x-factor} for three-point amplitudes (with momenta satisfying $p_1^2=p_2^2=m^2$ and $(p_1+p_2)^2=p_3^2=0$) as follows:
\begin{equation}
\begin{split}
&\xfactor =  i \frac{\langle q |  p_1|3] }{m \spa{q}.{3} }  = i \frac{ m \spb{q}.{3}}{\langle 3 | p_1 | q]} ,
\\
&\frac{\sqrt{2} i}{m} \varepsilon_3^{\pm} \cdot p_1 = \xfactor^{\pm 1}\,,
\end{split}
\end{equation}
and we give the following useful formulae, valid at three-point and assuming $\varepsilon_3 = \varepsilon_3^-$:
\begin{equation}
\begin{split}
& [\bm1 3]=i \xfactor \angle{ \bm1 3 }\, ,\\
&[\bm2 3]=-i \xfactor \angle{ \bm2 3 }\, ,\\
&\xfactor \angle{\bm2 3} \angle{3 \bm1} = i m (\angle{\bm2 \bm1} - [\bm2 \bm1])\, ,\\
&\bep_2 \cdot \bep_1 =   \angle{\bm1 \bm2} [\bm2 \bm1]  / m^2\, ,\\
&f_{3 \rho \sigma} \bep_2^\rho \bep_1^\sigma =  i \xfactor^{-1} ( \angle{\bm1 \bm2} [\bm2 \bm1] + [\bm1 \bm2]^2 ) / \sqrt{2}m\, ,\\
&i \bar{u}_2 \slashed{\varepsilon_3} v_1 =  \sqrt{2} \xfactor^{-1} [\bm1 \bm2]\, .
\end{split}
\end{equation}

 When studying Compton scattering, we will use massive momenta $p_1$, $p_2$ and massless momenta $p_3$, $p_4$ with the following Mandelstam variables:
\begin{equation}
\begin{split}
s_{1 2} &= (p_1 + p_2)^2 = 2 p_3 \cdot p_4\, ,\\
t_{1 3} &= (p_1 + p_3)^2 - m^2 = 2 p_1 \cdot p_3\, ,\\
t_{1 4} &= (p_1 + p_4)^2 - m^2 = 2 p_1 \cdot p_4\,,
\end{split}
\end{equation}
which satisfy $s_{1 2}+t_{1 3} +t_{1 4}=0$.

Our gamma matrices are in the Weyl basis,
\be
\gamma^\mu = 
\left(
\begin{matrix}
~ 0 && \sigma^\mu  \\ \bar{\sigma}^\mu && 0 ~
\end{matrix}
\right)\,,
\ee
and we define the higher-rank matrices as $\gamma^{\mu_1 \mu_2 \cdots \mu_n} = \gamma^{[\mu_1} \gamma^{\mu_2} \cdots \gamma^{\mu_n]}$, where the antisymmetrization includes an $1/n!$ factor. 
 Using $\gamma^5 = i \gamma^0 \gamma^1 \gamma^2 \gamma^3 =  -i \gamma^{\mu \nu \rho \sigma}\epsilon_{\mu \nu \rho \sigma}$, we list some gamma matrix identities that were convenient in our calculations,
\begin{equation}
\begin{split}
&\gamma^{a b c} \gamma_c = 2 \gamma^{a b}\, , \\
&\gamma^{a b} \gamma_b = 3 \gamma^{a}\,, \\
&\gamma^{a b c} A_c = \gamma^{a b} \slashed{A} - 2 \gamma^{[a} A^{b]}\, , \\
&\gamma (A,B,C) \equiv \gamma^{a b c} A_a B_b C_c = \slashed{A} \slashed{B} \slashed{C} - A \cdot B \slashed{C} - B \cdot C \slashed{A} + A \cdot C \slashed{B} \,, \\
&\gamma (A,B) \equiv \gamma^{a b} A_a B_b = \slashed{A} \slashed{B} - A \cdot B\, .
\end{split}
\end{equation}
Finally, the four-dimensional fully-antisymmetric Levi-Civita tensor $\epsilon^{\mu \nu \rho \sigma}$ is normalized such that $\epsilon^{0 1 2 3} = 1$.

\section{Details on Lagrangian for massive spin-$2$ fields \label{appspin2}}

After introducing the unitarity gauge through the $\xi\rightarrow \infty$ limit, the Lagrangian for massive Kaluza-Klein gravity reduced to four dimensions can be written as follows:
\begin{eqnarray}
{\cal L} &=& 
-{3 } \nabla_\mu \bH_{\nu \rho} \nabla^\mu H^{\nu \rho} - 2 \bH_{\nu \rho} \nabla_\mu \nabla^\mu H^{\nu \rho} - 2  H_{\nu \rho}   \nabla_\mu \nabla^\mu\bH^{\nu \rho}   + 2 \nabla_\mu \bH_{\nu \rho} \nabla^\rho H^{\mu \nu} 
 \no \\
&&
\null + 4 \bH_{\nu \rho} \nabla_\mu \nabla^\rho H^{\mu \nu}  + 4  H_{ \nu \rho}  \nabla_\mu \nabla^\rho \bH^{\mu \nu} + 4  \nabla^\mu \bH_{\nu \mu} \nabla_\rho H^{\rho \nu}  - 2 \nabla_\mu \bH_{\rho}^{\rho} \nabla_\nu H^{\mu \nu} 
\no \\
&& 
\null - 2 H^{\mu \nu} \nabla_\mu \nabla_\nu \bH_{\rho}^{\rho}  -   \bH^\rho_{ \rho} \nabla_\mu \nabla_\nu H^{\mu \nu} 
- 2 \nabla_\nu \bH^{\mu \nu} \nabla_\mu  H_{\rho}^{\rho} 
- 2 \bH^{\mu \nu} \nabla_\mu \nabla_\nu H_{\rho}^{\rho}  
\no \\
&&    
\null - H^\rho_{ \rho} \nabla_\mu \nabla_\nu \bH^{\mu \nu}  + \nabla_\mu \bH^{ \nu}_\nu \nabla^\mu  H_{\rho}^{\rho} + \bH^{ \nu}_\nu  \nabla_\mu \nabla^\mu  H_{\rho}^{\rho}  +  H^{ \nu}_\nu \nabla_\mu \nabla^\mu \bH_{\rho}^{\rho} - m^2  \bH_{\mu \nu} H^{\mu \nu}  
\no \\
&&  
 \null + m^2  \bH^{ \mu}_{\mu } H^{ \nu}_\nu + 4  R^{\mu \nu \rho \sigma} \bH_{\mu \rho} H_{\nu \sigma}  - 8 R_{\mu\nu} \bH^\mu_{\rho} H^{\nu\rho} +  R_{\mu\nu} \bH^{\mu\nu} H^{\rho}_\rho +  R_{\mu\nu} \bH^{\rho}_\rho H^{\mu\nu}  \no \\
&& 
\null+ R  \bH_{\mu \nu} H^{\mu \nu}  - {1\over 2} R    \bH^{ \mu}_{\mu } H^{ \nu}_\nu + {\cal O} (a_\mu,\phi) \, , \quad 
\end{eqnarray}
where again we are including terms with at most two massive fields (relying on the fact that the other terms will not contribute to the Compton amplitude we are interested in).
This expression can be further simplified by integrating by parts and using the expression for the Riemann tensor in terms of the commutator of two covariant derivatives,
\begin{eqnarray}{\cal L} &=&  \nabla_\mu \bH_{\nu \rho} \nabla^\mu H^{\nu \rho}   - 2 \nabla_\mu \bH_{\nu \rho} \nabla^\rho H^{\mu \nu}   -   \bH^\rho_{ \rho} \nabla_\mu \nabla_\nu H^{\mu \nu}  - H^\rho_{ \rho} \nabla_\mu \nabla_\nu \bH^{\mu \nu}  \no \\
&& \null  - \nabla_\mu \bH^{ \nu}_\nu \nabla^\mu  H_{\rho}^{\rho} - m^2  \bH_{\mu \nu} H^{\mu \nu}   +  m^2  \bH^{ \mu}_{\mu } H^{ \nu}_\nu   - 4 R_{\mu\nu} \bH^\mu_{\rho}
 H^{\nu\rho} +  R_{\mu\nu} \bH^{\mu\nu} H^{\rho}_\rho \no \\
&&\null +  R_{\mu\nu} \bH^{\rho}_\rho H^{\mu\nu} + R  \bH_{\mu \nu} H^{\mu \nu}  - {1 \over 2} R  \bH^{ \mu}_{\mu } H^{ \nu}_\nu  + \text{ Total derivative} + {\cal O} (a_\mu,\phi) \, . \ \
\end{eqnarray}
Furthermore, the explicit dependence from the Ricci tensor and scalar can be rewritten as
\begin{equation}
-4 \Big(R_{\mu\nu} - {1 \over 2} g_{\mu \nu} R \Big) \Big(\bH^\mu_{\rho} H^{\nu\rho} - {1 \over 4} g^{\mu\nu} \bH_{\rho \sigma} H^{\rho \sigma} - {1 \over 4}  \bH^{\mu\nu} H^{\rho}_\rho - {1 \over 4} \bH^{\rho}_\rho H^{\mu\nu}  + {1 \over 8} g^{\mu\nu} \bH_{\rho}^\rho H^{ \sigma}_\sigma \Big) \, .
\end{equation}
This suggests that the field redefinition 
\begin{equation}
g_{\mu\nu} \rightarrow g_{\mu\nu} - 2 \bH^\mu_{\rho} H^{\nu\rho} + {1 \over 2} g^{\mu\nu} \bH_{\rho \sigma} H^{\rho \sigma} + {1 \over 2}  \bH^{\mu\nu} H^{\rho}_\rho + {1 \over 2} \bH^{\rho}_\rho H^{\mu\nu}  - {1 \over 4} g^{\mu\nu} \bH_{\rho}^\rho H^{ \sigma}_\sigma  \label{fieldredef}
\end{equation} 
will remove the dependence from such terms in the Lagrangian. Additional terms with at least quartic dependence from the massive fields are added as an effect of this redefinition, but they do not contribute to the Compton amplitude. The Lagrangian is then rewritten as
\begin{eqnarray}{\cal L} &=&  \nabla_\mu \bH_{\nu \rho} \nabla^\mu H^{\nu \rho}   - 2 \nabla_\mu \bH_{\nu \rho} \nabla^\rho H^{\mu \nu}   -  \bH^\rho_{ \rho} \nabla_\mu \nabla_\nu H^{\mu \nu}  -  H^\rho_{ \rho} \nabla_\mu \nabla_\nu \bH^{\mu \nu}  \no \\
&& \null  -\nabla_\mu \bH^{ \nu}_\nu \nabla^\mu  H_{\rho}^{\rho} - m^2  \bH_{\mu \nu} H^{\mu \nu}   +  m^2  \bH^{ \mu}_{\mu } H^{ \nu}_\nu   \, .
\end{eqnarray} 
Using the expression for the Riemann tensor in terms of the commutator of two covariant derivatives, we finally rewrite the above as 
\begin{eqnarray}{\cal L} &=& \nabla_\mu \bH_{\nu \rho} \nabla^\mu H^{\nu \rho}   - 2 \nabla_\nu \bH_{\mu}^{\nu} \nabla^\rho H^{\mu}_{ \rho}   -  \bH^\rho_{ \rho} \nabla_\mu \nabla_\nu H^{\mu \nu}  -  H^\rho_{ \rho} \nabla_\mu \nabla_\nu \bH^{\mu \nu}  \no \\
&& \null  - \nabla_\mu \bH^{ \nu}_\nu \nabla^\mu  H_{\rho}^{\rho} - m^2  \bH_{\mu \nu} H^{\mu \nu}   +  m^2  \bH^{ \mu}_{\mu } H^{ \nu}_\nu   -
2 R^{\mu \nu \rho \sigma} \bH_{\mu \rho} H_{\nu \sigma} \, ,
\end{eqnarray} 
where terms involving the Ricci tensor can be removed with a field redefinition analogous to (\ref{fieldredef}).

\section{Details on Lagrangian for massive spin-$5/2$ fields}
\label{APP:5o2lagrangian}

There are various formulations of the Lagrangian of a massive spin-$5/2$ particle. Here we adopt the one in \SHFermion, in terms of a gamma-traceless symmetric tensor-spinor $\xi_{\mu\nu}$, a gamma-traceless vector-spinor $\chi_\mu$ and two auxiliary spinors $\Omega$ and $\lambda$:
\begin{multline}
\label{spin5o2freelagrangian}
\cL = \bar\xi_{\mu \nu}( i \slashed{\partial} - m) \xi^{\mu \nu}
- \frac{8}{5} (  \bar{\chi}^\mu (i \partial \cdot \xi)_\mu + (\bar{\xi} \cdot i \partial)^\mu \chi_\mu  )
+ \frac{8}{5} \bar{\chi}^\mu ( i \slashed{\partial} + \frac{3}{2} m) \chi_\mu \\
+ \frac{144}{25} \{
	\bar{\Omega} i \partial \cdot \chi + \bar{\chi} \cdot i \partial \Omega
	+  \bar{\Omega} ( - i \slashed{\partial} + 3 m) \Omega
	+ \frac{5}{3} m (  \bar{\Omega} \lambda + \bar{\lambda} \Omega  )
	- \frac{5}{3} \bar{\lambda} ( i \slashed{\partial} + 3 m) \lambda  
\} \, .
\end{multline}
In the $m \to 0$ limit, the above reduces to a massless spin-$5/2$ field, with gauge symmetry $\delta \xi_{\mu \nu} = \partial_{(\mu} \varepsilon_{\nu)} - (1/6) \gamma_{(\mu} \slashed{\partial}  \varepsilon_{\nu)}$ and $\delta \chi_\mu = (5/12) \slashed{\partial}  \varepsilon_{\mu}$. Note that this is a \textit{restricted} gauge symmetry, since the gauge variable $\varepsilon_{\mu}$ must satisfy the conditions $\gamma \cdot \varepsilon = \partial \cdot \varepsilon = 0$. This implies a modification of the $P \cdot J = {\cal O}(m)$ condition in the gravitational current, so we will come back to it later.

For convenience, we define a symmetric tensor-spinor $\psi_{\mu\nu}$ and reabsorb the fields $\xi$, $\chi$ and $\Omega$ into it. More precisely, one can write down projectors for irreducible representations of the Lorentz group:
\begin{subequations}
\begin{align}
\label{S5o2projectors}
&{(\cP_0)_{\mu \nu}}^{\rho \sigma} = \frac{1}{4} \eta_{\mu \nu} \eta^{\rho \sigma} ,\\
&{(\cP_1)_{\mu \nu}}^{\rho \sigma} = \frac{1}{3} {\eta_{(\mu}}^{(\rho} \gamma_{\nu)} \gamma^{\sigma)} - \frac{1}{12} \eta_{\mu \nu} \eta^{\rho \sigma} ,\\
&{(\cP_2)_{\mu \nu}}^{\rho \sigma} = {\eta_{(\mu}}^{(\rho} {\eta_{\nu)}}^{\sigma)} - {(\cP_0)_{\mu \nu}}^{\rho \sigma} - {(\cP_1)_{\mu \nu}}^{\rho \sigma} .
\end{align}
\end{subequations}
Then, we can set $\xi_{\mu\nu} = (\cP_2 \psi)_{\mu\nu}$, $\chi_{\mu} = (5/6) \gamma^\nu (\cP_1 \psi)_{\mu\nu}$ and $\Omega = (5/24) (\cP_0 \psi)^\mu_\mu$. The numerical factors are in principle arbitrary, but they were chosen such that the gauge transformation has now form $\delta \psi_{\mu \nu} = \partial_{(\mu} \varepsilon_{\nu)}$ with $\gamma \cdot \varepsilon = 0$. Substituting back into \ref{spin5o2freelagrangian} and rescaling $\lambda \to \lambda/2$ we obtain:
\begin{multline}
\label{spin5o2freelagrangian2}
\cL = \bar\psi_{\mu \nu}( i \slashed{\partial} - m) \psi^{\mu \nu}
+ 2 \bar\psi_{\mu \nu} \gamma^\nu ( i \slashed{\partial} + m) \gamma^\rho \psi_\rho^\mu
- \frac{1}{2} \bar\psi_\mu^\mu ( i \slashed{\partial} - m) \psi_\rho^\rho \\
- (  2  \bar\psi^{\rho \mu} i \partial_\rho \gamma^\nu \psi_{\mu \nu} + 2 \bar\psi_{\mu \nu} \gamma^\nu i \partial_\rho \psi^{\rho \mu})
+ (  \bar\psi_\mu^\mu i \partial_\rho \gamma_\sigma \psi^{\rho \sigma} +  \bar\psi^{\rho \sigma} \gamma_\sigma i \partial_\rho \psi_\mu^\mu ) \\
+ m (\bar\psi_\mu^\mu \lambda + \bar\lambda \psi_\mu^\mu)
- \frac{12}{5} \bar\lambda (i \slashed{\partial} + 3m) \lambda \, .
\end{multline}
Alternatively, we could have reabsorbed the fields $\xi$, $\chi$ and $\lambda$ into $\psi_{\mu\nu}$, leaving $\Omega$ as auxiliary field. This would give rise to a Lagrangian with an auxiliary kinetic term of form $\bar{\Omega} ( i \slashed{\partial} - 3 m) \Omega$ and an off-diagonal derivative coupling between $\Omega$ and $\psi$. A Lagrangian with similar features appears in \Berends, and we checked it is equivalent up to field redefinitions. The equations of motion from the Lagrangian \ref{spin5o2freelagrangian2} are:
\be
\label{spin5o2eoms}
( i \slashed{\partial} - m ) \psi_{\mu \nu} = \gamma^\mu \psi_{\mu \nu} = \partial^\mu \psi_{\mu \nu} = \lambda = 0 \, .
\ee
As expected, the auxiliary field vanishes on-shell and the $\psi_{\mu \nu}$ field becomes a transverse gamma-traceless symmetric tensor-spinor. The latter has six independent components, as expected from a spin-$5/2$ particle.

The Compton amplitude requires knowledge of the propagator for the $\psi$ field, which is defined as follows:
\be
\label{}
\langle \psi_{\mu \nu} (-P) \bar\psi_{\rho \sigma} (P) \rangle = \int \cD \psi \cD \bar\psi \cD \lambda \cD \bar\lambda \psi_{\mu \nu} (-P) \bar\psi_{\rho \sigma} (P) e^{i \int \cL_{\rm free}}\,  .
\ee
To compute this, we first integrate out $\lambda$ in \eqn{spin5o2freelagrangian2} to obtain an effective Lagrangian
\begin{multline}
\label{spin5o2efflagrangian}
\cL_{\rm free}^{\rm eff} =
\bar\psi_{\mu \nu}( i \slashed{\partial} - m) \psi^{\mu \nu}
+ 2 \bar\psi_{\mu \nu} \gamma^\nu ( i \slashed{\partial} + m) \gamma^\rho \psi_\rho^\mu
- \frac{1}{2} \bar\psi_\mu^\mu ( i \slashed{\partial} - m) \psi_\rho^\rho \\
- (  2  \bar\psi^{\rho \mu} i \partial_\rho \gamma^\nu \psi_{\mu \nu} + 2 \bar\psi_{\mu \nu} \gamma^\nu i \partial_\rho \psi^{\rho \mu})
+ (  \bar\psi_\mu^\mu i \partial_\rho \gamma_\sigma \psi^{\rho \sigma} +  \bar\psi^{\rho \sigma} \gamma_\sigma i \partial_\rho \psi_\mu^\mu )  \\
+ \frac{5}{12} m^2  \bar\psi_\mu^\mu \left( i \slashed{\partial} + 3m \right)^{-1} \psi_\mu^\mu \, .
\end{multline}
Now we have to invert the above expression to compute the propagator. One might worry about the nonlocality $( i \slashed{\partial} + 3m )^{-1}$ due to the kinetic term for $\lambda$, but as we have shown the equations of motion imply $\lambda = 0$ so we do not expect this to be an issue. Indeed, explicit computation reveals a valid propagator free of spurious poles. This is given by \eqn{prop5o2}.

\bibliographystyle{JHEP}
\bibliography{Compton5o2BIB}

\providecommand{\href}[2]{#2}\begingroup\raggedright\begin{thebibliography}{100}

\bibitem{LIGOScientific:2016aoc}
{\bf LIGO Scientific, Virgo} Collaboration, B.~P. Abbott et~al., {\it
  {Observation of Gravitational Waves from a Binary Black Hole Merger}},  {\em
  Phys. Rev. Lett.} {\bf 116} (2016), no.~6 061102,
  [\href{http://arxiv.org/abs/1602.03837}{{\tt arXiv:1602.03837}}].

\bibitem{LIGOScientific:2017vwq}
{\bf LIGO Scientific, Virgo} Collaboration, B.~P. Abbott et~al., {\it
  {GW170817: Observation of Gravitational Waves from a Binary Neutron Star
  Inspiral}},  {\em Phys. Rev. Lett.} {\bf 119} (2017), no.~16 161101,
  [\href{http://arxiv.org/abs/1710.05832}{{\tt arXiv:1710.05832}}].

\bibitem{BjerrumBohr:2002kt}
N.~E.~J. Bjerrum-Bohr, J.~F. Donoghue, and B.~R. Holstein, {\it Quantum
  gravitational corrections to the nonrelativistic scattering potential of two
  masses},  {\em Phys. Rev. D} {\bf 67} (2003) 084033,
  [\href{http://arxiv.org/abs/hep-th/0211072}{{\tt hep-th/0211072}}]. [Erratum:
  Phys.Rev.D 71, 069903 (2005)].

\bibitem{Bjerrum-Bohr:2013bxa}
N.~E.~J. Bjerrum-Bohr, J.~F. Donoghue, and P.~Vanhove, {\it {On-shell
  Techniques and Universal Results in Quantum Gravity}},  {\em JHEP} {\bf 02}
  (2014) 111, [\href{http://arxiv.org/abs/1309.0804}{{\tt arXiv:1309.0804}}].

\bibitem{Neill:2013wsa}
D.~Neill and I.~Z. Rothstein, {\it Classical space-times from the s matrix},
  {\em Nucl. Phys. B} {\bf 877} (2013) 177--189,
  [\href{http://arxiv.org/abs/1304.7263}{{\tt arXiv:1304.7263}}].

\bibitem{Akhoury:2013yua}
R.~Akhoury, R.~Saotome, and G.~Sterman, {\it High energy scattering in
  perturbative quantum gravity at next to leading power},  {\em Phys. Rev. D}
  {\bf 103} (2021), no.~6 064036, [\href{http://arxiv.org/abs/1308.5204}{{\tt
  arXiv:1308.5204}}].

\bibitem{Vaidya:2014kza}
V.~Vaidya, {\it {Gravitational spin Hamiltonians from the S matrix}},  {\em
  Phys. Rev. D} {\bf 91} (2015), no.~2 024017,
  [\href{http://arxiv.org/abs/1410.5348}{{\tt arXiv:1410.5348}}].

\bibitem{Luna:2016due}
A.~Luna, R.~Monteiro, I.~Nicholson, D.~O'Connell, and C.~D. White, {\it {The
  double copy: Bremsstrahlung and accelerating black holes}},  {\em JHEP} {\bf
  06} (2016) 023, [\href{http://arxiv.org/abs/1603.05737}{{\tt
  arXiv:1603.05737}}].

\bibitem{Luna:2016idw}
A.~Luna, S.~Melville, S.~G. Naculich, and C.~D. White, {\it {Next-to-soft
  corrections to high energy scattering in QCD and gravity}},  {\em JHEP} {\bf
  01} (2017) 052, [\href{http://arxiv.org/abs/1611.02172}{{\tt
  arXiv:1611.02172}}].

\bibitem{Luna:2017dtq}
A.~Luna, I.~Nicholson, D.~O'Connell, and C.~D. White, {\it {Inelastic Black
  Hole Scattering from Charged Scalar Amplitudes}},  {\em JHEP} {\bf 03} (2018)
  044, [\href{http://arxiv.org/abs/1711.03901}{{\tt arXiv:1711.03901}}].

\bibitem{Cachazo:2017jef}
F.~Cachazo and A.~Guevara, {\it Leading singularities and classical
  gravitational scattering},  {\em JHEP} {\bf 02} (2020) 181,
  [\href{http://arxiv.org/abs/1705.10262}{{\tt arXiv:1705.10262}}].

\bibitem{Damour:2017zjx}
T.~Damour, {\it High-energy gravitational scattering and the general
  relativistic two-body problem},  {\em Phys. Rev. D} {\bf 97} (2018), no.~4
  044038, [\href{http://arxiv.org/abs/1710.10599}{{\tt arXiv:1710.10599}}].

\bibitem{Bjerrum-Bohr:2018xdl}
N.~E.~J. Bjerrum-Bohr, P.~H. Damgaard, G.~Festuccia, L.~Plant\'e, and
  P.~Vanhove, {\it General relativity from scattering amplitudes},  {\em Phys.
  Rev. Lett.} {\bf 121} (2018), no.~17 171601,
  [\href{http://arxiv.org/abs/1806.04920}{{\tt arXiv:1806.04920}}].

\bibitem{Cheung:2018wkq}
C.~Cheung, I.~Z. Rothstein, and M.~P. Solon, {\it From scattering amplitudes to
  classical potentials in the post-minkowskian expansion},  {\em Phys. Rev.
  Lett.} {\bf 121} (2018), no.~25 251101,
  [\href{http://arxiv.org/abs/1808.02489}{{\tt arXiv:1808.02489}}].

\bibitem{Bern:2019nnu}
Z.~Bern, C.~Cheung, R.~Roiban, C.-H. Shen, M.~P. Solon, and M.~Zeng, {\it
  Scattering amplitudes and the conservative hamiltonian for binary systems at
  third post-minkowskian order},  {\em Phys. Rev. Lett.} {\bf 122} (2019),
  no.~20 201603, [\href{http://arxiv.org/abs/1901.04424}{{\tt
  arXiv:1901.04424}}].

\bibitem{KoemansCollado:2019ggb}
A.~Koemans~Collado, P.~Di~Vecchia, and R.~Russo, {\it Revisiting the second
  post-minkowskian eikonal and the dynamics of binary black holes},  {\em Phys.
  Rev. D} {\bf 100} (2019), no.~6 066028,
  [\href{http://arxiv.org/abs/1904.02667}{{\tt arXiv:1904.02667}}].

\bibitem{Arkani-Hamed:2019ymq}
N.~Arkani-Hamed, Y.-t. Huang, and D.~O'Connell, {\it {Kerr black holes as
  elementary particles}},  {\em JHEP} {\bf 01} (2020) 046,
  [\href{http://arxiv.org/abs/1906.10100}{{\tt arXiv:1906.10100}}].

\bibitem{Cristofoli:2019neg}
A.~Cristofoli, N.~E.~J. Bjerrum-Bohr, P.~H. Damgaard, and P.~Vanhove, {\it
  {Post-Minkowskian Hamiltonians in general relativity}},  {\em Phys. Rev. D}
  {\bf 100} (2019), no.~8 084040, [\href{http://arxiv.org/abs/1906.01579}{{\tt
  arXiv:1906.01579}}].

\bibitem{Bern:2019crd}
Z.~Bern, C.~Cheung, R.~Roiban, C.-H. Shen, M.~P. Solon, and M.~Zeng, {\it Black
  hole binary dynamics from the double copy and effective theory},  {\em JHEP}
  {\bf 10} (2019) 206, [\href{http://arxiv.org/abs/1908.01493}{{\tt
  arXiv:1908.01493}}].

\bibitem{Damgaard:2019lfh}
P.~H. Damgaard, K.~Haddad, and A.~Helset, {\it {Heavy Black Hole Effective
  Theory}},  {\em JHEP} {\bf 11} (2019) 070,
  [\href{http://arxiv.org/abs/1908.10308}{{\tt arXiv:1908.10308}}].

\bibitem{DiVecchia:2019kta}
P.~Di~Vecchia, S.~G. Naculich, R.~Russo, G.~Veneziano, and C.~D. White, {\it A
  tale of two exponentiations in $ \mathcal{N} $ = 8 supergravity at subleading
  level},  {\em JHEP} {\bf 03} (2020) 173,
  [\href{http://arxiv.org/abs/1911.11716}{{\tt arXiv:1911.11716}}].

\bibitem{Damour:2019lcq}
T.~Damour, {\it {Classical and quantum scattering in post-Minkowskian
  gravity}},  {\em Phys. Rev. D} {\bf 102} (2020), no.~2 024060,
  [\href{http://arxiv.org/abs/1912.02139}{{\tt arXiv:1912.02139}}].

\bibitem{Bjerrum-Bohr:2019kec}
N.~E.~J. Bjerrum-Bohr, A.~Cristofoli, and P.~H. Damgaard, {\it
  {Post-Minkowskian Scattering Angle in Einstein Gravity}},  {\em JHEP} {\bf
  08} (2020) 038, [\href{http://arxiv.org/abs/1910.09366}{{\tt
  arXiv:1910.09366}}].

\bibitem{Bern:2020gjj}
Z.~Bern, H.~Ita, J.~Parra-Martinez, and M.~S. Ruf, {\it {Universality in the
  classical limit of massless gravitational scattering}},  {\em Phys. Rev.
  Lett.} {\bf 125} (2020), no.~3 031601,
  [\href{http://arxiv.org/abs/2002.02459}{{\tt arXiv:2002.02459}}].

\bibitem{Chung:2020rrz}
M.-Z. Chung, Y.-t. Huang, J.-W. Kim, and S.~Lee, {\it {Complete Hamiltonian for
  spinning binary systems at first post-Minkowskian order}},  {\em JHEP} {\bf
  05} (2020) 105, [\href{http://arxiv.org/abs/2003.06600}{{\tt
  arXiv:2003.06600}}].

\bibitem{Parra-Martinez:2020dzs}
J.~Parra-Martinez, M.~S. Ruf, and M.~Zeng, {\it {Extremal black hole scattering
  at $\mathcal{O}(G^3)$: graviton dominance, eikonal exponentiation, and
  differential equations}},  {\em JHEP} {\bf 11} (2020) 023,
  [\href{http://arxiv.org/abs/2005.04236}{{\tt arXiv:2005.04236}}].

\bibitem{Aoude:2020mlg}
R.~Aoude, M.-Z. Chung, Y.-t. Huang, C.~S. Machado, and M.-K. Tam, {\it {Silence
  of Binary Kerr Black Holes}},  {\em Phys. Rev. Lett.} {\bf 125} (2020),
  no.~18 181602, [\href{http://arxiv.org/abs/2007.09486}{{\tt
  arXiv:2007.09486}}].

\bibitem{Bini:2020uiq}
D.~Bini, T.~Damour, A.~Geralico, S.~Laporta, and P.~Mastrolia, {\it
  {Gravitational dynamics at $O(G^6)$: perturbative gravitational scattering
  meets experimental mathematics}},
  \href{http://arxiv.org/abs/2008.09389}{{\tt arXiv:2008.09389}}.

\bibitem{DiVecchia:2020ymx}
P.~Di~Vecchia, C.~Heissenberg, R.~Russo, and G.~Veneziano, {\it {Universality
  of ultra-relativistic gravitational scattering}},  {\em Phys. Lett. B} {\bf
  811} (2020) 135924, [\href{http://arxiv.org/abs/2008.12743}{{\tt
  arXiv:2008.12743}}].

\bibitem{Damour:2020tta}
T.~Damour, {\it {Radiative contribution to classical gravitational scattering
  at the third order in $G$}},  {\em Phys. Rev. D} {\bf 102} (2020), no.~12
  124008, [\href{http://arxiv.org/abs/2010.01641}{{\tt arXiv:2010.01641}}].

\bibitem{Bern:2020uwk}
Z.~Bern, J.~Parra-Martinez, R.~Roiban, E.~Sawyer, and C.-H. Shen, {\it {Leading
  Nonlinear Tidal Effects and Scattering Amplitudes}},
  \href{http://arxiv.org/abs/2010.08559}{{\tt arXiv:2010.08559}}.

\bibitem{Aoude:2020ygw}
R.~Aoude, K.~Haddad, and A.~Helset, {\it {Tidal effects for spinning
  particles}},  {\em JHEP} {\bf 03} (2021) 097,
  [\href{http://arxiv.org/abs/2012.05256}{{\tt arXiv:2012.05256}}].

\bibitem{AccettulliHuber:2020dal}
M.~Accettulli~Huber, A.~Brandhuber, S.~De~Angelis, and G.~Travaglini, {\it
  {From amplitudes to gravitational radiation with cubic interactions and tidal
  effects}},  {\em Phys. Rev. D} {\bf 103} (2021), no.~4 045015,
  [\href{http://arxiv.org/abs/2012.06548}{{\tt arXiv:2012.06548}}].

\bibitem{Bini:2020rzn}
D.~Bini, T.~Damour, A.~Geralico, S.~Laporta, and P.~Mastrolia, {\it
  {Gravitational scattering at the seventh order in $G$: nonlocal contribution
  at the sixth post-Newtonian accuracy}},  {\em Phys. Rev. D} {\bf 103} (2021),
  no.~4 044038, [\href{http://arxiv.org/abs/2012.12918}{{\tt
  arXiv:2012.12918}}].

\bibitem{Bern:2021dqo}
Z.~Bern, J.~Parra-Martinez, R.~Roiban, M.~S. Ruf, C.-H. Shen, M.~P. Solon, and
  M.~Zeng, {\it {Scattering Amplitudes and Conservative Binary Dynamics at
  ${\cal O}(G^4)$}},  {\em Phys. Rev. Lett.} {\bf 126} (2021), no.~17 171601,
  [\href{http://arxiv.org/abs/2101.07254}{{\tt arXiv:2101.07254}}].

\bibitem{Herrmann:2021lqe}
E.~Herrmann, J.~Parra-Martinez, M.~S. Ruf, and M.~Zeng, {\it {Gravitational
  Bremsstrahlung from Reverse Unitarity}},  {\em Phys. Rev. Lett.} {\bf 126}
  (2021), no.~20 201602, [\href{http://arxiv.org/abs/2101.07255}{{\tt
  arXiv:2101.07255}}].

\bibitem{Kosmopoulos:2021zoq}
D.~Kosmopoulos and A.~Luna, {\it {Quadratic-in-spin Hamiltonian at $
  \mathcal{O} $(G$^{2}$) from scattering amplitudes}},  {\em JHEP} {\bf 07}
  (2021) 037, [\href{http://arxiv.org/abs/2102.10137}{{\tt arXiv:2102.10137}}].

\bibitem{Bjerrum-Bohr:2021vuf}
N.~E.~J. Bjerrum-Bohr, P.~H. Damgaard, L.~Plant\'e, and P.~Vanhove, {\it
  {Classical Gravity from Loop Amplitudes}},
  \href{http://arxiv.org/abs/2104.04510}{{\tt arXiv:2104.04510}}.

\bibitem{DiVecchia:2021bdo}
P.~Di~Vecchia, C.~Heissenberg, R.~Russo, and G.~Veneziano, {\it {The Eikonal
  Approach to Gravitational Scattering and Radiation at $\mathcal O(G^3)$}},
  \href{http://arxiv.org/abs/2104.03256}{{\tt arXiv:2104.03256}}.

\bibitem{Herrmann:2021tct}
E.~Herrmann, J.~Parra-Martinez, M.~S. Ruf, and M.~Zeng, {\it {Radiative
  Classical Gravitational Observables at $\mathcal O(G^3)$ from Scattering
  Amplitudes}},  \href{http://arxiv.org/abs/2104.03957}{{\tt
  arXiv:2104.03957}}.

\bibitem{Cristofoli:2021vyo}
A.~Cristofoli, R.~Gonzo, D.~A. Kosower, and D.~O'Connell, {\it {Waveforms from
  Amplitudes}},  \href{http://arxiv.org/abs/2107.10193}{{\tt
  arXiv:2107.10193}}.

\bibitem{Bautista:2021wfy}
Y.~F. Bautista, A.~Guevara, C.~Kavanagh, and J.~Vines, {\it {From Scattering in
  Black Hole Backgrounds to Higher-Spin Amplitudes: Part I}},
  \href{http://arxiv.org/abs/2107.10179}{{\tt arXiv:2107.10179}}.

\bibitem{Chen:2021huj}
B.-T. Chen, M.-Z. Chung, Y.-t. Huang, and M.~K. Tam, {\it {Minimal spin
  deflection of Kerr-Newman and Supersymmetric black hole}},
  \href{http://arxiv.org/abs/2106.12518}{{\tt arXiv:2106.12518}}.

\bibitem{Iwa}
Y.~Iwasaki, {\it {Quantum Theory of Gravitation vs. Classical Theory*):
  Fourth-Order Potential}},  {\em Progress of Theoretical Physics} {\bf 46}
  (11, 1971) 1587--1609.

\bibitem{Oka}
H.~Okamura, T.~Ohta, T.~Kimura, and K.~Hiida, {\it {Perturbation Calculation of
  Gravitational Potentials}},  {\em Progress of Theoretical Physics} {\bf 50}
  (12, 1973) 2066--2079.

\bibitem{Donoghue:1993eb}
J.~F. Donoghue, {\it {Leading quantum correction to the Newtonian potential}},
  {\em Phys. Rev. Lett.} {\bf 72} (1994) 2996--2999,
  [\href{http://arxiv.org/abs/gr-qc/9310024}{{\tt gr-qc/9310024}}].

\bibitem{Donoghue:1994dn}
J.~F. Donoghue, {\it {General relativity as an effective field theory: The
  leading quantum corrections}},  {\em Phys. Rev. D} {\bf 50} (1994)
  3874--3888, [\href{http://arxiv.org/abs/gr-qc/9405057}{{\tt gr-qc/9405057}}].

\bibitem{Goldberger:2004jt}
W.~D. Goldberger and I.~Z. Rothstein, {\it An effective field theory of gravity
  for extended objects},  {\em Phys. Rev. D} {\bf 73} (2006) 104029,
  [\href{http://arxiv.org/abs/hep-th/0409156}{{\tt hep-th/0409156}}].

\bibitem{Goldberger:2007hy}
W.~D. Goldberger, {\it Les houches lectures on effective field theories and
  gravitational radiation},  in {\em Les Houches Summer School - Session 86:
  Particle Physics and Cosmology: The Fabric of Spacetime}, 1, 2007.
\newblock \href{http://arxiv.org/abs/hep-ph/0701129}{{\tt hep-ph/0701129}}.

\bibitem{Kol:2007bc}
B.~Kol and M.~Smolkin, {\it {Non-Relativistic Gravitation: From Newton to
  Einstein and Back}},  {\em Class. Quant. Grav.} {\bf 25} (2008) 145011,
  [\href{http://arxiv.org/abs/0712.4116}{{\tt arXiv:0712.4116}}].

\bibitem{Goldberger:2009qd}
W.~D. Goldberger and A.~Ross, {\it Gravitational radiative corrections from
  effective field theory},  {\em Phys. Rev. D} {\bf 81} (2010) 124015,
  [\href{http://arxiv.org/abs/0912.4254}{{\tt arXiv:0912.4254}}].

\bibitem{Foffa:2013qca}
S.~Foffa and R.~Sturani, {\it {Effective field theory methods to model compact
  binaries}},  {\em Class. Quant. Grav.} {\bf 31} (2014), no.~4 043001,
  [\href{http://arxiv.org/abs/1309.3474}{{\tt arXiv:1309.3474}}].

\bibitem{Porto:2016pyg}
R.~A. Porto, {\it {The effective field theorist\textquoteright{}s approach to
  gravitational dynamics}},  {\em Phys. Rept.} {\bf 633} (2016) 1--104,
  [\href{http://arxiv.org/abs/1601.04914}{{\tt arXiv:1601.04914}}].

\bibitem{Foffa:2016rgu}
S.~Foffa, P.~Mastrolia, R.~Sturani, and C.~Sturm, {\it {Effective field theory
  approach to the gravitational two-body dynamics, at fourth post-Newtonian
  order and quintic in the Newton constant}},  {\em Phys. Rev. D} {\bf 95}
  (2017), no.~10 104009, [\href{http://arxiv.org/abs/1612.00482}{{\tt
  arXiv:1612.00482}}].

\bibitem{Levi:2018nxp}
M.~Levi, {\it {Effective Field Theories of Post-Newtonian Gravity: A
  comprehensive review}},  {\em Rept. Prog. Phys.} {\bf 83} (2020), no.~7
  075901, [\href{http://arxiv.org/abs/1807.01699}{{\tt arXiv:1807.01699}}].

\bibitem{Britto:2005fq}
R.~Britto, F.~Cachazo, B.~Feng, and E.~Witten, {\it {Direct proof of tree-level
  recursion relation in Yang-Mills theory}},  {\em Phys. Rev. Lett.} {\bf 94}
  (2005) 181602, [\href{http://arxiv.org/abs/hep-th/0501052}{{\tt
  hep-th/0501052}}].

\bibitem{Bern:1994zx}
Z.~Bern, L.~J. Dixon, D.~C. Dunbar, and D.~A. Kosower, {\it {One loop n point
  gauge theory amplitudes, unitarity and collinear limits}},  {\em Nucl. Phys.
  B} {\bf 425} (1994) 217--260,
  [\href{http://arxiv.org/abs/hep-ph/9403226}{{\tt hep-ph/9403226}}].

\bibitem{Bern:1994cg}
Z.~Bern, L.~J. Dixon, D.~C. Dunbar, and D.~A. Kosower, {\it {Fusing gauge
  theory tree amplitudes into loop amplitudes}},  {\em Nucl. Phys. B} {\bf 435}
  (1995) 59--101, [\href{http://arxiv.org/abs/hep-ph/9409265}{{\tt
  hep-ph/9409265}}].

\bibitem{Bern:1995db}
Z.~Bern and A.~G. Morgan, {\it {Massive loop amplitudes from unitarity}},  {\em
  Nucl. Phys. B} {\bf 467} (1996) 479--509,
  [\href{http://arxiv.org/abs/hep-ph/9511336}{{\tt hep-ph/9511336}}].

\bibitem{Bern:1997sc}
Z.~Bern, L.~J. Dixon, and D.~A. Kosower, {\it {One loop amplitudes for e+ e- to
  four partons}},  {\em Nucl. Phys. B} {\bf 513} (1998) 3--86,
  [\href{http://arxiv.org/abs/hep-ph/9708239}{{\tt hep-ph/9708239}}].

\bibitem{Kawai:1985xq}
H.~Kawai, D.~C. Lewellen, and S.~H.~H. Tye, {\it {A Relation Between Tree
  Amplitudes of Closed and Open Strings}},  {\em Nucl. Phys. B} {\bf 269}
  (1986) 1--23.

\bibitem{Bern:2008qj}
Z.~Bern, J.~J.~M. Carrasco, and H.~Johansson, {\it {New Relations for
  Gauge-Theory Amplitudes}},  {\em Phys. Rev. D} {\bf 78} (2008) 085011,
  [\href{http://arxiv.org/abs/0805.3993}{{\tt arXiv:0805.3993}}].

\bibitem{Bern:2010ue}
Z.~Bern, J.~J.~M. Carrasco, and H.~Johansson, {\it {Perturbative Quantum
  Gravity as a Double Copy of Gauge Theory}},  {\em Phys. Rev. Lett.} {\bf 105}
  (2010) 061602, [\href{http://arxiv.org/abs/1004.0476}{{\tt
  arXiv:1004.0476}}].

\bibitem{Bern:2019prr}
Z.~Bern, J.~J. Carrasco, M.~Chiodaroli, H.~Johansson, and R.~Roiban, {\it {The
  Duality Between Color and Kinematics and its Applications}},
  \href{http://arxiv.org/abs/1909.01358}{{\tt arXiv:1909.01358}}.

\bibitem{Bini:2019nra}
D.~Bini, T.~Damour, and A.~Geralico, {\it {Novel approach to binary dynamics:
  application to the fifth post-Newtonian level}},  {\em Phys. Rev. Lett.} {\bf
  123} (2019), no.~23 231104, [\href{http://arxiv.org/abs/1909.02375}{{\tt
  arXiv:1909.02375}}].

\bibitem{Blumlein:2020znm}
J.~Bl\"umlein, A.~Maier, P.~Marquard, and G.~Sch\"afer, {\it {Testing binary
  dynamics in gravity at the sixth post-Newtonian level}},  {\em Phys. Lett. B}
  {\bf 807} (2020) 135496, [\href{http://arxiv.org/abs/2003.07145}{{\tt
  arXiv:2003.07145}}].

\bibitem{Bini:2020wpo}
D.~Bini, T.~Damour, and A.~Geralico, {\it {Binary dynamics at the fifth and
  fifth-and-a-half post-Newtonian orders}},  {\em Phys. Rev. D} {\bf 102}
  (2020), no.~2 024062, [\href{http://arxiv.org/abs/2003.11891}{{\tt
  arXiv:2003.11891}}].

\bibitem{Bini:2020nsb}
D.~Bini, T.~Damour, and A.~Geralico, {\it {Sixth post-Newtonian local-in-time
  dynamics of binary systems}},  {\em Phys. Rev. D} {\bf 102} (2020), no.~2
  024061, [\href{http://arxiv.org/abs/2004.05407}{{\tt arXiv:2004.05407}}].

\bibitem{Antonelli:2019ytb}
A.~Antonelli, A.~Buonanno, J.~Steinhoff, M.~van~de Meent, and J.~Vines, {\it
  {Energetics of two-body Hamiltonians in post-Minkowskian gravity}},  {\em
  Phys. Rev. D} {\bf 99} (2019), no.~10 104004,
  [\href{http://arxiv.org/abs/1901.07102}{{\tt arXiv:1901.07102}}].

\bibitem{Buonanno:2014aza}
A.~Buonanno and B.~S. Sathyaprakash, {\em {Sources of Gravitational Waves:
  Theory and Observations}}.
\newblock 10, 2014.
\newblock \href{http://arxiv.org/abs/1410.7832}{{\tt arXiv:1410.7832}}.

\bibitem{Porto:2005ac}
R.~A. Porto, {\it Post-newtonian corrections to the motion of spinning bodies
  in nrgr},  {\em Phys. Rev. D} {\bf 73} (2006) 104031,
  [\href{http://arxiv.org/abs/gr-qc/0511061}{{\tt gr-qc/0511061}}].

\bibitem{Porto:2006bt}
R.~A. Porto and I.~Z. Rothstein, {\it {The Hyperfine Einstein-Infeld-Hoffmann
  potential}},  {\em Phys. Rev. Lett.} {\bf 97} (2006) 021101,
  [\href{http://arxiv.org/abs/gr-qc/0604099}{{\tt gr-qc/0604099}}].

\bibitem{Porto:2008tb}
R.~A. Porto and I.~Z. Rothstein, {\it {Spin(1)Spin(2) Effects in the Motion of
  Inspiralling Compact Binaries at Third Order in the Post-Newtonian
  Expansion}},  {\em Phys. Rev. D} {\bf 78} (2008) 044012,
  [\href{http://arxiv.org/abs/0802.0720}{{\tt arXiv:0802.0720}}]. [Erratum:
  Phys.Rev.D 81, 029904 (2010)].

\bibitem{Porto:2008jj}
R.~A. Porto and I.~Z. Rothstein, {\it {Next to Leading Order Spin(1)Spin(1)
  Effects in the Motion of Inspiralling Compact Binaries}},  {\em Phys. Rev. D}
  {\bf 78} (2008) 044013, [\href{http://arxiv.org/abs/0804.0260}{{\tt
  arXiv:0804.0260}}]. [Erratum: Phys.Rev.D 81, 029905 (2010)].

\bibitem{Levi:2008nh}
M.~Levi, {\it {Next to Leading Order gravitational Spin1-Spin2 coupling with
  Kaluza-Klein reduction}},  {\em Phys. Rev. D} {\bf 82} (2010) 064029,
  [\href{http://arxiv.org/abs/0802.1508}{{\tt arXiv:0802.1508}}].

\bibitem{Porto:2010tr}
R.~A. Porto, {\it {Next to leading order spin-orbit effects in the motion of
  inspiralling compact binaries}},  {\em Class. Quant. Grav.} {\bf 27} (2010)
  205001, [\href{http://arxiv.org/abs/1005.5730}{{\tt arXiv:1005.5730}}].

\bibitem{Porto:2010zg}
R.~A. Porto, A.~Ross, and I.~Z. Rothstein, {\it {Spin induced multipole moments
  for the gravitational wave flux from binary inspirals to third Post-Newtonian
  order}},  {\em JCAP} {\bf 03} (2011) 009,
  [\href{http://arxiv.org/abs/1007.1312}{{\tt arXiv:1007.1312}}].

\bibitem{Levi:2010zu}
M.~Levi, {\it {Next to Leading Order gravitational Spin-Orbit coupling in an
  Effective Field Theory approach}},  {\em Phys. Rev. D} {\bf 82} (2010)
  104004, [\href{http://arxiv.org/abs/1006.4139}{{\tt arXiv:1006.4139}}].

\bibitem{Levi:2011eq}
M.~Levi, {\it {Binary dynamics from spin1-spin2 coupling at fourth
  post-Newtonian order}},  {\em Phys. Rev. D} {\bf 85} (2012) 064043,
  [\href{http://arxiv.org/abs/1107.4322}{{\tt arXiv:1107.4322}}].

\bibitem{Porto:2012as}
R.~A. Porto, A.~Ross, and I.~Z. Rothstein, {\it {Spin induced multipole moments
  for the gravitational wave amplitude from binary inspirals to 2.5
  Post-Newtonian order}},  {\em JCAP} {\bf 09} (2012) 028,
  [\href{http://arxiv.org/abs/1203.2962}{{\tt arXiv:1203.2962}}].

\bibitem{Levi:2014gsa}
M.~Levi and J.~Steinhoff, {\it {Leading order finite size effects with spins
  for inspiralling compact binaries}},  {\em JHEP} {\bf 06} (2015) 059,
  [\href{http://arxiv.org/abs/1410.2601}{{\tt arXiv:1410.2601}}].

\bibitem{Levi:2014sba}
M.~Levi and J.~Steinhoff, {\it {Equivalence of ADM Hamiltonian and Effective
  Field Theory approaches at next-to-next-to-leading order spin1-spin2 coupling
  of binary inspirals}},  {\em JCAP} {\bf 12} (2014) 003,
  [\href{http://arxiv.org/abs/1408.5762}{{\tt arXiv:1408.5762}}].

\bibitem{Levi:2015msa}
M.~Levi and J.~Steinhoff, {\it {Spinning gravitating objects in the effective
  field theory in the post-Newtonian scheme}},  {\em JHEP} {\bf 09} (2015) 219,
  [\href{http://arxiv.org/abs/1501.04956}{{\tt arXiv:1501.04956}}].

\bibitem{Levi:2015uxa}
M.~Levi and J.~Steinhoff, {\it {Next-to-next-to-leading order gravitational
  spin-orbit coupling via the effective field theory for spinning objects in
  the post-Newtonian scheme}},  {\em JCAP} {\bf 01} (2016) 011,
  [\href{http://arxiv.org/abs/1506.05056}{{\tt arXiv:1506.05056}}].

\bibitem{Levi:2015ixa}
M.~Levi and J.~Steinhoff, {\it {Next-to-next-to-leading order gravitational
  spin-squared potential via the effective field theory for spinning objects in
  the post-Newtonian scheme}},  {\em JCAP} {\bf 01} (2016) 008,
  [\href{http://arxiv.org/abs/1506.05794}{{\tt arXiv:1506.05794}}].

\bibitem{Levi:2016ofk}
M.~Levi and J.~Steinhoff, {\it {Complete conservative dynamics for inspiralling
  compact binaries with spins at fourth post-Newtonian order}},
  \href{http://arxiv.org/abs/1607.04252}{{\tt arXiv:1607.04252}}.

\bibitem{Maia:2017yok}
N.~T. Maia, C.~R. Galley, A.~K. Leibovich, and R.~A. Porto, {\it {Radiation
  reaction for spinning bodies in effective field theory II: Spin-spin
  effects}},  {\em Phys. Rev. D} {\bf 96} (2017), no.~8 084065,
  [\href{http://arxiv.org/abs/1705.07938}{{\tt arXiv:1705.07938}}].

\bibitem{Maia:2017gxn}
N.~T. Maia, C.~R. Galley, A.~K. Leibovich, and R.~A. Porto, {\it {Radiation
  reaction for spinning bodies in effective field theory I: Spin-orbit
  effects}},  {\em Phys. Rev. D} {\bf 96} (2017), no.~8 084064,
  [\href{http://arxiv.org/abs/1705.07934}{{\tt arXiv:1705.07934}}].

\bibitem{Levi:2017oqx}
M.~Levi, {\it {Effective Field Theory of Post-Newtonian Gravity Including
  Spins}},  in {\em {52nd Rencontres de Moriond on Gravitation}}, 5, 2017.
\newblock \href{http://arxiv.org/abs/1705.07515}{{\tt arXiv:1705.07515}}.

\bibitem{Levi:2017kzq}
M.~Levi and J.~Steinhoff, {\it {EFTofPNG: A package for high precision
  computation with the Effective Field Theory of Post-Newtonian Gravity}},
  {\em Class. Quant. Grav.} {\bf 34} (2017), no.~24 244001,
  [\href{http://arxiv.org/abs/1705.06309}{{\tt arXiv:1705.06309}}].

\bibitem{Levi:2019kgk}
M.~Levi, S.~Mougiakakos, and M.~Vieira, {\it {Gravitational cubic-in-spin
  interaction at the next-to-leading post-Newtonian order}},  {\em JHEP} {\bf
  01} (2021) 036, [\href{http://arxiv.org/abs/1912.06276}{{\tt
  arXiv:1912.06276}}].

\bibitem{Kalin:2019rwq}
G.~K\"alin and R.~A. Porto, {\it {From Boundary Data to Bound States}},  {\em
  JHEP} {\bf 01} (2020) 072, [\href{http://arxiv.org/abs/1910.03008}{{\tt
  arXiv:1910.03008}}].

\bibitem{Kalin:2019inp}
G.~K\"alin and R.~A. Porto, {\it {From boundary data to bound states. Part II.
  Scattering angle to dynamical invariants (with twist)}},  {\em JHEP} {\bf 02}
  (2020) 120, [\href{http://arxiv.org/abs/1911.09130}{{\tt arXiv:1911.09130}}].

\bibitem{Levi:2020kvb}
M.~Levi, A.~J. Mcleod, and M.~Von~Hippel, {\it {N$^3$LO gravitational
  spin-orbit coupling at order $G^4$}},
  \href{http://arxiv.org/abs/2003.02827}{{\tt arXiv:2003.02827}}.

\bibitem{Levi:2020uwu}
M.~Levi, A.~J. Mcleod, and M.~Von~Hippel, {\it {NNNLO gravitational
  quadratic-in-spin interactions at the quartic order in G}},
  \href{http://arxiv.org/abs/2003.07890}{{\tt arXiv:2003.07890}}.

\bibitem{Levi:2020lfn}
M.~Levi and F.~Teng, {\it {NLO gravitational quartic-in-spin interaction}},
  {\em JHEP} {\bf 01} (2021) 066, [\href{http://arxiv.org/abs/2008.12280}{{\tt
  arXiv:2008.12280}}].

\bibitem{Liu:2021zxr}
Z.~Liu, R.~A. Porto, and Z.~Yang, {\it {Spin Effects in the Effective Field
  Theory Approach to Post-Minkowskian Conservative Dynamics}},
  \href{http://arxiv.org/abs/2102.10059}{{\tt arXiv:2102.10059}}.

\bibitem{Cho:2021mqw}
G.~Cho, B.~Pardo, and R.~A. Porto, {\it {Gravitational radiation from
  inspiralling compact objects: Spin-spin effects completed at the
  next-to-leading post-Newtonian order}},
  \href{http://arxiv.org/abs/2103.14612}{{\tt arXiv:2103.14612}}.

\bibitem{Jakobsen:2021lvp}
G.~U. Jakobsen, G.~Mogull, J.~Plefka, and J.~Steinhoff, {\it {Gravitational
  Bremsstrahlung and Hidden Supersymmetry of Spinning Bodies}},
  \href{http://arxiv.org/abs/2106.10256}{{\tt arXiv:2106.10256}}.

\bibitem{Arkani-Hamed:2017jhn}
N.~Arkani-Hamed, T.-C. Huang, and Y.-t. Huang, {\it {Scattering Amplitudes For
  All Masses and Spins}},  \href{http://arxiv.org/abs/1709.04891}{{\tt
  arXiv:1709.04891}}.

\bibitem{Guevara:2017csg}
A.~Guevara, {\it Holomorphic classical limit for spin effects in gravitational
  and electromagnetic scattering},  {\em JHEP} {\bf 04} (2019) 033,
  [\href{http://arxiv.org/abs/1706.02314}{{\tt arXiv:1706.02314}}].

\bibitem{Vines:2017hyw}
J.~Vines, {\it {Scattering of two spinning black holes in post-Minkowskian
  gravity, to all orders in spin, and effective-one-body mappings}},  {\em
  Class. Quant. Grav.} {\bf 35} (2018), no.~8 084002,
  [\href{http://arxiv.org/abs/1709.06016}{{\tt arXiv:1709.06016}}].

\bibitem{Vines:2018gqi}
J.~Vines, J.~Steinhoff, and A.~Buonanno, {\it {Spinning-black-hole scattering
  and the test-black-hole limit at second post-Minkowskian order}},  {\em Phys.
  Rev. D} {\bf 99} (2019), no.~6 064054,
  [\href{http://arxiv.org/abs/1812.00956}{{\tt arXiv:1812.00956}}].

\bibitem{Guevara:2018wpp}
A.~Guevara, A.~Ochirov, and J.~Vines, {\it Scattering of spinning black holes
  from exponentiated soft factors},  {\em JHEP} {\bf 09} (2019) 056,
  [\href{http://arxiv.org/abs/1812.06895}{{\tt arXiv:1812.06895}}].

\bibitem{Chung:2018kqs}
M.-Z. Chung, Y.-T. Huang, J.-W. Kim, and S.~Lee, {\it The simplest massive
  s-matrix: from minimal coupling to black holes},  {\em JHEP} {\bf 04} (2019)
  156, [\href{http://arxiv.org/abs/1812.08752}{{\tt arXiv:1812.08752}}].

\bibitem{Guevara:2019fsj}
A.~Guevara, A.~Ochirov, and J.~Vines, {\it {Black-hole scattering with general
  spin directions from minimal-coupling amplitudes}},  {\em Phys. Rev. D} {\bf
  100} (2019), no.~10 104024, [\href{http://arxiv.org/abs/1906.10071}{{\tt
  arXiv:1906.10071}}].

\bibitem{Chung:2019duq}
M.-Z. Chung, Y.-T. Huang, and J.-W. Kim, {\it {Classical potential for general
  spinning bodies}},  {\em JHEP} {\bf 09} (2020) 074,
  [\href{http://arxiv.org/abs/1908.08463}{{\tt arXiv:1908.08463}}].

\bibitem{Siemonsen:2019dsu}
N.~Siemonsen and J.~Vines, {\it {Test black holes, scattering amplitudes and
  perturbations of Kerr spacetime}},  {\em Phys. Rev. D} {\bf 101} (2020),
  no.~6 064066, [\href{http://arxiv.org/abs/1909.07361}{{\tt
  arXiv:1909.07361}}].

\bibitem{Guevara:2020xjx}
A.~Guevara, B.~Maybee, A.~Ochirov, D.~O'connell, and J.~Vines, {\it {A
  worldsheet for Kerr}},  {\em JHEP} {\bf 03} (2021) 201,
  [\href{http://arxiv.org/abs/2012.11570}{{\tt arXiv:2012.11570}}].

\bibitem{Aoude:2020onz}
R.~Aoude, K.~Haddad, and A.~Helset, {\it {On-shell heavy particle effective
  theories}},  {\em JHEP} {\bf 05} (2020) 051,
  [\href{http://arxiv.org/abs/2001.09164}{{\tt arXiv:2001.09164}}].

\bibitem{Kosower:2018adc}
D.~A. Kosower, B.~Maybee, and D.~O'Connell, {\it Amplitudes, observables, and
  classical scattering},  {\em JHEP} {\bf 02} (2019) 137,
  [\href{http://arxiv.org/abs/1811.10950}{{\tt arXiv:1811.10950}}].

\bibitem{Maybee:2019jus}
B.~Maybee, D.~O'Connell, and J.~Vines, {\it Observables and amplitudes for
  spinning particles and black holes},  {\em JHEP} {\bf 12} (2019) 156,
  [\href{http://arxiv.org/abs/1906.09260}{{\tt arXiv:1906.09260}}].

\bibitem{Huang:2019cja}
Y.-T. Huang, U.~Kol, and D.~O'Connell, {\it {Double copy of electric-magnetic
  duality}},  {\em Phys. Rev. D} {\bf 102} (2020), no.~4 046005,
  [\href{http://arxiv.org/abs/1911.06318}{{\tt arXiv:1911.06318}}].

\bibitem{Chung:2019yfs}
M.-Z. Chung, Y.-T. Huang, and J.-W. Kim, {\it {Kerr-Newman stress-tensor from
  minimal coupling}},  {\em JHEP} {\bf 12} (2020) 103,
  [\href{http://arxiv.org/abs/1911.12775}{{\tt arXiv:1911.12775}}].

\bibitem{Haddad:2020tvs}
K.~Haddad and A.~Helset, {\it {The double copy for heavy particles}},  {\em
  Phys. Rev. Lett.} {\bf 125} (2020) 181603,
  [\href{http://arxiv.org/abs/2005.13897}{{\tt arXiv:2005.13897}}].

\bibitem{Emond:2020lwi}
W.~T. Emond, Y.-T. Huang, U.~Kol, N.~Moynihan, and D.~O'Connell, {\it
  {Amplitudes from Coulomb to Kerr-Taub-NUT}},
  \href{http://arxiv.org/abs/2010.07861}{{\tt arXiv:2010.07861}}.

\bibitem{Brandhuber:2021kpo}
A.~Brandhuber, G.~Chen, G.~Travaglini, and C.~Wen, {\it {A new gauge-invariant
  double copy for heavy-mass effective theory}},  {\em JHEP} {\bf 07} (2021)
  047, [\href{http://arxiv.org/abs/2104.11206}{{\tt arXiv:2104.11206}}].

\bibitem{Johansson:2019dnu}
H.~Johansson and A.~Ochirov, {\it {Double copy for massive quantum particles
  with spin}},  {\em JHEP} {\bf 09} (2019) 040,
  [\href{http://arxiv.org/abs/1906.12292}{{\tt arXiv:1906.12292}}].

\bibitem{Ferrara:1992yc}
S.~Ferrara, M.~Porrati, and V.~L. Telegdi, {\it {$g=2$ as the natural value of
  the tree-level gyromagnetic ratio of elementary particles}},  {\em Phys. Rev.
  D} {\bf 46} (1992) 3529--3537.

\bibitem{Cucchieri:1994tx}
A.~Cucchieri, M.~Porrati, and S.~Deser, {\it {Tree level unitarity constraints
  on the gravitational couplings of higher spin massive fields}},  {\em Phys.
  Rev. D} {\bf 51} (1995) 4543--4549,
  [\href{http://arxiv.org/abs/hep-th/9408073}{{\tt hep-th/9408073}}].

\bibitem{Klishevich:1997pd}
S.~M. Klishevich and Y.~M. Zinovev, {\it {On electromagnetic interaction of
  massive spin-2 particle}},  {\em Phys. Atom. Nucl.} {\bf 61} (1998)
  1527--1537, [\href{http://arxiv.org/abs/hep-th/9708150}{{\tt
  hep-th/9708150}}].

\bibitem{Giannakis:1998wi}
I.~Giannakis, J.~T. Liu, and M.~Porrati, {\it {Massive higher spin states in
  string theory and the principle of equivalence}},  {\em Phys. Rev. D} {\bf
  59} (1999) 104013, [\href{http://arxiv.org/abs/hep-th/9809142}{{\tt
  hep-th/9809142}}].

\bibitem{Buchbinder:2000ta}
I.~L. Buchbinder and V.~D. Pershin, {\it {Gravitational interaction of higher
  spin massive fields and string theory}},  in {\em {Conference on Geometrical
  Aspects of Quantum Fields}}, 4, 2000.
\newblock \href{http://arxiv.org/abs/hep-th/0009026}{{\tt hep-th/0009026}}.

\bibitem{Buchbinder:1999be}
I.~L. Buchbinder, V.~A. Krykhtin, and V.~D. Pershin, {\it {On consistent
  equations for massive spin two field coupled to gravity in string theory}},
  {\em Phys. Lett. B} {\bf 466} (1999) 216--226,
  [\href{http://arxiv.org/abs/hep-th/9908028}{{\tt hep-th/9908028}}].

\bibitem{Deser:2001dt}
S.~Deser and A.~Waldron, {\it {Inconsistencies of massive charged gravitating
  higher spins}},  {\em Nucl. Phys. B} {\bf 631} (2002) 369--387,
  [\href{http://arxiv.org/abs/hep-th/0112182}{{\tt hep-th/0112182}}].

\bibitem{Metsaev:2007rn}
R.~R. Metsaev, {\it {Cubic interaction vertices for fermionic and bosonic
  arbitrary spin fields}},  {\em Nucl. Phys. B} {\bf 859} (2012) 13--69,
  [\href{http://arxiv.org/abs/0712.3526}{{\tt arXiv:0712.3526}}].

\bibitem{Francia:2007ee}
D.~Francia, {\it {Geometric Lagrangians for massive higher-spin fields}},  {\em
  Nucl. Phys. B} {\bf 796} (2008) 77--122,
  [\href{http://arxiv.org/abs/0710.5378}{{\tt arXiv:0710.5378}}].

\bibitem{Porrati:2008ha}
M.~Porrati and R.~Rahman, {\it {A Model Independent Ultraviolet Cutoff for
  Theories with Charged Massive Higher Spin Fields}},  {\em Nucl. Phys. B} {\bf
  814} (2009) 370--404, [\href{http://arxiv.org/abs/0812.4254}{{\tt
  arXiv:0812.4254}}].

\bibitem{Porrati:2008gv}
M.~Porrati and R.~Rahman, {\it {Intrinsic Cutoff and Acausality for Massive
  Spin 2 Fields Coupled to Electromagnetism}},  {\em Nucl. Phys. B} {\bf 801}
  (2008) 174--186, [\href{http://arxiv.org/abs/0801.2581}{{\tt
  arXiv:0801.2581}}].

\bibitem{Francia:2008ac}
D.~Francia, {\it {Geometric massive higher spins and current exchanges}},  {\em
  Fortsch. Phys.} {\bf 56} (2008) 800--808,
  [\href{http://arxiv.org/abs/0804.2857}{{\tt arXiv:0804.2857}}].

\bibitem{Sagnotti:2010at}
A.~Sagnotti and M.~Taronna, {\it {String Lessons for Higher-Spin
  Interactions}},  {\em Nucl. Phys. B} {\bf 842} (2011) 299--361,
  [\href{http://arxiv.org/abs/1006.5242}{{\tt arXiv:1006.5242}}].

\bibitem{Hassan:2012wr}
S.~F. Hassan, A.~Schmidt-May, and M.~von Strauss, {\it {On Consistent Theories
  of Massive Spin-2 Fields Coupled to Gravity}},  {\em JHEP} {\bf 05} (2013)
  086, [\href{http://arxiv.org/abs/1208.1515}{{\tt arXiv:1208.1515}}].

\bibitem{Buchbinder:2012iz}
I.~L. Buchbinder, T.~V. Snegirev, and Y.~M. Zinoviev, {\it {Cubic interaction
  vertex of higher-spin fields with external electromagnetic field}},  {\em
  Nucl. Phys. B} {\bf 864} (2012) 694--721,
  [\href{http://arxiv.org/abs/1204.2341}{{\tt arXiv:1204.2341}}].

\bibitem{Cortese:2013lda}
I.~Cortese, R.~Rahman, and M.~Sivakumar, {\it {Consistent Non-Minimal Couplings
  of Massive Higher-Spin Particles}},  {\em Nucl. Phys. B} {\bf 879} (2014)
  143--161, [\href{http://arxiv.org/abs/1307.7710}{{\tt arXiv:1307.7710}}].

\bibitem{Bernard:2015uic}
L.~Bernard, C.~Deffayet, A.~Schmidt-May, and M.~von Strauss, {\it {Linear
  spin-2 fields in most general backgrounds}},  {\em Phys. Rev. D} {\bf 93}
  (2016), no.~8 084020, [\href{http://arxiv.org/abs/1512.03620}{{\tt
  arXiv:1512.03620}}].

\bibitem{Rahman:2015pzl}
R.~Rahman and M.~Taronna, {\it {From Higher Spins to Strings: A Primer}},
  \href{http://arxiv.org/abs/1512.07932}{{\tt arXiv:1512.07932}}.

\bibitem{Fukuma:2016rru}
M.~Fukuma, H.~Kawai, K.~Sakai, and J.~Yamamoto, {\it {Massive higher spin
  fields in curved spacetime and necessity of non-minimal couplings}},  {\em
  PTEP} {\bf 2016} (2016), no.~7 073B02,
  [\href{http://arxiv.org/abs/1605.03363}{{\tt arXiv:1605.03363}}].

\bibitem{Bonifacio:2018aon}
J.~Bonifacio and K.~Hinterbichler, {\it {Universal bound on the strong coupling
  scale of a gravitationally coupled massive spin-2 particle}},  {\em Phys.
  Rev. D} {\bf 98} (2018), no.~8 085006,
  [\href{http://arxiv.org/abs/1806.10607}{{\tt arXiv:1806.10607}}].

\bibitem{Afkhami-Jeddi:2018apj}
N.~Afkhami-Jeddi, S.~Kundu, and A.~Tajdini, {\it {A Bound on Massive Higher
  Spin Particles}},  {\em JHEP} {\bf 04} (2019) 056,
  [\href{http://arxiv.org/abs/1811.01952}{{\tt arXiv:1811.01952}}].

\bibitem{Kaplan:2020ldi}
J.~Kaplan and S.~Kundu, {\it {Closed Strings and Weak Gravity from Higher-Spin
  Causality}},  {\em JHEP} {\bf 02} (2021) 145,
  [\href{http://arxiv.org/abs/2008.05477}{{\tt arXiv:2008.05477}}].

\bibitem{Rahman:2013sta}
R.~Rahman, {\it {Higher Spin Theory - Part I}},  {\em PoS} {\bf ModaveVIII}
  (2012) 004, [\href{http://arxiv.org/abs/1307.3199}{{\tt arXiv:1307.3199}}].

\bibitem{Bern:2020buy}
Z.~Bern, A.~Luna, R.~Roiban, C.-H. Shen, and M.~Zeng, {\it {Spinning Black Hole
  Binary Dynamics, Scattering Amplitudes and Effective Field Theory}},
  \href{http://arxiv.org/abs/2005.03071}{{\tt arXiv:2005.03071}}.

\bibitem{Craig:2011ws}
N.~Craig, H.~Elvang, M.~Kiermaier, and T.~Slatyer, {\it {Massive amplitudes on
  the Coulomb branch of N=4 SYM}},  {\em JHEP} {\bf 12} (2011) 097,
  [\href{http://arxiv.org/abs/1104.2050}{{\tt arXiv:1104.2050}}].

\bibitem{Dobrev:1975ru}
V.~K. Dobrev, V.~B. Petkova, S.~G. Petrova, and I.~T. Todorov, {\it {Dynamical
  Derivation of Vacuum Operator Product Expansion in Euclidean Conformal
  Quantum Field Theory}},  {\em Phys. Rev. D} {\bf 13} (1976) 887.

\bibitem{Chiodaroli:2017ehv}
M.~Chiodaroli, M.~Gunaydin, H.~Johansson, and R.~Roiban, {\it {Gauged
  Supergravities and Spontaneous Supersymmetry Breaking from the Double Copy
  Construction}},  {\em Phys. Rev. Lett.} {\bf 120} (2018), no.~17 171601,
  [\href{http://arxiv.org/abs/1710.08796}{{\tt arXiv:1710.08796}}].

\bibitem{Chiodaroli:2018dbu}
M.~Chiodaroli, M.~G\"unaydin, H.~Johansson, and R.~Roiban, {\it {Non-Abelian
  gauged supergravities as double copies}},  {\em JHEP} {\bf 06} (2019) 099,
  [\href{http://arxiv.org/abs/1812.10434}{{\tt arXiv:1812.10434}}].

\bibitem{Chiodaroli:2015rdg}
M.~Chiodaroli, M.~Gunaydin, H.~Johansson, and R.~Roiban, {\it {Spontaneously
  Broken Yang-Mills-Einstein Supergravities as Double Copies}},  {\em JHEP}
  {\bf 06} (2017) 064, [\href{http://arxiv.org/abs/1511.01740}{{\tt
  arXiv:1511.01740}}].

\bibitem{Momeni:2020vvr}
A.~Momeni, J.~Rumbutis, and A.~J. Tolley, {\it {Massive Gravity from Double
  Copy}},  {\em JHEP} {\bf 12} (2020) 030,
  [\href{http://arxiv.org/abs/2004.07853}{{\tt arXiv:2004.07853}}].

\bibitem{Johnson:2020pny}
L.~A. Johnson, C.~R.~T. Jones, and S.~Paranjape, {\it {Constraints on a Massive
  Double-Copy and Applications to Massive Gravity}},  {\em JHEP} {\bf 02}
  (2021) 148, [\href{http://arxiv.org/abs/2004.12948}{{\tt arXiv:2004.12948}}].

\bibitem{Momeni:2020hmc}
A.~Momeni, J.~Rumbutis, and A.~J. Tolley, {\it {Kaluza-Klein from
  Colour-Kinematics Duality for Massive Fields}},
  \href{http://arxiv.org/abs/2012.09711}{{\tt arXiv:2012.09711}}.

\bibitem{Hang:2021fmp}
Y.-F. Hang and H.-J. He, {\it {Structure of Kaluza-Klein Graviton Scattering
  Amplitudes from Gravitational Equivalence Theorem and Double-Copy}},
  \href{http://arxiv.org/abs/2106.04568}{{\tt arXiv:2106.04568}}.

\bibitem{Chi:2021mio}
H.-H. Chi, H.~Elvang, A.~Herderschee, C.~R.~T. Jones, and S.~Paranjape, {\it
  {Generalizations of the Double-Copy: the KLT Bootstrap}},
  \href{http://arxiv.org/abs/2106.12600}{{\tt arXiv:2106.12600}}.

\bibitem{Britto:2004ap}
R.~Britto, F.~Cachazo, and B.~Feng, {\it {New recursion relations for tree
  amplitudes of gluons}},  {\em Nucl. Phys. B} {\bf 715} (2005) 499--522,
  [\href{http://arxiv.org/abs/hep-th/0412308}{{\tt hep-th/0412308}}].

\bibitem{Monteiro:2014cda}
R.~Monteiro, D.~O'Connell, and C.~D. White, {\it {Black holes and the double
  copy}},  {\em JHEP} {\bf 12} (2014) 056,
  [\href{http://arxiv.org/abs/1410.0239}{{\tt arXiv:1410.0239}}].

\bibitem{Deser:2000dz}
S.~Deser, V.~Pascalutsa, and A.~Waldron, {\it {Massive spin 3/2
  electrodynamics}},  {\em Phys. Rev. D} {\bf 62} (2000) 105031,
  [\href{http://arxiv.org/abs/hep-th/0003011}{{\tt hep-th/0003011}}].

\bibitem{Rarita:1941mf}
W.~Rarita and J.~Schwinger, {\it {On a theory of particles with half integral
  spin}},  {\em Phys. Rev.} {\bf 60} (1941) 61.

\bibitem{Freedman:1976aw}
D.~Z. Freedman and A.~K. Das, {\it {Gauge Internal Symmetry in Extended
  Supergravity}},  {\em Nucl. Phys. B} {\bf 120} (1977) 221--230.

\bibitem{Johnson:1960vt}
K.~Johnson and E.~C.~G. Sudarshan, {\it {Inconsistency of the local field
  theory of charged spin 3/2 particles}},  {\em Annals Phys.} {\bf 13} (1961)
  126--145.

\bibitem{Velo:1969bt}
G.~Velo and D.~Zwanziger, {\it {Propagation and quantization of
  Rarita-Schwinger waves in an external electromagnetic potential}},  {\em
  Phys. Rev.} {\bf 186} (1969) 1337--1341.

\bibitem{Weinberg:1980kq}
S.~Weinberg and E.~Witten, {\it {Limits on Massless Particles}},  {\em Phys.
  Lett. B} {\bf 96} (1980) 59--62.

\bibitem{Das:1976ct}
A.~K. Das and D.~Z. Freedman, {\it {Gauge Quantization for Spin 3/2 Fields}},
  {\em Nucl. Phys. B} {\bf 114} (1976) 271--296.

\bibitem{Deser:1977uq}
S.~Deser and B.~Zumino, {\it {Broken Supersymmetry and Supergravity}},  {\em
  Phys. Rev. Lett.} {\bf 38} (1977) 1433--1436.

\bibitem{Arkani-Hamed:2008bsc}
N.~Arkani-Hamed and J.~Kaplan, {\it {On Tree Amplitudes in Gauge Theory and
  Gravity}},  {\em JHEP} {\bf 04} (2008) 076,
  [\href{http://arxiv.org/abs/0801.2385}{{\tt arXiv:0801.2385}}].

\bibitem{Cohen:2010mi}
T.~Cohen, H.~Elvang, and M.~Kiermaier, {\it {On-shell constructibility of tree
  amplitudes in general field theories}},  {\em JHEP} {\bf 04} (2011) 053,
  [\href{http://arxiv.org/abs/1010.0257}{{\tt arXiv:1010.0257}}].

\bibitem{Ochirov:2018uyq}
A.~Ochirov, {\it {Helicity amplitudes for QCD with massive quarks}},  {\em
  JHEP} {\bf 04} (2018) 089, [\href{http://arxiv.org/abs/1802.06730}{{\tt
  arXiv:1802.06730}}].

\bibitem{Bautista:2019evw}
Y.~F. Bautista and A.~Guevara, {\it {On the Double Copy for Spinning Matter}},
  \href{http://arxiv.org/abs/1908.11349}{{\tt arXiv:1908.11349}}.

\bibitem{Edison:2020ehu}
A.~Edison and F.~Teng, {\it {Efficient Calculation of Crossing Symmetric BCJ
  Tree Numerators}},  {\em JHEP} {\bf 12} (2020) 138,
  [\href{http://arxiv.org/abs/2005.03638}{{\tt arXiv:2005.03638}}].

\bibitem{Johansson:2015oia}
H.~Johansson and A.~Ochirov, {\it {Color-Kinematics Duality for QCD
  Amplitudes}},  {\em JHEP} {\bf 01} (2016) 170,
  [\href{http://arxiv.org/abs/1507.00332}{{\tt arXiv:1507.00332}}].

\bibitem{Freedman:1976xh}
D.~Z. Freedman, P.~van Nieuwenhuizen, and S.~Ferrara, {\it {Progress Toward a
  Theory of Supergravity}},  {\em Phys. Rev. D} {\bf 13} (1976) 3214--3218.

\bibitem{Deser:1976eh}
S.~Deser and B.~Zumino, {\it {Consistent Supergravity}},  {\em Phys. Lett. B}
  {\bf 62} (1976) 335.

\bibitem{Fierz:1939ix}
M.~Fierz and W.~Pauli, {\it {On relativistic wave equations for particles of
  arbitrary spin in an electromagnetic field}},  {\em Proc. Roy. Soc. Lond. A}
  {\bf 173} (1939) 211--232.

\bibitem{Singh:1974rc}
L.~P.~S. Singh and C.~R. Hagen, {\it {Lagrangian formulation for arbitrary
  spin. 2. The fermion case}},  {\em Phys. Rev. D} {\bf 9} (1974) 910--920.

\bibitem{Falkowski:2020aso}
A.~Falkowski and C.~S. Machado, {\it {Soft Matters, or the Recursions with
  Massive Spinors}},  \href{http://arxiv.org/abs/2005.08981}{{\tt
  arXiv:2005.08981}}.

\bibitem{Porrati:1993in}
M.~Porrati, {\it {Massive spin 5/2 fields coupled to gravity: Tree level
  unitarity versus the equivalence principle}},  {\em Phys. Lett. B} {\bf 304}
  (1993) 77--80, [\href{http://arxiv.org/abs/gr-qc/9301012}{{\tt
  gr-qc/9301012}}].

\bibitem{Poisson:2020mdi}
E.~Poisson, {\it {Gravitomagnetic Love tensor of a slowly rotating body:
  post-Newtonian theory}},  {\em Phys. Rev. D} {\bf 102} (2020), no.~6 064059,
  [\href{http://arxiv.org/abs/2007.01678}{{\tt arXiv:2007.01678}}].

\bibitem{Chia:2020yla}
H.~S. Chia, {\it {Tidal deformation and dissipation of rotating black holes}},
  {\em Phys. Rev. D} {\bf 104} (2021), no.~2 024013,
  [\href{http://arxiv.org/abs/2010.07300}{{\tt arXiv:2010.07300}}].

\bibitem{LeTiec:2020bos}
A.~Le~Tiec, M.~Casals, and E.~Franzin, {\it {Tidal Love Numbers of Kerr Black
  Holes}},  {\em Phys. Rev. D} {\bf 103} (2021), no.~8 084021,
  [\href{http://arxiv.org/abs/2010.15795}{{\tt arXiv:2010.15795}}].

\bibitem{Goldberger:2020fot}
W.~D. Goldberger, J.~Li, and I.~Z. Rothstein, {\it {Non-conservative effects on
  spinning black holes from world-line effective field theory}},  {\em JHEP}
  {\bf 06} (2021) 053, [\href{http://arxiv.org/abs/2012.14869}{{\tt
  arXiv:2012.14869}}].

\bibitem{Bjerrum-Bohr:2020syg}
N.~E.~J. Bjerrum-Bohr, T.~V. Brown, and H.~Gomez, {\it {Scattering of Gravitons
  and Spinning Massive States from Compact Numerators}},  {\em JHEP} {\bf 04}
  (2021) 234, [\href{http://arxiv.org/abs/2011.10556}{{\tt arXiv:2011.10556}}].

\bibitem{Berends:1979rv}
F.~A. Berends, J.~W. van Holten, P.~van Nieuwenhuizen, and B.~de~Wit, {\it {On
  Field Theory for Massive and Massless Spin 5/2 Particles}},  {\em Nucl. Phys.
  B} {\bf 154} (1979) 261--282.

\end{thebibliography}\endgroup

\end{document}